\documentclass[11pt,reqno]{amsart}
\usepackage[a4paper,margin=2.5cm]{geometry}
\PassOptionsToPackage{hyperfootnotes=false}{hyperref}
\usepackage{amsmath,amssymb,amsthm,mathtools,graphicx,hyperref}
\usepackage{tikz}
\usetikzlibrary{arrows.meta,decorations.pathreplacing,calc}
\graphicspath{{./}{../}}
\numberwithin{equation}{section}
\newtheorem{theorem}{Theorem}[section]
\newtheorem{corollary}{Corollary}[section]
\newtheorem{proposition}[theorem]{Proposition}
\newtheorem{lemma}[theorem]{Lemma}
\theoremstyle{definition}
\newtheorem{assumption}{Assumption}[section]
\newtheorem{rhp}{RH problem}[section]
\theoremstyle{remark}
\newtheorem{remark}[theorem]{Remark}
\newcommand{\e}{\mathrm{e}}
\newcommand{\di}{\,\mathrm{d}}
\newcommand{\C}{\mathbb{C}}
\newcommand{\R}{\mathbb{R}}
\newcommand{\cO}{\mathcal{O}}
\newcommand{\TIIm}{\mathcal T_{\rm II}^-}

\makeatletter
\renewcommand{\footnoterule}{\kern-3\p@\hrule\@width.4\columnwidth\kern2.6\p@}
\newcommand{\NoTopFloatOnThisPage}{\global\@topnum\z@}
\newcommand{\ArticleTitleFootnotes}{%
  \renewcommand{\thefootnote}{\fnsymbol{footnote}}%
  \def\@makefnmark{\rlap{\@textsuperscript{\normalfont\@thefnmark}}}%
  \long\def\@makefntext##1{\parindent 1em\noindent
    \hb@xt@1.8em{\hss\@textsuperscript{\normalfont\@thefnmark}}##1}}
\makeatother

\begin{document}

\thispagestyle{plain}
\NoTopFloatOnThisPage
\markboth{WENHAO LIU, YUFENG ZHANG, AND ZHENBO WANG}{TRANSITION REGIONS FOR THE DEFOCUSING HIROTA EQUATION}
\begingroup
\ArticleTitleFootnotes
\centering
\null\vskip 2em
{\fontsize{17.28}{22}\selectfont\bfseries PAINLEV\'E XXXIV ASYMPTOTICS FOR THE DEFOCUSING HIROTA EQUATION WITH STEP-LIKE INITIAL DATA IN TRANSITION REGIONS\par}
\vskip 22pt
{\large WENHAO LIU, YUFENG ZHANG\footnotemark, ZHENBO WANG\par}
\footnotetext{Corresponding author. {\slshape Email address}: zyfusf@163.com}
\vskip 4.8pt
{\small\slshape\raggedright
\hspace*{0.7cm}School of Mathematics, China University of Mining and Technology, 1 Daxue Road,\\[4.8pt]
\hspace*{0.7cm}Xuzhou, Jiangsu 221116, People's Republic of China\par}
\endgroup
\setcounter{footnote}{0}

\vskip 41pt
{\small
{\centering\bfseries Abstract\par}
\vskip 8pt
\begin{list}{}{\setlength{\leftmargin}{2.5em}\setlength{\rightmargin}{\leftmargin}%
  \setlength{\listparindent}{1.5em}\setlength{\itemindent}{\listparindent}%
  \setlength{\parsep}{0pt plus 1pt}\setlength{\topsep}{0pt}\setlength{\partopsep}{0pt}}
\item\relax
We study the long-time asymptotics of the solution of the Cauchy problem for the defocusing Hirota equation with real step-like initial data. Because of the term $\alpha(q_{xx}-2|q|^2q)$, the backgrounds rotate in time, and the symmetry $k\mapsto-k$ of the associated Riemann--Hilbert problem is lost. As a consequence, the branch points $\pm c_l$ and $\pm c_r$ become soft edges along different critical lines. We consider the two transition regions in which the edge $-c_r$, respectively $c_l$, becomes soft, namely the $t^{-2/3}$-neighbourhoods of the lines $x/t=-6\beta c_r^2+2\alpha c_r$ and $x/t=-6\beta c_l^2-2\alpha c_l$. By the Deift--Zhou nonlinear steepest descent method, we show that in both regions the leading correction to the rotating background is of order $t^{-1/3}$ and is a phase correction, in contrast with the order $t^{-2/3}$ known for the defocusing mKdV equation. In the first region, the correction is given by an integral of a solution of the Painlev\'e XXXIV equation. In the second region, the relevant Painlev\'e XXXIV model problem reduces to Airy functions, and the correction is a phase shift which is quadratic in $x/t$ and independent of the initial data; in addition, a stationary point of the phase function produces a contribution of order $t^{-1/2}$, which is expressed through parabolic cylinder functions.

\par\vspace{1.2em}
\end{list}
}
\vskip 1em

\section{Introduction}\label{sec:intro}
We consider the Cauchy problem for the defocusing Hirota equation
\begin{equation}\label{eq:Hirota}
iq_t+\alpha\big(q_{xx}-2|q|^2q\big)+i\beta\big(q_{xxx}-6|q|^2q_x\big)=0,\qquad x\in\R,\ t>0,
\end{equation}
where $\alpha\in\R$ and $\beta>0$, with real step-like initial data
\begin{equation}\label{eq:ID}
q(x,0)=q_0(x)\to\begin{cases}c_l, & x\to-\infty,\\ c_r, & x\to+\infty,\end{cases}\qquad c_l>c_r>0 .
\end{equation}
Equation \eqref{eq:Hirota} was introduced by Hirota \cite{Hirota73} in the course of his studies
of envelope solitons \cite{Hirota71}. It is a completely integrable extension of the nonlinear
Schr\"odinger (NLS) equation, obtained by adjoining to the group velocity dispersion and the
cubic nonlinearity a third-order dispersion term and a nonlinear term of derivative type; it
arises in the description of ultrashort pulses in optical fibers \cite{Kodama85,KH87}. Its
complete integrability, like that of the NLS and mKdV equations \cite{ZS72,AKNS73}, rests on a
Lax pair \cite{Lax68} with a rational dependence on the spectral parameter; for $\beta=0$ equation
\eqref{eq:Hirota} reduces to the defocusing NLS equation, and for $\alpha=0$ and real-valued
$q$ to the defocusing modified Korteweg--de Vries (mKdV) equation.

Since a nonzero constant is not a solution of \eqref{eq:Hirota} when $\alpha\neq0$, the natural boundary conditions for $t>0$ are the rotating plane waves
\begin{equation}\label{eq:BC}
q(x,t)\to\begin{cases}c_l\,\e^{-2i\alpha c_l^2t}, & x\to-\infty,\\ c_r\,\e^{-2i\alpha c_r^2t}, & x\to+\infty.\end{cases}
\end{equation}
If $q$ solves \eqref{eq:Hirota} with parameter $\alpha$, then $\bar q$ solves \eqref{eq:Hirota} with $\alpha$ replaced by $-\alpha$. Since $q_0$ is real-valued, uniqueness of the solution of \eqref{eq:Hirota}--\eqref{eq:ID} gives $q_{-\alpha}(x,t)=\overline{q_{\alpha}(x,t)}$. Hence it suffices to consider $\alpha>0$; the case $\alpha=0$ is the defocusing mKdV equation, which is treated in \cite{XZ,FWZ}. Throughout the paper we assume that $\alpha>0$ and $\beta>0$, and that the initial data satisfy the following conditions.

\begin{assumption}\label{ass:1}
The initial data $q_0$ is real-valued and
\begin{enumerate}
\item[(a)] $x^m(q_0-c_l)|_{\R^-}\in L^1(\R^-)$, $x^m(q_0-c_r)|_{\R^+}\in L^1(\R^+)$, $m=0,\dots,8$;
\item[(b)] $q_0\in C^4(\R)$ and $\partial_x^nq_0\in L^\infty(\R)$, $n=0,\dots,4$.
\end{enumerate}
\end{assumption}

\subsection{Background}
The long-time behaviour of solutions of integrable equations can be studied by means of the Riemann--Hilbert (RH) formulation of the inverse scattering transform \cite{BealsCoifman84,Mus} and the nonlinear steepest descent method of Deift and Zhou \cite{DZ93}; see also \cite{DeiftItsZhou93,DZ02}; for the $\bar\partial$ and
low-regularity extensions of the method we refer to \cite{DKMVZ99,DKMVZ99b}. For step-like initial data, several regions with qualitatively different behaviour, such as plane wave regions, rarefaction waves and dispersive shock waves, develop in the $(x,t)$ half-plane \cite{GurevichPitaevskii73,Khruslov76,Whitham74,El05}, and the nonlinear steepest descent method requires an appropriate $g$-function in each of them; see, e.g., \cite{BV07,Jenkins15,EGKT13,FLQ25}, and \cite{BoutetItsKotlyarov09,BoutetKotlyarovShepelsky11,BLS21,BLS25} for the focusing case. For the defocusing mKdV equation with step-like initial data, Xu and Zhang \cite{XZ} obtained the asymptotics in the main regions of the $(x,t)$ half-plane, and Fan, Wang and Zhang \cite{FWZ} derived the asymptotics in the transition regions between them; the rarefaction-wave region is further studied in \cite{GravaMinakov20,KotlyarovMinakov10,Minakov11,WangWen26}, and transitional asymptotics of Laguerre type in \cite{BertolaMinakov19}. In these transition regions, a branch point of the $g$-function changes from a hard edge to a soft edge, the local behaviour is described by a Painlev\'e XXXIV parametrix \cite{IKO08,IKO09,FWZ}, and the leading correction to the background is of order $t^{-2/3}$. Parametrices of Painlev\'e type govern the asymptotics of the mKdV equation also in other regimes \cite{CharlierLenells20,HuangZhang22}, of the KdV equation in the small dispersion limit \cite{ClaeysGrava09,ClaeysGrava10,DeiftVenakidesZhou94,Grava01,GravaTian02}, and of NLS-type equations near the gradient catastrophe \cite{BertolaTovbis13}. The long-time asymptotics of the Hirota equation itself have been studied for other classes of initial data; see \cite{PengChen23} and the references therein. For Painlev\'e asymptotics of the mKdV and NLS equations with nonzero backgrounds we refer to \cite{WangXuFan25,FanLiYangZhang26}.

\subsection{Transition regions}
The RH problem associated with \eqref{eq:Hirota}--\eqref{eq:ID} is formulated in Section~\ref{sec:IST}. Its jump matrix is supported on $\R$, has the branch points $\pm c_l$ and $\pm c_r$, and depends on $(x,t)$ through the phase $t\theta(k)$, where $\theta(k)=4\beta k^3+2\alpha k^2+\xi k$ and $\xi=x/t$. For the mKdV equation, the RH problem is invariant under $k\mapsto-k$. For $\alpha\neq0$ this symmetry is broken by the term $2\alpha k^2$, and only the symmetry $M(k)=\sigma_1\overline{M(\bar k)}\sigma_1$ survives (Proposition~\ref{prop:symM}). As a consequence, the branch points $c_j$ and $-c_j$, $j\in\{l,r\}$, have to be analysed separately.

\begin{figure}[htbp]
\centering
\includegraphics[width=0.98\textwidth]{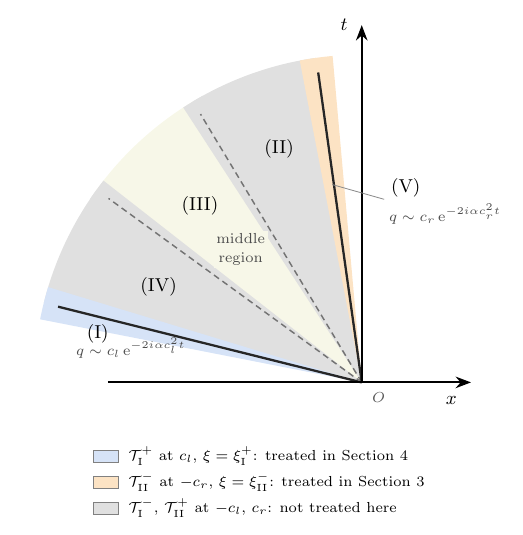}
\caption{The different asymptotic regions of the $(x,t)$-half plane.}
\label{fig:regions}
\end{figure}

In the left (resp.\ right) background region, the asymptotic analysis is based on a $g$-function with branch cut $[-c_l,c_l]$ (resp.\ $[-c_r,c_r]$); see \eqref{eq:gI} and \eqref{eq:g}. For $\alpha>0$, the left background region is bounded by the critical line $\xi=\xi_{\rm I}^+$, on which the branch point $c_l$ becomes a soft edge, and the right background region is bounded by the critical line $\xi=\xi_{\rm II}^-$, on which $-c_r$ becomes a soft edge, where
\begin{equation}\label{eq:crit-lines}
\xi_{\rm I}^+=-6\beta c_l^2-2\alpha c_l,\qquad \xi_{\rm II}^-=-6\beta c_r^2+2\alpha c_r .
\end{equation}
The complete partition of the $(x,t)$ half-plane is described in Figure~\ref{fig:regions}: the two critical lines \eqref{eq:crit-lines}, together with the two further lines $\xi=\xi_{\rm I}^-=-6\beta c_l^2+2\alpha c_l+\frac{\alpha^2}{6\beta}$ and $\xi=\xi_{\rm II}^+=-6\beta c_r^2-2\alpha c_r+\frac{\alpha^2}{6\beta}$ on which the edges $-c_l$ and $c_r$ become soft, divide the half-plane into five sectors and four layers of width $O(t^{-2/3})$; the transition regions treated here are the layers around $\xi_{\rm I}^+$ and $\xi_{\rm II}^-$.
In contrast with the mKdV case, the opposite edges $-c_l$ and $c_r$ remain hard edges on these lines. In this paper we study the long-time asymptotics of $q$ in the transition regions
\begin{equation*}
\mathcal T_{\rm II}^-=\big\{(x,t):\ |\xi-\xi_{\rm II}^-|\,t^{2/3}<C\big\},\qquad
\mathcal T_{\rm I}^+=\big\{(x,t):\ |\xi-\xi_{\rm I}^+|\,t^{2/3}<C\big\},\qquad C>0 .
\end{equation*}
More precisely, as at the beginning of Sections~\ref{sec:TIIm} and~\ref{sec:TIp}, we restrict ourselves to the halves $0<t^{2/3}(\xi-\xi_{\rm II}^-)<C$ and $-C<t^{2/3}(\xi-\xi_{\rm I}^+)<0$, which are adjacent to the background regions. The transition regions attached to the edges $-c_l$ and $c_r$ are not considered in this paper.

\subsection{Main result}
Our main result is the following theorem.

\begin{theorem}\label{thm:main}
Let $\alpha>0$, $\beta>0$ and $c_l>c_r>0$. Let $q_0$ satisfy Assumption~\ref{ass:1}, and assume that the reflection coefficient $r$ defined in Section~\ref{sec:IST} satisfies \eqref{eq:r-ne-1}. Let $q$ be the solution of \eqref{eq:Hirota}--\eqref{eq:ID} satisfying \eqref{eq:BC} (see Remark~\ref{rem:global}), and fix $C>0$ and $\epsilon\in(\frac12,\frac23)$. Then the following statements hold as $t\to+\infty$, where $\xi=x/t$.
\begin{enumerate}
\item[(a)] For $\xi\in\mathcal T_{\rm II}^-$, we have
\begin{equation}\label{eq:main-a}
q(x,t)=\e^{-2i\alpha c_r^2t}\Big(c_r-i\,\kappa_{\rm II}\,a(s_{\rm II})\,t^{-\frac13}\Big)+\cO(t^{-\epsilon}),
\end{equation}
where $\kappa_{\rm II}=\big(\frac{2c_r}{3(4\beta c_r-\alpha)}\big)^{1/3}$, $s_{\rm II}=-\kappa_{\rm II}\,t^{2/3}(\xi-\xi_{\rm II}^-)$, and
\begin{equation*}
a(s)=\int_{-\infty}^{s}\Big(u(\tau)+\frac\tau2\Big)\di\tau ;
\end{equation*}
here $u$ is the solution of the Painlev\'e XXXIV equation with parameter $b=-\frac{\arg r(c_r)}{2\pi}$ specified in Theorem~\ref{thm:TIIm}.
\item[(b)] For $\xi\in\mathcal T_{\rm I}^+$, we obtain
\begin{equation}\label{eq:main-b}
q(x,t)=\e^{-2i\alpha c_l^2t-2i\varphi_{\rm I}(\xi)}\Big(c_l+\frac{i\kappa_{\rm I}\,s_{\rm I}^2}{4}\,t^{-\frac13}+q_{\rm PC}(x,t)\Big)+\cO(t^{-\epsilon}),
\end{equation}
where $\kappa_{\rm I}=\big(\frac{2c_l}{3(4\beta c_l+\alpha)}\big)^{1/3}$, $s_{\rm I}=\kappa_{\rm I}\,t^{2/3}(\xi_{\rm I}^+-\xi)$,
\begin{equation*}
\varphi_{\rm I}(\xi)=\frac{1}{2\pi}\int_{-\hat\eta_l}^{-c_l}\frac{\log\big(1-|r(s)|^2\big)}{-\sqrt{s^2-c_l^2}}\,\di s>0,
\end{equation*}
and $q_{\rm PC}(x,t)=\cO(t^{-1/2})$ is the parabolic cylinder term
\begin{gather*}
q_{\rm PC}(x,t)=-\sqrt{\frac{2\nu}{t\,|g_{\rm I}''(-\hat\eta_l)|}}\,
\Big(\frac{\hat\eta_l+R}{2R}\,\e^{i\Theta}-\frac{\hat\eta_l-R}{2R}\,\e^{-i\Theta}\Big),\\
R=\sqrt{\hat\eta_l^{\,2}-c_l^2},\qquad \nu=-\frac{1}{2\pi}\log\big(1-|r(-\hat\eta_l)|^2\big).
\end{gather*}
Here $g_{\rm I}(k)=(4\beta k^2+2\alpha k+\xi+2\beta c_l^2)\sqrt{k^2-c_l^2}$, $-\hat\eta_l=-\hat\eta_l(\xi)$ is the unique critical point of $g_{\rm I}$ in $(-\infty,-c_l)$, and the real phase $\Theta=\Theta(x,t)$ is given in \eqref{eq:Theta-TIp}.
\end{enumerate}
In particular, $|q(x,t)|=c_r+\cO(t^{-\epsilon})$ in case (a), whereas in case (b)
\begin{equation}\label{eq:main-b-phase}
q(x,t)=c_l\exp\Big\{-2i\alpha c_l^2t-2i\varphi_{\rm I}(\xi)+\frac{i\,(x-\xi_{\rm I}^+t)^2}{6(4\beta c_l+\alpha)\,t}\Big\}+\cO(t^{-\frac12}),
\end{equation}
and $|q(x,t)|=c_l+\cO(t^{-1/2})$.
\end{theorem}

Let us comment on Theorem~\ref{thm:main}. In both regions, the leading correction to the rotating background is of order $t^{-1/3}$, and to this order it is a pure phase correction. In the transition regions of the mKdV equation, the leading correction is of order $t^{-2/3}$ \cite{FWZ}. The reason for this difference is the following. For $\alpha=0$, the two edges $c_j$ and $-c_j$ are critical simultaneously, and their contributions of order $t^{-1/3}$ cancel by the symmetry $k\mapsto-k$. For $\alpha\neq0$, only one edge is critical in each of the regions considered here. In $\TIIm$, the parameter of the Painlev\'e XXXIV parametrix at the noncritical edge $c_r$ tends to $-\infty$ at the rate $t^{2/3}$, and this edge contributes only $\cO(t^{-1})$. In $\mathcal T_{\rm I}^+$, the edge $-c_l$ is a hard edge, and no local parametrix is needed there.

The two regions differ in the local models at the critical edges. At $-c_r$, the model is the Painlev\'e XXXIV parametrix $M^{P34}(\zeta;b,\omega,s)$ with $b=-\frac{\arg r(c_r)}{2\pi}$ and $\omega=0$, and the coefficient $a(s)$ in \eqref{eq:main-a} is an integral of the corresponding Painlev\'e XXXIV transcendent. At $c_l$, the jump matrix has a nonvanishing $(2,2)$-entry on the band $(c_r,c_l)$, which tends to $1$ at $c_l$. This leads to the parameters $(b,\omega)=(0,1)$, for which the model problem is solved by Airy functions. Consequently, the coefficient of $t^{-1/3}$ in \eqref{eq:main-b} is an explicit polynomial in $s_{\rm I}$ and does not depend on the initial data. The initial data enter \eqref{eq:main-b} through the phase $\varphi_{\rm I}$ and through the term $q_{\rm PC}$. The latter comes from the stationary point $-\hat\eta_l$ of the phase function, which stays at a positive distance from $-c_l$ and has no counterpart in the mKdV case.

\subsection{Outline of the proof and organization of the paper}
In $\TIIm$, the lenses can be opened without an auxiliary scalar function, and the local analysis at $-c_r$ is reduced to an analysis at $c_r$ by a reflection principle which changes $\alpha$ into $-\alpha$ (Lemma~\ref{lem:reflection}). The local parametrix at the noncritical edge $c_r$ requires asymptotics of the Painlev\'e XXXIV parametrix which are uniform as $s\to-\infty$ (Lemma~\ref{lem:uniform}).

In $\mathcal T_{\rm I}^+$, the interval $(-\hat\eta_l,-c_l)$ between the stationary point and the hard edge does not shrink as $t\to+\infty$. The factor $1-|r|^2$ on this interval is removed by a scalar function $\delta_{\rm I}$ which contains the factor $\sqrt{k^2-c_l^2}$, so that it preserves the jump on $(-c_l,c_l)$; it produces the phase $\varphi_{\rm I}$. The lenses emanating from $-\hat\eta_l$ are extended across the hard edge $-c_l$ into the band, which avoids a local parametrix at $-c_l$ (Lemma~\ref{lem:hard-edge-TIp}). The local parametrix at $c_l$ contains an error function, which is defined by a small norm RH problem in the local variable (Proposition~\ref{prop:E-TIp}). This is necessary because the holomorphic prefactor of the local parametrix is of order $t^{1/6}$, so that conjugation by it enlarges the mismatch between the exact and the model jumps by a factor of order $t^{1/3}$.

The paper is organized as follows. In Section~\ref{sec:IST} we set up the inverse scattering transform and formulate the basic RH problem. Sections~\ref{sec:TIIm} and~\ref{sec:TIp} are devoted to the asymptotic analysis in $\TIIm$ and $\mathcal T_{\rm I}^+$, respectively. The Painlev\'e XXXIV parametrix and its uniform asymptotics are discussed in Appendix~\ref{app:P34}.

\section{Spectral analysis and inverse scattering transform}\label{sec:IST}
In this section we set up the inverse scattering transform for the Cauchy problem
\eqref{eq:Hirota}--\eqref{eq:ID}. Since the $x$-part of the Lax pair coincides with that of
the defocusing mKdV equation and the initial data are real, the direct scattering problem at
$t=0$ is identical to the one studied in \cite{XZ}; we therefore only state the relevant results
and give details where the presence of the dispersive term $\alpha(q_{xx}-2|q|^2q)$ requires
modifications, namely in the background solutions, the time evolution, and the symmetries of
the associated RH problem.

\subsection{Lax pair and background solutions}
The defocusing Hirota equation \eqref{eq:Hirota} is the compatibility condition
$U_t-V_x+[U,V]=0$ of the Lax pair
\begin{equation}\label{eq:Lax}
\begin{cases}
\Phi_x+ik\sigma_3\Phi=Q\Phi,\\
\Phi_t+i(4\beta k^3+2\alpha k^2)\sigma_3\Phi=\mathcal V\Phi,
\end{cases}
\end{equation}
where $\Phi=\Phi(k;x,t)$ is a $2\times2$ matrix-valued function, $k\in\C$ is the spectral
parameter, and
\begin{equation*}
Q=\begin{pmatrix}0&q\\ \bar q&0\end{pmatrix},\qquad
\mathcal V=4\beta k^2Q+kV_1+V_0,
\end{equation*}
with
\begin{align*}
V_1&=\begin{pmatrix}-2i\beta|q|^2 & 2i\beta q_x+2\alpha q\\ -2i\beta\bar q_x+2\alpha\bar q & 2i\beta|q|^2\end{pmatrix},\\
V_0&=\begin{pmatrix}-i\alpha|q|^2+\beta(q_x\bar q-q\bar q_x) & i\alpha q_x-\beta(q_{xx}-2|q|^2q)\\
-i\alpha\bar q_x-\beta(\bar q_{xx}-2|q|^2\bar q) & i\alpha|q|^2-\beta(q_x\bar q-q\bar q_x)\end{pmatrix}.
\end{align*}

For $j\in\{l,r\}$, substituting the background $q=c_j\e^{-2i\alpha c_j^2t}$ into
\eqref{eq:Lax}, we obtain the background Lax pair
\begin{equation}\label{eq:bLax}
\begin{cases}
\Phi^b_{j,x}+ik\sigma_3\Phi^b_j=Q^b_j\Phi^b_j,\\
\Phi^b_{j,t}+i(4\beta k^3+2\alpha k^2)\sigma_3\Phi^b_j=V^b_j\Phi^b_j,
\end{cases}
\end{equation}
where
\begin{equation}\label{eq:QbVb}
Q^b_j=\begin{pmatrix}0&c_j\e^{-2i\alpha c_j^2t}\\ c_j\e^{2i\alpha c_j^2t}&0\end{pmatrix},\qquad
V^b_j=\begin{pmatrix}-ic_j^2(2\beta k+\alpha) & p_j(k)\,c_j\e^{-2i\alpha c_j^2t}\\
p_j(k)\,c_j\e^{2i\alpha c_j^2t} & ic_j^2(2\beta k+\alpha)\end{pmatrix},
\end{equation}
and $p_j(k):=4\beta k^2+2\alpha k+2\beta c_j^2$. In contrast with the mKdV case, $Q^b_j$ and
$V^b_j$ depend on $t$.

\begin{lemma}\label{lem:background}
For $j\in\{l,r\}$, the background Lax pair \eqref{eq:bLax} admits the explicit solution
\begin{equation}\label{eq:Phib}
\Phi^b_j(k;x,t)=\e^{-i\alpha c_j^2t\sigma_3}\,\Delta_j(k)\,\e^{-i(X_j(k)x+\Omega_j(k)t)\sigma_3},
\end{equation}
where
\begin{equation}\label{eq:XOmega}
X_j(k)=\sqrt{k^2-c_j^2},\qquad \Omega_j(k)=p_j(k)X_j(k)=(4\beta k^2+2\alpha k+2\beta c_j^2)X_j(k),
\end{equation}
\begin{equation}\label{eq:Delta}
\Delta_j(k)=\frac12\begin{pmatrix}\chi_j(k)+\chi_j^{-1}(k) & i\big(\chi_j(k)-\chi_j^{-1}(k)\big)\\
-i\big(\chi_j(k)-\chi_j^{-1}(k)\big) & \chi_j(k)+\chi_j^{-1}(k)\end{pmatrix},\qquad
\chi_j(k)=\Big(\frac{k-c_j}{k+c_j}\Big)^{1/4},
\end{equation}
and the branches are fixed on $\C\setminus[-c_j,c_j]$ by $X_j(k)=k+\cO(k^{-1})$ and
$\chi_j(k)=1+\cO(k^{-1})$ as $k\to\infty$. Moreover, the following hold.
\begin{enumerate}
\item[(i)] $\det\Phi^b_j\equiv1$.
\item[(ii)] As $k\to\infty$,
\begin{equation}\label{eq:Phib-infty}
\Phi^b_j(k;x,t)\,\e^{it\theta(k)\sigma_3}=I+\cO(k^{-1}),\qquad
\theta(k)=\theta(k;\xi):=4\beta k^3+2\alpha k^2+\xi k,\quad \xi:=\frac{x}{t}.
\end{equation}
\item[(iii)] For $k\in(-c_j,c_j)$,
\begin{equation}\label{eq:Phib-jump}
\Phi^b_{j,+}(k;x,t)=\Phi^b_{j,-}(k;x,t)\begin{pmatrix}0&-1\\1&0\end{pmatrix}.
\end{equation}
\end{enumerate}
\end{lemma}

\begin{proof}
Write $\Phi^b_j=\e^{-i\alpha c_j^2t\sigma_3}\Psi_j$. Since
$Q^b_j=\e^{-i\alpha c_j^2t\sigma_3}Q^b_j|_{t=0}\,\e^{i\alpha c_j^2t\sigma_3}$, a direct
computation shows that \eqref{eq:bLax} is equivalent to
\begin{equation*}
\Psi_{j,x}=\widetilde U_j\Psi_j,\qquad \Psi_{j,t}=p_j(k)\,\widetilde U_j\Psi_j,\qquad
\widetilde U_j:=-ik\sigma_3+\begin{pmatrix}0&c_j\\c_j&0\end{pmatrix},
\end{equation*}
where the term $i\alpha c_j^2\sigma_3$ produced by differentiating the gauge factor exactly
cancels the constant diagonal part $\mp i\alpha c_j^2$ of $V^b_j$. As $\widetilde U_j^2=-X_j^2I$
and $\widetilde U_j\Delta_j=\Delta_j(-iX_j\sigma_3)$, we obtain
$\Psi_j=\Delta_j\e^{-i(X_jx+\Omega_jt)\sigma_3}$, which gives \eqref{eq:Phib}. Item (i) follows
from $\det\Delta_j=1$. For (ii), we note that
\begin{equation*}
X_j(k)x+\Omega_j(k)t=kx+(4\beta k^3+2\alpha k^2)t-\alpha c_j^2t+\cO(k^{-1}),\qquad k\to\infty,
\end{equation*}
so that the constant $-\alpha c_j^2t$ cancels the gauge factor $\e^{-i\alpha c_j^2t\sigma_3}$;
together with $\Delta_j=I+\cO(k^{-1})$ this yields \eqref{eq:Phib-infty}. Finally, for
$k\in(-c_j,c_j)$ we have $X_{j,+}=-X_{j,-}$ and hence $\Omega_{j,+}=-\Omega_{j,-}$, since $p_j$ is
a polynomial, whereas $\Delta_{j,+}=\Delta_{j,-}\left(\begin{smallmatrix}0&-1\\1&0\end{smallmatrix}\right)$.
Since an off-diagonal matrix $K$ satisfies $K\e^{i\phi\sigma_3}=\e^{-i\phi\sigma_3}K$, and the
gauge factor is independent of $k$, \eqref{eq:Phib-jump} follows.
\end{proof}

\subsection{Direct scattering transform}
For $j\in\{l,r\}$, we denote
\begin{equation*}
\Phi^b_j(k;x):=\Phi^b_j(k;x,0)=\Delta_j(k)\e^{-iX_j(k)x\sigma_3},\qquad
Q^b_j:=Q^b_j|_{t=0},\qquad Q_0(x):=Q(x,0).
\end{equation*}
The Jost functions
$\Phi_j(k;x):=\Phi_j(k;x,0)$ are defined by the Volterra integral equations
\begin{equation}\label{eq:Volterra}
\Phi_j(k;x)=\Phi^b_j(k;x)+\int_{\infty_j}^x\Phi^b_j(k;x)\big(\Phi^b_j\big)^{-1}(k;y)\big(Q_0(y)-Q^b_j\big)\Phi_j(k;y)\di y,
\end{equation}
where $\infty_r:=+\infty$ and $\infty_l:=-\infty$. Since the gauge factor in \eqref{eq:Phib}
equals $I$ at $t=0$ and the $x$-part of \eqref{eq:Lax} coincides with that of the defocusing
mKdV equation, \eqref{eq:Volterra} is exactly the integral equation studied in \cite{XZ}.
Consequently, the following two propositions hold verbatim; see \cite{XZ,FWZ}.

\begin{proposition}\label{prop:Jost}
Under Assumption~\ref{ass:1}, the Jost functions $\Phi_l$ and $\Phi_r$ defined by
\eqref{eq:Volterra} have the following properties for $j\in\{l,r\}$.
\begin{enumerate}
\item[(a)] For each $x\in\R$, $[\Phi_r(k;x)]_1$ and $[\Phi_l(k;x)]_2$ are holomorphic for
$k\in\C^-$, while $[\Phi_r(k;x)]_2$ and $[\Phi_l(k;x)]_1$ are holomorphic for $k\in\C^+$; moreover,
$[\Phi_j(k;x)]_1$, $[\Phi_j(k;x)]_2$ extend continuously to the corresponding closed half-plane
with the points $\pm c_j$ removed.
\item[(b)] $\det\Phi_j(k;x)=1$ for $k\in\R\setminus[-c_j,c_j]$.
\item[(c)] As $k\to\infty$,
\begin{align*}
\big([\Phi_l(k;x)]_1,[\Phi_r(k;x)]_2\big)\e^{ikx\sigma_3}&=I+\cO(k^{-1}), && k\in\C^+,\\
\big([\Phi_r(k;x)]_1,[\Phi_l(k;x)]_2\big)\e^{ikx\sigma_3}&=I+\cO(k^{-1}), && k\in\C^-.
\end{align*}
\item[(d)] $\sigma_1\overline{\Phi_j(\bar k;x)}\sigma_1=\Phi_j(k;x)$,\quad
$[\Phi_j(k;x)]_1=\sigma_1[\Phi_j(-k;x)]_2$,\quad $\overline{\Phi_j(-\bar k;x)}=\Phi_j(k;x)$.
\item[(e)] For $k\in(-c_j,c_j)$, the boundary values of $\Phi_j$ satisfy the relation in
\cite[Proposition~2.1(e)]{FWZ}.
\end{enumerate}
\end{proposition}

The scattering matrix $S(k)$ is defined by $S(k)=\Phi_r^{-1}(k;x)\Phi_l(k;x)$ for
$k\in\R\setminus\{\pm c_l,\pm c_r\}$, and by Proposition~\ref{prop:Jost}(d) it takes the form
\begin{equation*}
S(k)=\begin{pmatrix}a(k)&\overline{b(\bar k)}\\ b(k)&\overline{a(\bar k)}\end{pmatrix},\qquad
a(k)=\det\big([\Phi_l]_1,[\Phi_r]_2\big),\quad b(k)=\det\big([\Phi_r]_1,[\Phi_l]_1\big).
\end{equation*}
We define the reflection coefficient $r(k):=b(k)/a(k)$ for $k\in\R\setminus\{\pm c_l,\pm c_r\}$.

\begin{proposition}\label{prop:scattering}
The functions $a$, $b$ and $r$ have the following properties.
\begin{enumerate}
\item[(a)] $a(k)$ is holomorphic in $\C^+$ and extends continuously to
$\R\setminus\{\pm c_l,\pm c_r\}$; $a(k)=\cO((k\mp c_j)^{-1/4})$ as $k\to\pm c_j$, $j\in\{l,r\}$;
$a(k)=1+\cO(k^{-1})$ as $k\to\infty$ in $\C^+$; $b(k)$ is continuous on
$\R\setminus\{\pm c_l,\pm c_r\}$, and $a,b\in C^8(\R\setminus\{\pm c_l,\pm c_r\})$.
\item[(b)] $a(k)$ has no zeros in $\C$.
\item[(c)] In their domains of definition,
\begin{equation}\label{eq:sym-r}
a(k)=\overline{a(-\bar k)},\qquad b(k)=\overline{b(-\bar k)},\qquad r(k)=\overline{r(-\bar k)}.
\end{equation}
\item[(d)] Writing $f^*(k):=\overline{f(\bar k)}$, we have
\begin{align*}
&a_+=a^*_-,\quad b_+=-b^*_-,\quad r_+=-r^*_-, && k\in(-c_r,c_r),\\
&a_+=b^*_-,\quad b_+=a^*_-,\quad r_+=(r^*_-)^{-1}, && k\in(-c_l,-c_r)\cup(c_r,c_l),\\
&a_+=a_-,\quad b_+=b_-,\quad r_+=r_-, && k\in(-\infty,-c_l)\cup(c_l,+\infty).
\end{align*}
In particular, $|r(k)|=1$ for $k\in(-c_l,-c_r)\cup(c_r,c_l)$ and $|r(k)|<1$ on the remaining
intervals of $\R$.
\item[(e)] $r\in C^8(\R\setminus\{\pm c_l,\pm c_r\})$ and, near the branch points,
\begin{equation}\label{eq:r-endpoints}
r(k)=\begin{cases}
\sum_{m=0}^7 b_{l,m}(k-c_l)^{m/2}+o((k-c_l)^{7/2}), & k\searrow c_l,\\
\sum_{m=0}^7 i^m b_{l,m}(c_l-k)^{m/2}+o((c_l-k)^{7/2}), & k\nearrow c_l,\\
\sum_{m=0}^7 i^m b_{r,m}(k-c_r)^{m/2}+o((k-c_r)^{7/2}), & k\searrow c_r,\\
\sum_{m=0}^7 (-1)^m b_{r,m}(c_r-k)^{m/2}+o((c_r-k)^{7/2}), & k\nearrow c_r,\\
\sum_{m=0}^7 (-i)^m \overline{b_{l,m}}(k+c_l)^{m/2}+o((k+c_l)^{7/2}), & k\searrow -c_l,\\
\sum_{m=0}^7 \overline{b_{l,m}}(-c_l-k)^{m/2}+o((-c_l-k)^{7/2}), & k\nearrow -c_l,\\
\sum_{m=0}^7 (-1)^m \overline{b_{r,m}}(k+c_r)^{m/2}+o((k+c_r)^{7/2}), & k\searrow -c_r,\\
\sum_{m=0}^7 (-i)^m \overline{b_{r,m}}(-c_r-k)^{m/2}+o((-c_r-k)^{7/2}), & k\nearrow -c_r,
\end{cases}
\end{equation}
where the coefficients $b_{j,m}\in\C$ satisfy
\begin{equation}\label{eq:b-cond}
|b_{j,0}|=1,\qquad b_{j,1}\neq0,\qquad \sum_{m=0}^{n}i^{n-m}(-i)^m b_{j,n-m}\overline{b_{j,m}}=0,\qquad j\in\{l,r\},\ n=1,\dots,7.
\end{equation}
In particular, $r(-c_j)=\overline{r(c_j)}$, so that $\arg r(-c_r)=-\arg r(c_r)$.
\item[(f)] $\partial_k^m r(k)=\cO(k^{-5})$ as $k\to\pm\infty$, $m=0,\dots,8$.
\end{enumerate}
\end{proposition}

\subsection{Time evolution and the basic RH problem}
Assume that the solution $q(x,t)$ of \eqref{eq:Hirota}--\eqref{eq:ID} exists for $t\ge0$ and
satisfies \eqref{eq:BC}. For $t\ge0$, let $\Phi_l(k;x,t)$ and $\Phi_r(k;x,t)$ be the solutions of
the $x$-part of \eqref{eq:Lax} fixed by
\begin{align}
\Phi_l(k;x,t)&=\Phi^b_l(k;x,t)\big(I+o(1)\big), && k\in\R,\ x\to-\infty,\label{eq:asy-l}\\
\Phi_r(k;x,t)&=\Phi^b_r(k;x,t)\big(I+o(1)\big), && k\in\R,\ x\to+\infty,\label{eq:asy-r}
\end{align}
where $\Phi^b_j(k;x,t)$ is given by \eqref{eq:Phib}. These functions reduce to the Jost
functions of \eqref{eq:Volterra} at $t=0$.

\begin{lemma}\label{lem:Jost-t}
For every fixed $t\ge0$, the functions $\Phi_l(k;x,t)$ and $\Phi_r(k;x,t)$ satisfy items
\textup{(a)}, \textup{(b)} and \textup{(e)} of Proposition~\ref{prop:Jost}, and item
\textup{(c)} is replaced by
\begin{align}
\big([\Phi_l(k;x,t)]_1,[\Phi_r(k;x,t)]_2\big)\e^{it\theta(k)\sigma_3}&=I+\cO(k^{-1}), && k\to\infty,\ k\in\C^+,\label{eq:Jost-t-inf+}\\
\big([\Phi_r(k;x,t)]_1,[\Phi_l(k;x,t)]_2\big)\e^{it\theta(k)\sigma_3}&=I+\cO(k^{-1}), && k\to\infty,\ k\in\C^-,\label{eq:Jost-t-inf-}
\end{align}
where $\theta$ is defined in \eqref{eq:Phib-infty}. Moreover, the first symmetry in
Proposition~\ref{prop:Jost}(d) holds for all $t\ge0$, whereas the last two hold only at $t=0$.
\end{lemma}

\begin{proof}
For fixed $t$, $\Phi_j(\cdot\,;x,t)$ solves the Volterra equation obtained from
\eqref{eq:Volterra} by replacing $\Phi^b_j(k;\cdot)$, $Q^b_j$ and $Q_0$ with
$\Phi^b_j(k;\cdot,t)$, $Q^b_j(t)$ and $Q(\cdot,t)$, respectively. Its kernel is
\begin{equation*}
\Phi^b_j(k;x,t)\big(\Phi^b_j\big)^{-1}(k;y,t)
=\e^{-i\alpha c_j^2t\sigma_3}\Big[\Delta_j(k)\e^{-iX_j(k)(x-y)\sigma_3}\Delta_j^{-1}(k)\Big]\e^{i\alpha c_j^2t\sigma_3},
\end{equation*}
since the factor $\e^{-i\Omega_jt\sigma_3}$ cancels. The outer factors are constant matrices
independent of $k$, so the bracket is the same kernel as at $t=0$. Consequently, the branch
points $\pm c_j$, the half-planes of analyticity of the columns, and the jump relation across
$(-c_j,c_j)$ (cf.\ \eqref{eq:Phib-jump}) are the same as at $t=0$, which gives (a) and (e);
(b) follows from $\operatorname{tr}(-ik\sigma_3+Q)=0$. The standard large-$k$ analysis of the
Volterra equation shows that, column-wise in the respective half-planes,
$\big(\Phi_j(k;x,t)-\Phi^b_j(k;x,t)\big)\e^{it\theta(k)\sigma_3}=\cO(k^{-1})$, and
\eqref{eq:Jost-t-inf+}--\eqref{eq:Jost-t-inf-} follow from
\eqref{eq:Phib-infty}.
\end{proof}

The functions $\Phi_l(k;x,t)$ and $\Phi_r(k;x,t)$ are simultaneous solutions of both equations
in \eqref{eq:Lax}, and
\begin{equation}\label{eq:S-t}
\Phi_l(k;x,t)=\Phi_r(k;x,t)S(k),\qquad k\in\R\setminus\{\pm c_l,\pm c_r\},
\end{equation}
where $S(k)$ is the scattering matrix defined at $t=0$. In particular, the scattering data $a(k)$, $b(k)$ and $r(k)$ are independent of $t$. We now define the sectionally holomorphic function
\begin{equation}\label{eq:M}
M(k)=M(k;x,t):=\begin{cases}
\Big(\dfrac{[\Phi_l(k;x,t)]_1}{a(k)},\ [\Phi_r(k;x,t)]_2\Big)\e^{it\theta(k)\sigma_3}, & k\in\C^+,\\[10pt]
\Big([\Phi_r(k;x,t)]_1,\ \dfrac{[\Phi_l(k;x,t)]_2}{a^*(k)}\Big)\e^{it\theta(k)\sigma_3}, & k\in\C^-,
\end{cases}
\end{equation}
where $\theta$ is defined in \eqref{eq:Phib-infty}. It follows from
Proposition~\ref{prop:scattering}, Lemma~\ref{lem:background} and Lemma~\ref{lem:Jost-t} that $M$ solves the following RH problem.

\begin{rhp}\label{rhp:M}
Find a $2\times2$ matrix-valued function $M(k)=M(k;x,t)$ with the following properties.
\begin{enumerate}
\item[$\bullet$] $M(k)$ is holomorphic for $k\in\C\setminus\R$.
\item[$\bullet$] $M(k)$ has continuous boundary values $M_\pm(k)$ on
$\R\setminus\{\pm c_l,\pm c_r\}$ satisfying $M_+(k)=M_-(k)V(k)$, where
\begin{equation}\label{eq:V}
V(k)=\begin{cases}
\begin{pmatrix}1-r(k)r^*(k) & -r^*(k)\e^{-2it\theta(k)}\\ r(k)\e^{2it\theta(k)} & 1\end{pmatrix},
& k\in(-\infty,-c_l)\cup(c_l,+\infty),\\[10pt]
\begin{pmatrix}0 & -r^*_-(k)\e^{-2it\theta(k)}\\ r_+(k)\e^{2it\theta(k)} & 1\end{pmatrix},
& k\in(-c_l,-c_r)\cup(c_r,c_l),\\[10pt]
\begin{pmatrix}0 & -\e^{-2it\theta(k)}\\ \e^{2it\theta(k)} & 0\end{pmatrix},
& k\in(-c_r,c_r).
\end{cases}
\end{equation}
\item[$\bullet$] $M(k)=I+\cO(k^{-1})$ as $k\to\infty$ in $\C\setminus\R$.
\item[$\bullet$] $M(k)=\cO(1)$ as $k\to\pm c_l$.
\item[$\bullet$] As $k\to\pm c_r$,
\begin{align*}
M(k)&=\cO\begin{pmatrix}(k\mp c_r)^{1/4} & (k\mp c_r)^{-1/4}\\ (k\mp c_r)^{1/4} & (k\mp c_r)^{-1/4}\end{pmatrix}, && k\in\C^+,\\
M(k)&=\cO\begin{pmatrix}(k\mp c_r)^{-1/4} & (k\mp c_r)^{1/4}\\ (k\mp c_r)^{-1/4} & (k\mp c_r)^{1/4}\end{pmatrix}, && k\in\C^-.
\end{align*}
\end{enumerate}
\end{rhp}

The jumps in \eqref{eq:V} across the three groups of intervals are determined by \eqref{eq:S-t}, by the constant jump \eqref{eq:Phib-jump} of the background solutions, and by Proposition~\ref{prop:scattering}(d), none of which involves the time variable. The normalization at infinity follows from \eqref{eq:Phib-infty} and $a(k)=1+\cO(k^{-1})$, and the local behaviour near the branch points is the same as in \cite{XZ}. Since $\det M\equiv1$ and the singularities at $\pm c_l,\pm c_r$ are removable for
$M\widetilde M^{-1}$ whenever $M$ and $\widetilde M$ are two solutions, Liouville's theorem implies that RH problem~\ref{rhp:M} has at most one solution.

\begin{proposition}\label{prop:symM}
The solution of RH problem~\ref{rhp:M} satisfies
\begin{equation}\label{eq:symM}
M(k)=\sigma_1\overline{M(\bar k)}\sigma_1,\qquad k\in\C\setminus\R .
\end{equation}
\end{proposition}

\begin{proof}
Since $\theta$ has real coefficients, $\overline{\theta(\bar k)}=\theta(k)$, and
$\sigma_1\overline{V(\bar k)}^{-1}\sigma_1$ coincides with $V(k)$ on each interval by
Proposition~\ref{prop:scattering}(d). Thus both sides of \eqref{eq:symM} solve
RH problem~\ref{rhp:M}, and the claim follows from uniqueness.
\end{proof}

The solution of the Cauchy problem \eqref{eq:Hirota}--\eqref{eq:ID} is given by
\begin{equation}\label{eq:recon}
q(x,t)=2i\lim_{k\to\infty}\big(kM(k;x,t)\big)_{12}.
\end{equation}
Finally, we record two factorizations of the jump matrix \eqref{eq:V} which will be used to open
lenses. Let $r_2(k):=r^*(k)/\big(1-r(k)r^*(k)\big)$ for $k\in\R\setminus[-c_l,c_l]$. Then, for
$k\in(-\infty,-c_l)\cup(c_l,+\infty)$,
\begin{equation}\label{eq:fac1}
\begin{aligned}
V&=\begin{pmatrix}1&-r^*\e^{-2it\theta}\\0&1\end{pmatrix}
\begin{pmatrix}1&0\\ r\e^{2it\theta}&1\end{pmatrix}\\
&=\begin{pmatrix}1&0\\ r_2^*\e^{2it\theta}&1\end{pmatrix}
(1-rr^*)^{\sigma_3}
\begin{pmatrix}1&-r_2\e^{-2it\theta}\\0&1\end{pmatrix};
\end{aligned}
\end{equation}
for $k\in(-c_l,-c_r)\cup(c_r,c_l)$,
\begin{equation}\label{eq:fac2}
\begin{aligned}
V&=\begin{pmatrix}1&-r^*_-\e^{-2it\theta}\\0&1\end{pmatrix}
\begin{pmatrix}1&0\\ r_+\e^{2it\theta}&1\end{pmatrix}\\
&=\begin{pmatrix}1&0\\ r^*_{2,-}\e^{2it\theta}&1\end{pmatrix}
\begin{pmatrix}0&-r^*_-\e^{-2it\theta}\\ r_+\e^{2it\theta}&0\end{pmatrix}
\begin{pmatrix}1&-r_{2,+}\e^{-2it\theta}\\0&1\end{pmatrix};
\end{aligned}
\end{equation}
and for $k\in(-c_r,c_r)$,
\begin{equation}\label{eq:fac3}
\begin{aligned}
V&=\begin{pmatrix}1&-r^*_-\e^{-2it\theta}\\0&1\end{pmatrix}
\begin{pmatrix}0&-\e^{-2it\theta}\\ \e^{2it\theta}&0\end{pmatrix}
\begin{pmatrix}1&0\\ r\e^{2it\theta}&1\end{pmatrix}\\
&=\begin{pmatrix}1&0\\ r^*_{2,-}\e^{2it\theta}&1\end{pmatrix}
\begin{pmatrix}0&-\e^{-2it\theta}\\ \e^{2it\theta}&0\end{pmatrix}
\begin{pmatrix}1&-r_{2,+}\e^{-2it\theta}\\0&1\end{pmatrix}.
\end{aligned}
\end{equation}

\section{Asymptotic analysis in the transition region $\TIIm$}\label{sec:TIIm}
This section is devoted to the long-time asymptotic analysis of RH problem~\ref{rhp:M} in the
transition region
\begin{equation*}
\TIIm:=\Big\{(x,t):\ \big|\xi-\xi_{\rm II}^-\big|\,t^{2/3}<C\Big\},\qquad
\xi_{\rm II}^-:=-6\beta c_r^2+2\alpha c_r,\qquad C>0 .
\end{equation*}
Throughout this section we assume that $0<t^{2/3}(\xi-\xi_{\rm II}^-)<C$, since the analysis in
the other half of $\TIIm$ is similar. We also assume $0<\alpha<4\beta c_r$.

\subsection{First transformation: $M\to M^{(1)}$}
For $\xi\in\TIIm$, we introduce the $g$-function
\begin{equation}\label{eq:g}
g(k)=g(k;\xi):=\big(4\beta k^2+2\alpha k+\xi+2\beta c_r^2\big)X_r(k),
\end{equation}
where $X_r(k)=\sqrt{k^2-c_r^2}$ is fixed by $X_r(k)=k+\cO(k^{-1})$ as $k\to\infty$.

\begin{proposition}\label{prop:g}
The function $g$ defined in \eqref{eq:g} has the following properties.
\begin{itemize}
\item $g(k)$ is holomorphic for $k\in\C\setminus[-c_r,c_r]$, and $g_+(k)+g_-(k)=0$ for
$k\in(-c_r,c_r)$.
\item As $k\to\infty$ in $\C\setminus[-c_r,c_r]$, we have $g(k)=\theta(k)-\alpha c_r^2+\cO(k^{-1})$.
\item For $\epsilon\in\{\pm1\}$, as $k\to\epsilon c_r$, we have
\begin{equation}\label{eq:g-exp}
g(k)=g^{(1)}_\epsilon(\epsilon k-c_r)^{\frac12}+g^{(2)}_\epsilon(\epsilon k-c_r)^{\frac32}
+\cO\big((\epsilon k-c_r)^{\frac52}\big),
\end{equation}
where the principal branch is taken and
\begin{equation}\label{eq:g12}
g^{(1)}_\epsilon=\epsilon(2c_r)^{\frac12}\big(\xi+6\beta c_r^2+2\epsilon\alpha c_r\big),\qquad
g^{(2)}_\epsilon=\epsilon(2c_r)^{\frac12}\Big[\frac{\xi+6\beta c_r^2+2\epsilon\alpha c_r}{4c_r}
+8\beta c_r+2\epsilon\alpha\Big].
\end{equation}
\end{itemize}
\end{proposition}

\begin{figure}[htbp]
\centering
\includegraphics[width=0.98\textwidth]{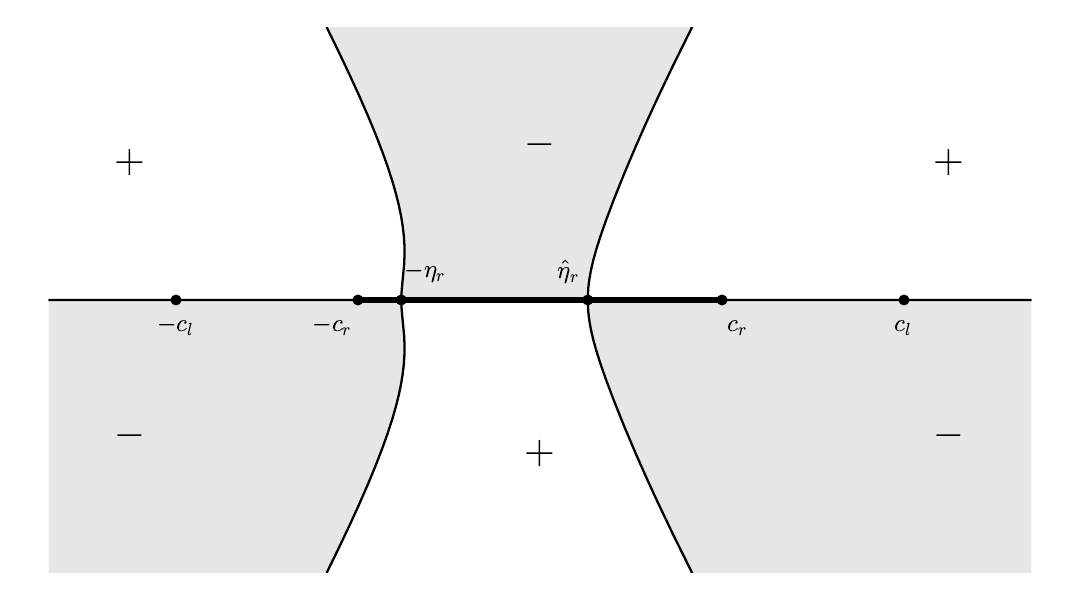}
\caption{The signature table of $\operatorname{Im} g$ for $0<t^{2/3}(\xi-\xi_{\rm II}^-)<C$.
The shaded and white regions correspond to $\operatorname{Im} g<0$ and $\operatorname{Im} g>0$,
respectively; the thick segment is the branch cut $[-c_r,c_r]$, and $-\eta_r$, $\hat\eta_r$
are the two real zeros of $g$.}
\label{fig:sign}
\end{figure}

The function $g$ has two real zeros $-\eta_r(\xi)$ and $\hat\eta_r(\xi)$, namely
\begin{equation*}
4\beta k^2+2\alpha k+\xi+2\beta c_r^2=4\beta\big(k+\eta_r\big)\big(k-\hat\eta_r\big),\qquad
-\eta_r=-c_r+\frac{\xi-\xi_{\rm II}^-}{8\beta c_r-2\alpha}+\cO\big((\xi-\xi_{\rm II}^-)^2\big),
\end{equation*}
and $\hat\eta_r\to c_r-\frac{\alpha}{2\beta}$ as $t\to+\infty$. Hence, for $\xi\in\TIIm$,
$-\eta_r\in(-c_r,c_r)$ tends to $-c_r$ at the rate $\cO(t^{-2/3})$, whereas $\hat\eta_r$ stays at a
positive distance from $c_r$. Correspondingly, $g^{(1)}_-=\cO(t^{-2/3})$ and
$g^{(2)}_-=-2(2c_r)^{1/2}(4\beta c_r-\alpha)+\cO(t^{-2/3})\neq0$, while
$g^{(1)}_+=4\alpha c_r(2c_r)^{1/2}+\cO(t^{-2/3})\neq0$. Thus only the branch point $-c_r$ is
critical in $\TIIm$. The signature table of $\operatorname{Im}g$ is illustrated in
Figure~\ref{fig:sign}.

\subsubsection{RH problem for $M^{(1)}$}
In view of Proposition~\ref{prop:g}, $g(k)-\theta(k)\to-\alpha c_r^2$ as $k\to\infty$. To keep the
normalization at infinity, we define
\begin{equation}\label{eq:M1}
M^{(1)}(k)=M^{(1)}(k;x,t):=\e^{i\alpha c_r^2t\sigma_3}\,M(k)\,\e^{it(g(k)-\theta(k))\sigma_3}.
\end{equation}
Since the left factor is independent of $k$, it affects neither the jumps nor the local
behaviour of $M^{(1)}$. Then $M^{(1)}$ satisfies the following RH problem.

\begin{rhp}\label{rhp:M1}\leavevmode
\begin{itemize}
\item $M^{(1)}(k)$ is holomorphic for $k\in\C\setminus\R$.
\item $M^{(1)}(k)$ has continuous boundary values $M^{(1)}_\pm(k)$ on
$\R\setminus\{\pm c_l,\pm c_r\}$ satisfying $M^{(1)}_+(k)=M^{(1)}_-(k)V^{(1)}(k)$, where
\begin{equation}\label{eq:V1}
V^{(1)}(k)=\begin{cases}
\begin{pmatrix}1-r(k)r^*(k) & -r^*(k)\e^{-2itg(k)}\\ r(k)\e^{2itg(k)} & 1\end{pmatrix},
& k\in(-\infty,-c_l)\cup(c_l,+\infty),\\[10pt]
\begin{pmatrix}0 & -r^*_-(k)\e^{-2itg(k)}\\ r_+(k)\e^{2itg(k)} & 1\end{pmatrix},
& k\in(-c_l,-c_r)\cup(c_r,c_l),\\[10pt]
\begin{pmatrix}0&-1\\1&0\end{pmatrix}, & k\in(-c_r,c_r).
\end{cases}
\end{equation}
\item As $k\to\infty$ in $\C\setminus\R$, we have $M^{(1)}(k)=I+\cO(k^{-1})$.
\item $M^{(1)}(k)$ admits the same singular behaviour as $M(k)$ at the branch points
$\pm c_l$ and $\pm c_r$; see RH problem~\ref{rhp:M}.
\item $M^{(1)}(k)=\sigma_1\overline{M^{(1)}(\bar k)}\sigma_1$ for $k\in\C\setminus\R$.
\end{itemize}
\end{rhp}

Indeed, since $g_+=g_-$ on $\R\setminus[-c_r,c_r]$, the jumps on the first two groups of
intervals are obtained from \eqref{eq:V} by replacing $\theta$ with $g$. For $k\in(-c_r,c_r)$ we have
$g_+(k)+g_-(k)=0$, hence
\begin{equation*}
\e^{-it(g_--\theta)\sigma_3}\begin{pmatrix}0&-\e^{-2it\theta}\\ \e^{2it\theta}&0\end{pmatrix}
\e^{it(g_+-\theta)\sigma_3}=\begin{pmatrix}0&-1\\1&0\end{pmatrix}.
\end{equation*}
The normalization follows from
$\e^{it(g-\theta)\sigma_3}=\e^{-i\alpha c_r^2t\sigma_3}\big(I+\cO(k^{-1})\big)$, and the symmetry
follows from $\overline{g(\bar k)}=g(k)$ and Proposition~\ref{prop:symM}. By the reconstruction formula
\eqref{eq:recon} and \eqref{eq:M1}, we have
\begin{equation}\label{eq:recon1}
q(x,t)=2i\,\e^{-2i\alpha c_r^2t}\lim_{k\to\infty}\big(kM^{(1)}(k;x,t)\big)_{12}.
\end{equation}

\subsection{Second transformation: $M^{(1)}\to M^{(2)}$}
Since the branch cut of $g$ is $[-c_r,c_r]$, we have $g_+=g_-\in\R$ on $(-c_l,-c_r)\cup(c_r,c_l)$, where the jump matrix \eqref{eq:V1}
admits the factorization
\begin{equation}\label{eq:fac-mid}
V^{(1)}=\begin{pmatrix}1&-r^*_-\e^{-2itg}\\0&1\end{pmatrix}
\begin{pmatrix}1&0\\ r_+\e^{2itg}&1\end{pmatrix}
\end{equation}
by virtue of $r_+r^*_-=1$; see Proposition~\ref{prop:scattering}(d). The same factorization, with $r_\pm$ replaced
by $r$, holds on $(-\infty,-c_l)\cup(c_l,+\infty)$. Together with the signature table in
Figure~\ref{fig:sign}, this indicates that it is not necessary to introduce an auxiliary
$D$-function to open lenses. Nevertheless, we still need an analytic approximation of the
reflection coefficient. To this end, we set
\begin{equation}\label{eq:rtilde}
\tilde r(k):=\begin{cases}r_+(k), & k\in(-c_l,-c_r)\cup(c_r,c_l),\\
r(k), & k\in(-\infty,-c_l)\cup(c_l,+\infty).\end{cases}
\end{equation}
By \eqref{eq:r-endpoints}, $\tilde r$ is continuous on $(-\infty,-c_r)\cup(c_r,+\infty)$ and admits expansions in
powers of $(k\mp c_l)^{1/2}$ near $\pm c_l$; moreover, $\tilde r(c_r)=b_{r,0}$ and
$\tilde r(-c_r)=\overline{b_{r,0}}$.

\begin{figure}[htbp]
\centering
\includegraphics[width=0.98\textwidth]{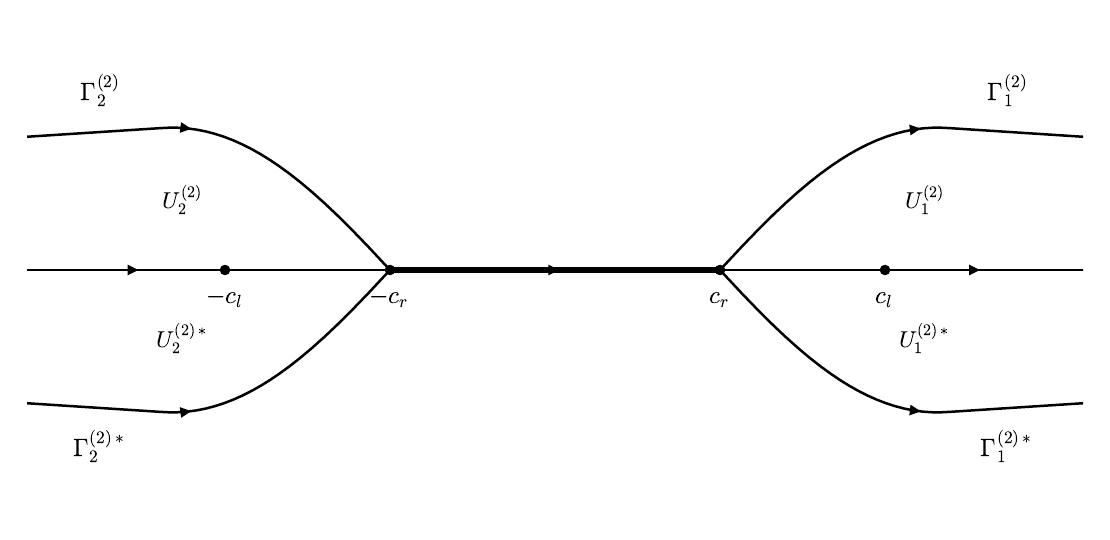}
\caption{The jump contour $\Gamma^{(2)}$ and the domains $U^{(2)}_j$, $U^{(2)*}_j$, $j=1,2$, of
RH problem~\ref{rhp:M2} for $\xi\in\TIIm$.}
\label{fig:contour}
\end{figure}

Let $U^{(2)}_1$ (resp.\ $U^{(2)}_2$) be the domain in $\C^+$ bounded by $(c_r,+\infty)$ (resp.\
$(-\infty,-c_r)$) and a contour $\Gamma^{(2)}_1$ (resp.\ $\Gamma^{(2)}_2$) emanating from $c_r$
(resp.\ $-c_r$) to infinity, as shown in Figure~\ref{fig:contour}. The contours are chosen such
that $U^{(2)}_1\cup U^{(2)}_2\subset\{k\in\C^+:\operatorname{Im}g(k)>0\}$ outside the disk
$D_\varrho(-c_r)$ introduced in Section~\ref{subsubsec:TIIm-local}, and such that, inside $D_\varrho(-c_r)$,
$\Gamma^{(2)}_2$ is mapped onto a ray of the Painlev\'e XXXIV model by the local conformal map
defined there. This is possible since $\hat\eta_r$ is bounded away from $c_r$, while the zero
$-\eta_r$ lies within distance $\cO(t^{-2/3})$ of $-c_r$; see Figure~\ref{fig:sign}. We denote
by $U^{(2)*}_j$ and $\Gamma^{(2)*}_j$, $j=1,2$, the complex conjugates of $U^{(2)}_j$ and
$\Gamma^{(2)}_j$.

\begin{proposition}\label{prop:ra}
There exist continuous functions
\begin{equation*}
r_a:\ \overline{U^{(2)}_1}\cup \overline{U^{(2)}_2}\times\TIIm\to\C\qquad\text{and}\qquad
r_r:\ \big((-\infty,-c_r)\cup(c_r,+\infty)\big)\times\TIIm\to\C,
\end{equation*}
which satisfy the following properties.
\begin{enumerate}
\item[(a)] $\tilde r(k)=r_a(k)+r_r(k)$ for all
$(k;\xi)\in\big((-\infty,-c_r)\cup(c_r,+\infty)\big)\times\TIIm$, where $r_a(k)=r_a(k;\xi)$ and
$r_r(k)=r_r(k;\xi)$.
\item[(b)] For $\xi\in\TIIm$, $r_a(k)$ is holomorphic in $U^{(2)}_1\cup U^{(2)}_2$. Moreover, for
$\epsilon\in\{\pm1\}$ and $(k;\xi)\in U^{(2)}_{j_\epsilon}\times\TIIm$, where $j_{+}=1$ and
$j_{-}=2$, we have
\begin{equation*}
\big|r_a(k)-\tilde r(\epsilon c_r)\big|\lesssim|k-\epsilon c_r|^{1/2}\,\e^{\frac t4|\operatorname{Im}g(k)|}
\qquad\text{and}\qquad
|r_a(k)|\lesssim\frac{\e^{\frac t4|\operatorname{Im}g(k)|}}{1+|k|^2}.
\end{equation*}
\item[(c)] For $\xi\in\TIIm$, $r_r\in L^p\big((-\infty,-c_r)\cup(c_r,+\infty)\big)$,
$p\in[1,+\infty]$, and as $t\to+\infty$,
\begin{equation*}
\|r_r\|_{L^p((-\infty,-c_r)\cup(c_r,+\infty))}=\cO(t^{-1}).
\end{equation*}
\end{enumerate}
\end{proposition}

\begin{proof}
The construction is analogous to those in Lemma~5.2 of \cite{FLQ25} and Proposition~4.2 of \cite{FWZ}.
\end{proof}

Now we are ready to open lenses. We introduce the transformation
\begin{equation}\label{eq:M2}
M^{(2)}(k)=M^{(2)}(k;x,t):=M^{(1)}(k)G(k),
\end{equation}
where
\begin{equation}\label{eq:G}
G(k):=\begin{cases}
\begin{pmatrix}1&0\\ -r_a(k)\e^{2itg(k)}&1\end{pmatrix}, & k\in U^{(2)}_1\cup U^{(2)}_2,\\[10pt]
\begin{pmatrix}1&-r_a^*(k)\e^{-2itg(k)}\\0&1\end{pmatrix}, & k\in U^{(2)*}_1\cup U^{(2)*}_2,\\[10pt]
I, & \text{elsewhere},
\end{cases}
\end{equation}
with $r_a^*(k):=\overline{r_a(\bar k)}$. Then $M^{(2)}$ satisfies the following RH problem.

\begin{rhp}\label{rhp:M2}\leavevmode
\begin{itemize}
\item $M^{(2)}(k)$ is holomorphic for $k\in\C\setminus\Gamma^{(2)}$, where
$\Gamma^{(2)}:=\bigcup_{j=1}^2\big(\Gamma^{(2)}_j\cup\Gamma^{(2)*}_j\big)\cup\R$; see
Figure~\ref{fig:contour}.
\item $M^{(2)}(k)$ has continuous boundary values $M^{(2)}_\pm(k)$ on $\Gamma^{(2)}$ with
the jump condition
\begin{equation*}
M^{(2)}_+(k)=M^{(2)}_-(k)V^{(2)}(k),
\end{equation*}
where
\begin{equation}\label{eq:V2}
V^{(2)}(k)=\begin{cases}
\begin{pmatrix}1&0\\ r_a(k)\e^{2itg(k)}&1\end{pmatrix}, & k\in\Gamma^{(2)}_1\cup\Gamma^{(2)}_2,\\[10pt]
\begin{pmatrix}1&-r_a^*(k)\e^{-2itg(k)}\\0&1\end{pmatrix}, & k\in\Gamma^{(2)*}_1\cup\Gamma^{(2)*}_2,\\[10pt]
\begin{pmatrix}1-r_r(k)r_r^*(k)&-r_r^*(k)\e^{-2itg(k)}\\ r_r(k)\e^{2itg(k)}&1\end{pmatrix},
& k\in(-\infty,-c_r)\cup(c_r,+\infty),\\[10pt]
\begin{pmatrix}0&-1\\1&0\end{pmatrix}, & k\in(-c_r,c_r),
\end{cases}
\end{equation}
with $r_r^*(k):=\overline{r_r(k)}$ for $k\in\R$.
\item As $k\to\infty$ in $\C\setminus\Gamma^{(2)}$, we have $M^{(2)}(k)=I+\cO(k^{-1})$.
\item As $k\to\pm c_r$, $M^{(2)}(k)=\cO\big((k\mp c_r)^{-1/4}\big)$; as $k\to\pm c_l$,
$M^{(2)}(k)=\cO(1)$.
\item $M^{(2)}(k)=\sigma_1\overline{M^{(2)}(\bar k)}\sigma_1$ for $k\in\C\setminus\Gamma^{(2)}$.
\end{itemize}
\end{rhp}

\subsection{Analysis of RH problem for $M^{(2)}$}
By Proposition~\ref{prop:ra} and the signature table of $\operatorname{Im}g$, $V^{(2)}(k)\to I$ as
$t\to+\infty$ on $\Gamma^{(2)}_j\cup\Gamma^{(2)*}_j$, $j=1,2$, except near $\pm c_r$, and on
$(-\infty,-c_r)\cup(c_r,+\infty)$ in the $L^p$ sense. Hence, to leading order, $M^{(2)}$ is
approximated by the solution of a RH problem with the constant jump on $(-c_r,c_r)$ only.

\subsubsection{Global parametrix}
\begin{rhp}\label{rhp:Minf}\leavevmode
\begin{itemize}
\item $M^{(\infty)}(k)$ is holomorphic for $k\in\C\setminus[-c_r,c_r]$.
\item $M^{(\infty)}_+(k)=M^{(\infty)}_-(k)\left(\begin{smallmatrix}0&-1\\1&0\end{smallmatrix}\right)$ for
$k\in(-c_r,c_r)$.
\item As $k\to\infty$ in $\C\setminus[-c_r,c_r]$, we have $M^{(\infty)}(k)=I+\cO(k^{-1})$.
\item As $k\to\pm c_r$, $M^{(\infty)}(k)=\cO\big((k\mp c_r)^{-1/4}\big)$.
\end{itemize}
\end{rhp}
The unique solution is
\begin{equation}\label{eq:Minf}
M^{(\infty)}(k)=\Delta_r(k),
\end{equation}
with $\Delta_r$ given in \eqref{eq:Delta}. It satisfies $M^{(\infty)}(k)=\sigma_1M^{(\infty)}(-k)\sigma_1$ and,
as $k\to c_r$,
\begin{multline}\label{eq:Minf-exp}
M^{(\infty)}(k)=\frac{(2c_r)^{1/4}}{2(k-c_r)^{1/4}}\bigg[
\begin{pmatrix}1&-i\\ i&1\end{pmatrix}
+\frac{(k-c_r)^{1/2}}{(2c_r)^{1/2}}\begin{pmatrix}1&i\\ -i&1\end{pmatrix}
+\frac{k-c_r}{8c_r}\begin{pmatrix}1&-i\\ i&1\end{pmatrix}\\
+\frac{(k-c_r)^{3/2}}{4(2c_r)^{3/2}}\begin{pmatrix}-1&-i\\ i&-1\end{pmatrix}
+\cO\big((k-c_r)^2\big)\bigg].
\end{multline}
The expansion near $-c_r$ follows from the symmetry $M^{(\infty)}(k)=\sigma_1M^{(\infty)}(-k)\sigma_1$.

\subsubsection{A reflection principle}
The local analysis near $-c_r$ can be reduced to an analysis near $c_r$ by the following observation. Let
\begin{equation}\label{eq:gcheck}
\check g(k):=-g(-k)=\big(4\beta k^2-2\alpha k+\xi+2\beta c_r^2\big)X_r(k),
\end{equation}
i.e., $\check g$ is the $g$-function \eqref{eq:g} with $\alpha$ replaced by $-\alpha$.

\begin{lemma}\label{lem:reflection}
Let $\check M^{(j)}(k):=\sigma_1M^{(j)}(-k)\sigma_1$, $j=1,2$. Then $\check M^{(1)}$ solves
RH problem~\ref{rhp:M1} with $g$ replaced by $\check g$, and $\check M^{(2)}$ solves
RH problem~\ref{rhp:M2} with $g$, $r_a$, $r_r$, $U^{(2)}_j$, $\Gamma^{(2)}_j$ replaced by
$\check g$, $\check r_a(k):=\overline{r_a(-\bar k)}$, $\check r_r(k):=\overline{r_r(-k)}$,
$-\overline{U^{(2)}_{3-j}}$, $-\overline{\Gamma^{(2)}_{3-j}}$, respectively. In particular,
near $c_r$, the jumps of $\check M^{(2)}$ are of the same form as those of $M^{(2)}$ near $c_r$,
with the same value $\tilde r(c_r)=b_{r,0}$.
\end{lemma}

\begin{proof}
The map $k\mapsto-k$ reverses the orientation of $\R$ and interchanges $\C^\pm$, so that
$\check M^{(1)}_+=\check M^{(1)}_-\big(\sigma_1V^{(1)}(-k)\sigma_1\big)^{-1}$ on $\R$. On
$(-\infty,-c_l)\cup(c_l,+\infty)$ we use $r(-k)=r^*(k)$, which follows from \eqref{eq:sym-r},
and $\e^{2itg(-k)}=\e^{-2it\check g(k)}$, to obtain
\begin{equation*}
\big(\sigma_1V^{(1)}(-k)\sigma_1\big)^{-1}
=\begin{pmatrix}1&r^*(k)\e^{-2it\check g(k)}\\ -r(k)\e^{2it\check g(k)}&1-r(k)r^*(k)\end{pmatrix}^{-1}
=\begin{pmatrix}1-r(k)r^*(k)&-r^*(k)\e^{-2it\check g(k)}\\ r(k)\e^{2it\check g(k)}&1\end{pmatrix}.
\end{equation*}
On $(-c_l,-c_r)\cup(c_r,c_l)$ the same computation applies, using $r_+(-k)=r^*_-(k)$,
$r^*_-(-k)=r_+(k)$ and $r_+r^*_-=1$. On $(-c_r,c_r)$,
$\big(\sigma_1\left(\begin{smallmatrix}0&-1\\1&0\end{smallmatrix}\right)\sigma_1\big)^{-1}
=\left(\begin{smallmatrix}0&-1\\1&0\end{smallmatrix}\right)$. The normalization, the local
behaviour and the $\sigma_1$-symmetry are clearly preserved. The statement for $\check M^{(2)}$
follows in the same way, noting that $\sigma_1$ interchanges lower and upper triangular factors
and that $k\mapsto-k$ maps $U^{(2)}_2$ onto $-U^{(2)}_2\subset\C^-$; the value at $c_r$ is
$\overline{\tilde r(-c_r)}=b_{r,0}$ by Proposition~\ref{prop:scattering}(e).
\end{proof}

\subsubsection{Local parametrices near $\pm c_r$}\label{subsubsec:TIIm-local}
Since $-c_r$ is critical while $c_r$ is not, we use disks of different sizes:
\begin{equation*}
D_l:=\{k:|k+c_r|<\varrho_l\},\quad \varrho_l<\tfrac13\min\big\{c_l-c_r,\ t^{-\frac23+\epsilon}\big\},
\ \tfrac12<\epsilon<\tfrac23;\qquad
D_r:=\{k:|k-c_r|<\varrho_r\},
\end{equation*}
where $\varrho_r<\frac13\min\{c_r-\hat\eta_r,\,c_l-c_r\}$ is a fixed number, independent of $t$,
which will be further restricted in Lemma~\ref{lem:local}. Then $-\eta_r\in D_l$ for $t$ large,
whereas $\hat\eta_r\notin D_r$. For $j\in\{r,l\}$, we seek $M^{(j)}$ holomorphic in
$D_j\setminus\Gamma^{(j)}$, where $\Gamma^{(j)}:=D_j\cap\Gamma^{(2)}$, with
$M^{(j)}_+=M^{(j)}_-V^{(2)}$ on $\Gamma^{(j)}$.

\paragraph{\it A model construction at $c_r$.}
Let $h$ denote either $g$ or $\check g$, and let $D$ be a disk centred at $c_r$ of radius $\varrho\in\{\varrho_l,\varrho_r\}$. By Proposition~\ref{prop:g}, as $k\to c_r$,
\begin{equation*}
h(k)=h^{(1)}(k-c_r)^{\frac12}+h^{(2)}(k-c_r)^{\frac32}+\cO\big((k-c_r)^{\frac52}\big),
\end{equation*}
where
\begin{align}
&g^{(1)}=g^{(1)}_+=(2c_r)^{\frac12}(\xi+6\beta c_r^2+2\alpha c_r),
&& g^{(2)}=g^{(2)}_+=(2c_r)^{\frac12}\Big[\tfrac{\xi+6\beta c_r^2+2\alpha c_r}{4c_r}+8\beta c_r+2\alpha\Big],\label{eq:g12-plus}\\
&\check g^{(1)}=-g^{(1)}_-=(2c_r)^{\frac12}(\xi-\xi_{\rm II}^-),
&& \check g^{(2)}=-g^{(2)}_-=(2c_r)^{\frac12}\Big[\tfrac{\xi-\xi_{\rm II}^-}{4c_r}+8\beta c_r-2\alpha\Big].\label{eq:g12-check}
\end{align}
Note that $h^{(2)}>0$ for $\xi\in\TIIm$ (recall $0<\alpha<4\beta c_r$). Since
$h(k)-h^{(1)}(k-c_r)^{\frac12}=h^{(2)}(k-c_r)^{\frac32}+\cO\big((k-c_r)^{\frac52}\big)$, the
bracket has a zero of order $\frac32$ at $k=c_r$; taking its $\frac23$-power directly is ambiguous,
because the principal branch of $[\,\cdot\,]^{2/3}$ is not determined by the requirement that
$\zeta_h$ be the local Painlev\'e coordinate once $|\arg(k-c_r)|>\frac{2\pi}{3}$. We therefore
factor out an analytic function first: set
\begin{equation}\label{eq:zeta-H}
H_h(k):=\frac{h(k)-h^{(1)}(k-c_r)^{\frac12}}{(k-c_r)^{\frac32}},\qquad k\in D\setminus\{c_r\},
\end{equation}
which extends holomorphically to $D$ with $H_h(c_r)=h^{(2)}>0$ and $H_h\neq0$ on $D$, and then
\begin{equation}\label{eq:zeta}
\zeta_h(k):=-\Big(\frac32\Big)^{\frac23}t^{\frac23}\,(k-c_r)\,\big(H_h(k)\big)^{\frac23},
\end{equation}
where $\big(H_h\big)^{2/3}:=\exp\big(\tfrac23\log H_h\big)$ is defined by the principal branch of
the logarithm, which is legitimate because $H_h(D)$ does not meet $(-\infty,0]$. Then $\zeta_h$ is
\emph{single-valued and analytic} on $D$, and
$\zeta_h'(c_r)=-\big(\frac32\big)^{2/3}t^{2/3}\big(h^{(2)}\big)^{2/3}<0$, so that $\zeta_h$ is a
conformal map of $D$ onto a neighbourhood of the origin. Equivalently, $\zeta_h$ is characterized
by the single-valued relation
\begin{equation}\label{eq:zeta-cubic}
\zeta_h(k)^3=-\Big(\frac32\Big)^2t^2\Big[h(k)-h^{(1)}(k-c_r)^{\frac12}\Big]^2,
\end{equation}
which shows that \eqref{eq:zeta} coincides with the naive definition wherever the latter is
unambiguous, and extends it to all of $D$. Moreover
\begin{equation*}
S_h(k):=\begin{cases}
it\,h^{(1)}(k-c_r)^{\frac12}\zeta_h(k)^{-\frac12}, & k\in D\cap\C^+,\\
-it\,h^{(1)}(k-c_r)^{\frac12}\zeta_h(k)^{-\frac12}, & k\in D\cap\C^-,
\end{cases}
\end{equation*}
is analytic in $D$, the two signs compensating the change of sign of $(k-c_r)^{\frac12}$ across
$\R\cap D$, while $\zeta_h$ is analytic on $D$ and real on $\R$ near $c_r$. Then
\begin{equation}\label{eq:phase-id}
\tfrac43\zeta_h(k)^{\frac32}+2S_h(k)\zeta_h(k)^{\frac12}=\pm2it\,h(k),\qquad
k\in D\cap\C^\pm,
\end{equation}
and we set $\hat\theta_h(\zeta_h):=\frac23\zeta_h^{3/2}+S_h(k)\zeta_h^{1/2}$ and
\begin{equation}\label{eq:s-h}
s_h:=S_h(c_r)=-t^{\frac23}\frac{h^{(1)}}{\big(\frac32\big)^{1/3}\big(h^{(2)}\big)^{1/3}} .
\end{equation}
Following \cite[(4.10)--(4.12)]{FWZ}, we define
\begin{equation}\label{eq:Mtilde-h}
\widetilde M_h(k):=P_h(k)\begin{pmatrix}1&0\\ ia(S_h(k))&1\end{pmatrix}
M^{P34}\Big(\zeta_h(k);-\frac{\arg r(c_r)}{2\pi},0,S_h(k)\Big)\e^{\hat\theta_h(\zeta_h)\sigma_3}Q(k),
\end{equation}
where $r(c_r):=\tilde r(c_r)=b_{r,0}$, $M^{P34}$ and $a(s)$ are given in Appendix~A. Here the
parameter of the model is the analytic function $S_h(k)$ rather than the constant $s_h$: since
the jumps of $M^{P34}$ do not depend on $s$, $\widetilde M_h$ is still holomorphic off the
contour, and with this choice the exponential factors in \eqref{eq:A1} and in
\eqref{eq:Mtilde-h} cancel exactly, $\frac23\zeta_h^{3/2}+S_h(k)\zeta_h^{1/2}=\hat\theta_h(\zeta_h)$.
Moreover,
\begin{equation*}
Q(k):=\begin{cases}\sigma_3,&k\in\C^+\cap D,\\ \sigma_1,&k\in\C^-\cap D,\end{cases}
\qquad
P_h(k):=M^{(\infty)}(k)Q(k)^{-1}\frac{1}{\sqrt2}(I-i\sigma_1)\zeta_h(k)^{\frac{\sigma_3}{4}}.
\end{equation*}
As in \cite[Section~4.3.2]{FWZ}, $P_h$ is analytic in $D$ and, as $k\to c_r$,
\begin{equation}\label{eq:Ph-c}
P_h(k)=\begin{pmatrix}
\frac{1-i}{2}(2c_r)^{\frac14}\big(\frac32\big)^{\frac16}t^{\frac16}\big(h^{(2)}\big)^{\frac16} &
\frac{1-i}{2}(2c_r)^{-\frac14}\big(\frac32\big)^{-\frac16}t^{-\frac16}\big(h^{(2)}\big)^{-\frac16}\\[4pt]
\frac{1+i}{2}(2c_r)^{\frac14}\big(\frac32\big)^{\frac16}t^{\frac16}\big(h^{(2)}\big)^{\frac16} &
\frac{-1-i}{2}(2c_r)^{-\frac14}\big(\frac32\big)^{-\frac16}t^{-\frac16}\big(h^{(2)}\big)^{-\frac16}
\end{pmatrix}+\cO(|k-c_r|).
\end{equation}

\paragraph{\it Local parametrices.}
We define
\begin{equation}\label{eq:Mtilde-rl}
\widetilde M^{(r)}(k):=\widetilde M_g(k),\quad k\in D_r;\qquad
\widetilde M^{(l)}(k):=\sigma_1\widetilde M_{\check g}(-k)\sigma_1,\quad k\in D_l,
\end{equation}
where in $\widetilde M_{\check g}$ the lenses are those of Lemma~\ref{lem:reflection}. By
Lemma~\ref{lem:reflection}, $\widetilde M^{(l)}$ has, up to the replacement of $r_a$ by its value at
$-c_r$, the same jumps as $M^{(2)}$ in $D_l$. The two parameters are
\begin{align}
s_l&:=s_{\check g}=-\Big(\frac23\Big)^{\frac13}\Big(\frac{c_r}{4\beta c_r-\alpha}\Big)^{\frac13}
t^{\frac23}\big(\xi-\xi_{\rm II}^-\big)\big(1+\cO(t^{-\frac23})\big)=\cO(1),\label{eq:s-l}\\
s_r&:=s_g=-\frac{4\cdot6^{2/3}\,\alpha\,c_r^{4/3}}{3\,(8\beta c_r+3\alpha)^{1/3}}\,t^{\frac23}
\big(1+\cO(t^{-\frac23})\big)\to-\infty.\label{eq:s-r}
\end{align}
In particular, the Painlev\'e XXXIV parameter is $b=-\frac{\arg r(c_r)}{2\pi}$ at both
branch points. Throughout the construction we assume
\begin{equation}\label{eq:r-ne-1}
r(c_r)\neq-1,\qquad\text{i.e.}\qquad b\in\big(-\tfrac12,\tfrac12\big)\setminus\{0\}.
\end{equation}

\begin{lemma}\label{lem:local}
For $\xi\in\TIIm$ and $t>0$, the following hold.
\begin{enumerate}
\item[(a)] As $t\to+\infty$,
\begin{equation*}
\big\|\widetilde M^{(l)}(k)M^{(\infty)}(k)^{-1}-I\big\|_{L^\infty(\partial D_\varrho(-c_r))}
=\cO(t^{1/3-\epsilon}),
\end{equation*}
and the jump $\widetilde V^{(l)}$ of $\widetilde M^{(l)}$ satisfies
\begin{equation*}
\|V^{(2)}-\widetilde V^{(l)}\|_{L^p(\Gamma^{(l)})}=\begin{cases}\cO(t^{-1}),&p=1,\\ \cO(t^{-2/3}),&p=2,\\ \cO(t^{-1/3}),&p=\infty.\end{cases}
\end{equation*}
\item[(b)] If $\varrho_r$ is sufficiently small, then, as $t\to+\infty$,
\begin{equation*}
\big\|\widetilde M^{(r)}(k)M^{(\infty)}(k)^{-1}-I\big\|_{L^\infty(\partial D_r)}=\cO(t^{-1}),\qquad
\|V^{(2)}-\widetilde V^{(r)}\|_{L^p(\Gamma^{(r)})}=\cO(t^{-1}),\quad p=1,2,\infty .
\end{equation*}
\end{enumerate}
\end{lemma}

\begin{proof}
Since $s_l$ is bounded, (a) is proved as \cite[Lemma~3.5]{FWZ}, with $s$ replaced by $S_{\check g}(k)$,
which ranges over a compact set.

For (b), write $S_g(k)=t^{2/3}F(k)$, where, by \eqref{eq:zeta} and the definition of $S_g$,
$F$ is analytic in $D_r$, independent of $t$ up to a factor $1+\cO(t^{-2/3})$, and
$F(c_r)=t^{-2/3}s_r<0$. Choosing $\varrho_r$ small, we have $|\arg(-S_g(k))|\le\delta$ and
$|S_g(k)|\ge\frac12|s_r|\to\infty$ for $k\in D_r$, with $\delta$ as in Lemma~\ref{lem:uniform}.
Moreover, $|\zeta_g(k)|\ge c\,t^{2/3}$ for $k\in\partial D_r$. By
$\frac{1}{\sqrt2}(I-i\sigma_1)=U^{-1}$, $U:=\frac{1}{\sqrt2}\left(\begin{smallmatrix}1&i\\ i&1\end{smallmatrix}\right)$,
we have $P_g^{-1}=\zeta_g^{-\sigma_3/4}UQM^{(\infty)\,-1}$, and Lemma~\ref{lem:uniform} gives, for
$k\in\partial D_r$,
\begin{equation*}
\widetilde M^{(r)}(k)M^{(\infty)}(k)^{-1}=P_g(k)\big(I+\mathcal E(\zeta_g(k);S_g(k))\big)P_g(k)^{-1},
\qquad |\mathcal E|\lesssim\frac{1}{|S_g||\zeta_g|}\lesssim t^{-4/3}.
\end{equation*}
Since $|P_g^{\pm1}(k)|\lesssim|M^{(\infty)}(k)^{\pm1}|\,|\zeta_g(k)|^{1/4}\lesssim t^{1/6}$ on $\partial D_r$,
the first estimate follows. For the second, on $\Gamma^{(r)}\setminus\R$ the only nonzero entry of
$V^{(2)}-\widetilde V^{(r)}$ is $\big(r_a(k)-r(c_r)\big)\e^{\pm2itg(k)}$ (or its conjugate), and
$\operatorname{Im}g(k)\ge c|k-c_r|^{1/2}$ on $\Gamma^{(r)}\cap\C^+$ because $g^{(1)}_+$ is bounded away from
zero. By Proposition~\ref{prop:ra}(b), this entry is bounded by
$|k-c_r|^{1/2}\e^{-ct|k-c_r|^{1/2}}$, whose $L^p$ norms are $\cO(t^{-1})$ or smaller. On
$\Gamma^{(r)}\cap\R$ the difference is controlled by Proposition~\ref{prop:ra}(c).
\end{proof}

\subsection{Small norm RH problem for $M^{(err)}$}
We define
\begin{equation}\label{eq:Merr}
M^{(err)}(k):=\begin{cases}
M^{(2)}(k)M^{(\infty)}(k)^{-1}, & k\in\C\setminus\big(\overline{D_l}\cup\overline{D_r}\big),\\
M^{(2)}(k)\widetilde M^{(l)}(k)^{-1}, & k\in D_l,\\
M^{(2)}(k)\widetilde M^{(r)}(k)^{-1}, & k\in D_r .
\end{cases}
\end{equation}
\begin{rhp}\label{rhp:Merr}\leavevmode
\begin{itemize}
\item $M^{(err)}$ is holomorphic in $\C\setminus\Gamma^{(err)}$, where
$\Gamma^{(err)}:=\partial D_l\cup\partial D_r\cup\big(\Gamma^{(2)}\setminus[-c_r,c_r]\big)$, the
circles being oriented clockwise.
\item $M^{(err)}_+=M^{(err)}_-V^{(err)}$ on $\Gamma^{(err)}$, where
\begin{equation}\label{eq:Verr}
V^{(err)}=\begin{cases}
M^{(\infty)}V^{(2)}M^{(\infty)\,-1}, & k\in\Gamma^{(2)}\setminus\big([-c_r,c_r]\cup\overline{D_l}\cup\overline{D_r}\big),\\
\widetilde M^{(j)}M^{(\infty)\,-1}, & k\in\partial D_j,\ j\in\{l,r\},\\
\widetilde M^{(j)}_-V^{(2)}\big(\widetilde V^{(j)}\big)^{-1}\widetilde M^{(j)\,-1}_-, & k\in\Gamma^{(2)}\cap D_j,\ j\in\{l,r\}.
\end{cases}
\end{equation}
\item $M^{(err)}(k)=I+\cO(k^{-1})$ as $k\to\infty$, and $M^{(err)}(k)=\cO(1)$ as $k\to\pm c_r$.
\end{itemize}
\end{rhp}
Note that $M^{(err)}$ has no jump on $(-c_r,c_r)$ outside the disks, since $M^{(2)}$ and $M^{(\infty)}$ have the same jump there.

\begin{proposition}\label{prop:Verr}
For $\xi\in\TIIm$, as $t\to+\infty$, the function $V^{(err)}$ admits the following estimates
\begin{align}
&\|V^{(err)}-I\|_{L^\infty(\Gamma^{(err)}\setminus(\overline{D_l}\cup\overline{D_r}\cup\R))}
=\cO\big(\e^{-ct^{3\epsilon/2}}\big),\label{eq:Verr-lens}\\
&\|V^{(err)}-I\|_{L^1(\R\setminus(D_l\cup D_r))}=\cO(t^{-1}),\qquad
\|V^{(err)}-I\|_{L^\infty(\R\setminus(D_l\cup D_r))}=\cO(t^{-\frac23-\frac\epsilon2}),\label{eq:Verr-R}\\
&\|V^{(err)}-I\|_{L^\infty(\partial D_l)}=\cO(t^{\frac13-\epsilon}),\qquad
\|V^{(err)}-I\|_{L^2(\partial D_l)}=\cO(t^{-\frac\epsilon2}),\label{eq:Verr-l}\\
&\|V^{(err)}-I\|_{L^\infty(\partial D_r)}=\cO(t^{-1}),\label{eq:Verr-r}\\
&\|V^{(err)}-I\|_{L^p(\Gamma^{(err)}\cap D_r)}=\cO(t^{-1}),\qquad p=1,2,\infty,\label{eq:Verr-in-r}\\
\end{align}
Consequently
\begin{equation}\label{eq:Verr-global}
\|V^{(err)}-I\|_{L^\infty(\Gamma^{(err)})}=\cO(t^{\frac13-\epsilon}),\qquad
\|V^{(err)}-I\|_{L^2(\Gamma^{(err)})}=\cO(t^{-\frac\epsilon2}),
\end{equation}
and the small norm condition holds as soon as $\epsilon>\frac13$.
\end{proposition}

\begin{proof}
These conclusions follow from item (c) of Proposition~\ref{prop:ra}(c) together with items (a) and (b) of Lemma~\ref{lem:local}.
\end{proof}

By Proposition~\ref{prop:Verr}, the Cauchy operator $C_{V^{(err)}}f:=C_-\big(f(V^{(err)}-I)\big)$ satisfies
\begin{equation*}
\|C_{V^{(err)}}\|_{L^2(\Gamma^{(err)})\to L^2(\Gamma^{(err)})}\lesssim
\|V^{(err)}-I\|_{L^\infty(\Gamma^{(err)})}=\cO(t^{\frac13-\epsilon})\to0
\end{equation*}
since $\epsilon>\frac13$, so the small norm condition is satisfied for large $t$. Hence, by the
small norm theory \cite{DZ02}, the RH problem \eqref{eq:Merr} has a unique solution for $t$ large,
\begin{equation}\label{eq:Merr-int}
M^{(err)}(k)=I+\frac{1}{2\pi i}\int_{\Gamma^{(err)}}
\frac{(I+\varpi(z))(V^{(err)}(z)-I)}{z-k}\,\di z,
\end{equation}
where $\varpi\in L^2(\Gamma^{(err)})$ solves $(1-C_{V^{(err)}})\varpi=C_{V^{(err)}}I$ and satisfies
$\|\varpi\|_{L^2(\Gamma^{(err)})}\lesssim\|V^{(err)}-I\|_{L^2(\Gamma^{(err)})}
=\cO(t^{-\epsilon/2})$. In particular, $M^{(err)}(k)=I+M^{(err)}_1k^{-1}+\cO(k^{-2})$ with
\begin{equation}\label{eq:Merr1}
M^{(err)}_1=-\frac{1}{2\pi i}\int_{\Gamma^{(err)}}(I+\varpi(z))(V^{(err)}(z)-I)\,\di z .
\end{equation}

\begin{proposition}\label{prop:Merr1}
For $\xi\in\TIIm$, as $t\to+\infty$,
\begin{equation}\label{eq:Merr1-asy}
M^{(err)}_1=\frac{t^{-2/3}}{\big(\frac32\big)^{2/3}\big(\check g^{(2)}\big)^{2/3}}\,
\sigma_1P_{\check g}(c_r)\,M_1^{P34}(s_l)\,P_{\check g}(c_r)^{-1}\sigma_1+\cO(t^{-\epsilon}),
\end{equation}
where $P_{\check g}(c_r)$ is given by \eqref{eq:Ph-c} with $h=\check g$. In particular,
\begin{equation}\label{eq:Merr1-12}
\big(M^{(err)}_1\big)_{12}=-\frac{(2c_r)^{1/2}}{2\big(\frac32\big)^{1/3}\big(\check g^{(2)}\big)^{1/3}}\,
a(s_l)\,t^{-\frac13}
+\frac{i\,t^{-2/3}}{2\big(\frac32\big)^{2/3}\big(\check g^{(2)}\big)^{2/3}}\big(a(s_l)^2-a'(s_l)\big)
+\cO(t^{-\epsilon}).
\end{equation}
\end{proposition}

\begin{proof}
Split \eqref{eq:Merr1} into $I_1+I_2+I_3+I_4$ as in \cite[Proposition~3.7]{FWZ}, where $I_1$
contains $\varpi$, and $I_2$, $I_3$, $I_4$ are the integrals of $-(V^{(err)}-I)/(2\pi i)$ over
$\Gamma^{(err)}\setminus(\overline{D_l}\cup\overline{D_r})$, $\partial D_l\cup\partial D_r$ and
$\Gamma^{(err)}\cap(D_l\cup D_r)$, respectively. By the Cauchy--Schwarz inequality and
Proposition~\ref{prop:Verr},
$|I_1|\le\|\varpi\|_{L^2}\|V^{(err)}-I\|_{L^2}=\cO(t^{-\epsilon})$, and
$I_2=\cO(t^{-1})$. For
$I_4$ we use that by the lens $\Gamma^{(l)}$ carries no jump at all in the
corrected problem, so $I_4$ reduces to the integral over $\Gamma^{(r)}\cap D_r$, where the jump
is $\cO(t^{-1})$ in $L^1$ by \eqref{eq:Verr-in-r}; hence $I_4=\cO(t^{-1})$.
The contribution of $\partial D_r$ to $I_3$ is $\cO(t^{-1})$ by \eqref{eq:Verr-r}.

It remains to compute the contribution of $\partial D_l$. By \eqref{eq:Mtilde-rl},
$M^{(\infty)}(k)=\sigma_1M^{(\infty)}(-k)\sigma_1$, \eqref{eq:A1} with $s=S_{\check g}(-k)$, and
$P_{\check g}^{-1}=N(\zeta_{\check g})QM^{(\infty)\,-1}$,
\begin{equation*}
\widetilde M^{(l)}(k)M^{(\infty)}(k)^{-1}-I
=\sigma_1P_{\check g}(-k)\Big(\frac{M_1^{P34}(S_{\check g}(-k))}{\zeta_{\check g}(-k)}
+\cO\big(\zeta_{\check g}(-k)^{-2}\big)\Big)P_{\check g}(-k)^{-1}\sigma_1 ,
\end{equation*}
where the factors $\left(\begin{smallmatrix}1&0\\ \pm ia&1\end{smallmatrix}\right)$ cancel and the
exponentials cancel exactly because the model parameter is $S_{\check g}(-k)$. The first term
has a simple pole at $k=-c_r$, coming from the simple zero of $\zeta_{\check g}$ at $c_r$ with
$\zeta_{\check g}'(c_r)=-(\frac32)^{2/3}t^{2/3}(\check g^{(2)})^{2/3}$. Substituting $\kappa=-k$,
which preserves the orientation of the circle, its residue at $k=-c_r$ equals
$-\sigma_1P_{\check g}(c_r)M_1^{P34}(s_l)P_{\check g}(c_r)^{-1}\sigma_1/\zeta'_{\check g}(c_r)$,
which is the main term in \eqref{eq:Merr1-asy}. The term of order $\zeta^{-2}$ contributes
$\cO(t^{-4/3}\,\|\partial_k(P_{\check g}M_2^{P34}P_{\check g}^{-1})\|)=\cO(t^{-1})$, using
$|P_{\check g}^{\pm1}|,|\partial_kP_{\check g}^{\pm1}|\lesssim t^{1/6}$ and $S_{\check g}'=\cO(1)$; the
remainder is $\cO(t^{1/3}\varrho_l|\zeta|^{-3})=\cO(t^{-1/3-2\epsilon})$. This proves
\eqref{eq:Merr1-asy}.

For \eqref{eq:Merr1-12}, write $P_{\check g}(c_r)$ as in \eqref{eq:Ph-c} with
$A:=(2c_r)^{1/4}(\frac32)^{1/6}t^{1/6}(\check g^{(2)})^{1/6}$ in the first column and $A^{-1}$ in
the second. A direct computation gives, for any $2\times2$ matrix $m$,
\begin{equation*}
\big(\sigma_1P_{\check g}(c_r)\,m\,P_{\check g}(c_r)^{-1}\sigma_1\big)_{12}
=\frac i2\Big(A^2m_{12}+m_{11}-m_{22}-A^{-2}m_{21}\Big).
\end{equation*}
With $m=M_1^{P34}(s_l)$, $m_{12}=ia(s_l)$ and $m_{11}-m_{22}=a^2-a'$ by Lemma~\ref{lem:M1}, and
since $A^{-2}m_{21}$ contributes $\cO(t^{-1})$, we obtain \eqref{eq:Merr1-12}.
\end{proof}

\subsection{Long-time asymptotics in $\TIIm$}
Tracing back the transformations \eqref{eq:M1}, \eqref{eq:M2} and \eqref{eq:Merr}, we have
$M^{(1)}(k)=M^{(err)}(k)M^{(\infty)}(k)$ for $k$ large and outside the lenses. Since
$(M^{(\infty)}(k))_{12}=-\frac{ic_r}{2k}+\cO(k^{-2})$, the reconstruction formula
\eqref{eq:recon1} gives
\begin{equation*}
q(x,t)=\e^{-2i\alpha c_r^2t}\Big(c_r+2i\big(M^{(err)}_1\big)_{12}\Big).
\end{equation*}
Combining this with Proposition~\ref{prop:Merr1}, we arrive at the part (a) of Theorem 1.1.


\section{Asymptotic analysis in the transition region $\mathcal{T}_{\mathrm{I}}^{+}$}\label{sec:TIp}

This section is devoted to the long-time asymptotic analysis of RH problem~\ref{rhp:M} in the transition region $\mathcal{T}_{\mathrm{I}}^{+}$, that is, in a $t^{-2/3}$-neighbourhood of the critical line $\xi=\xi_{\mathrm{I}}^{+}$, where
\begin{equation}\label{eq:xiIp}
\xi_{\mathrm{I}}^{+}:=-6\beta c_l^{2}-2\alpha c_l .
\end{equation}
Throughout this section it is assumed that
\begin{equation}\label{eq:TIp-half}
-C<t^{2/3}\big(\xi-\xi_{\mathrm{I}}^{+}\big)<0,
\end{equation}
since the analysis on the other half of $\mathcal{T}_{\mathrm{I}}^{+}$ is similar. Recall that $\alpha>0$ and $\beta>0$.

\subsection{First transformation: $M\to M^{(1)}$}\label{subsec:TIp-M1}

In order to control the exponentially growing entries of the jump matrix \eqref{eq:V} on the interval $(-c_l,c_l)$, we introduce the $g$-function
\begin{equation}\label{eq:gI}
g_{\mathrm{I}}(k)=g_{\mathrm{I}}(k;\xi):=p_{\mathrm{I}}(k)\,X_l(k),\qquad
p_{\mathrm{I}}(k)=p_{\mathrm{I}}(k;\xi):=4\beta k^{2}+2\alpha k+\xi+2\beta c_l^{2},
\end{equation}
where $X_l(k)=\sqrt{k^{2}-c_l^{2}}$ is given in \eqref{eq:XOmega}, with the branch cut $[-c_l,c_l]$ and the branch fixed by $X_l(k)=k+\mathcal{O}(k^{-1})$ as $k\to\infty$.

\begin{proposition}\label{prop:gI}
The function $g_{\mathrm{I}}$ defined in \eqref{eq:gI} has the following properties.
\begin{itemize}
\item[\rm(a)] $g_{\mathrm{I}}(k)$ is analytic for $k\in\mathbb{C}\setminus[-c_l,c_l]$, and $\overline{g_{\mathrm{I}}(\bar k)}=g_{\mathrm{I}}(k)$.
\item[\rm(b)] As $k\to\infty$ in $\mathbb{C}\setminus[-c_l,c_l]$,
\begin{equation}\label{eq:gI-infty}
g_{\mathrm{I}}(k)=\theta(k)-\alpha c_l^{2}-\frac{c_l^{2}\left(\xi+3\beta c_l^{2}\right)}{2k}+\mathcal{O}(k^{-2}).
\end{equation}
\item[\rm(c)] For $k\in(-c_l,c_l)$, we have $g_{\mathrm{I},+}(k)+g_{\mathrm{I},-}(k)=0$; more precisely,
\begin{equation}\label{eq:gI-cut}
g_{\mathrm{I},\pm}(k)=\pm i\,p_{\mathrm{I}}(k)\sqrt{c_l^{2}-k^{2}},\qquad k\in(-c_l,c_l).
\end{equation}
\item[\rm(d)] Let $\varepsilon\in\{+,-\}$. As $k\to\varepsilon c_l$,
\begin{equation}\label{eq:gI-exp}
g_{\mathrm{I}}(k)=g^{(1)}_{\mathrm{I},\varepsilon}\,(\varepsilon k-c_l)^{\frac12}+g^{(2)}_{\mathrm{I},\varepsilon}\,(\varepsilon k-c_l)^{\frac32}+\mathcal{O}\big(|k-\varepsilon c_l|^{\frac52}\big),
\end{equation}
where the principal branches of the powers are taken, and
\begin{equation}\label{eq:gI-coef}
\begin{aligned}
g^{(1)}_{\mathrm{I},+}&=(2c_l)^{\frac12}\big(\xi-\xi_{\mathrm{I}}^{+}\big), &
g^{(2)}_{\mathrm{I},+}&=(2c_l)^{\frac12}\Big(8\beta c_l+2\alpha+\frac{\xi-\xi_{\mathrm{I}}^{+}}{4c_l}\Big),\\
g^{(1)}_{\mathrm{I},-}&=(2c_l)^{\frac12}\big(4\alpha c_l-(\xi-\xi_{\mathrm{I}}^{+})\big), &
g^{(2)}_{\mathrm{I},-}&=-(2c_l)^{\frac12}\Big(8\beta c_l-3\alpha+\frac{\xi-\xi_{\mathrm{I}}^{+}}{4c_l}\Big).
\end{aligned}
\end{equation}
\end{itemize}
\end{proposition}

\begin{proof}
Items (a)--(c) follow directly from \eqref{eq:gI}: for (b) we use $X_l(k)=k-\frac{c_l^{2}}{2k}-\frac{c_l^{4}}{8k^{3}}+\mathcal{O}(k^{-5})$, which yields
$g_{\mathrm{I}}(k)=4\beta k^{3}+2\alpha k^{2}+\xi k-\alpha c_l^{2}-\frac{c_l^{2}(\xi+3\beta c_l^{2})}{2k}+\mathcal{O}(k^{-2})$; for (c) we use $X_{l,\pm}(k)=\pm i\sqrt{c_l^{2}-k^{2}}$ for $k\in(-c_l,c_l)$.
For (d), note that
\begin{equation*}
X_l(k)=\varepsilon(2c_l)^{\frac12}(\varepsilon k-c_l)^{\frac12}\Big(1+\frac{\varepsilon k-c_l}{4c_l}+\mathcal{O}\big((k-\varepsilon c_l)^{2}\big)\Big),\qquad k\to\varepsilon c_l,
\end{equation*}
and $p_{\mathrm{I}}(k)=p_{\mathrm{I}}(\varepsilon c_l)+\varepsilon p_{\mathrm{I}}'(\varepsilon c_l)(\varepsilon k-c_l)+4\beta(\varepsilon k-c_l)^{2}$. Hence \eqref{eq:gI-exp} holds with
$g^{(1)}_{\mathrm{I},\varepsilon}=\varepsilon(2c_l)^{\frac12}p_{\mathrm{I}}(\varepsilon c_l)$ and
$g^{(2)}_{\mathrm{I},\varepsilon}=\varepsilon(2c_l)^{\frac12}\big(\varepsilon p_{\mathrm{I}}'(\varepsilon c_l)+\frac{p_{\mathrm{I}}(\varepsilon c_l)}{4c_l}\big)$.
Since
\begin{equation}\label{eq:pI-endpoints}
p_{\mathrm{I}}(c_l)=\xi-\xi_{\mathrm{I}}^{+},\qquad p_{\mathrm{I}}(-c_l)=\xi-\xi_{\mathrm{I}}^{+}-4\alpha c_l,\qquad p_{\mathrm{I}}'(\pm c_l)=\pm8\beta c_l+2\alpha,
\end{equation}
we arrive at \eqref{eq:gI-coef}.
\end{proof}

It follows from \eqref{eq:gI-coef} and \eqref{eq:TIp-half} that $g^{(1)}_{\mathrm{I},+}=\mathcal{O}(t^{-2/3})$, whereas $g^{(1)}_{\mathrm{I},-}=4\alpha c_l(2c_l)^{\frac12}+\mathcal{O}(t^{-2/3})$ is bounded away from zero. Thus, as $t\to+\infty$, the branch point $c_l$ becomes a soft edge of $g_{\mathrm{I}}$, while $-c_l$ remains a hard edge; this asymmetry is caused by the term $2\alpha k^{2}$ in $\theta$.

It is readily seen that the $k$-derivative of $g_{\mathrm{I}}$ is given by
\begin{equation}\label{eq:gI-prime}
g_{\mathrm{I}}'(k)=\frac{n_{\mathrm{I}}(k)}{X_l(k)},\qquad
n_{\mathrm{I}}(k):=12\beta k^{3}+4\alpha k^{2}+\big(\xi-6\beta c_l^{2}\big)k-2\alpha c_l^{2}.
\end{equation}

\begin{lemma}\label{lem:TIp-saddle}
Let $\xi$ satisfy \eqref{eq:TIp-half}. Then, for $t$ large enough, the cubic polynomial $n_{\mathrm{I}}$ has three simple real zeros: one in $(c_l,+\infty)$, denoted by $\eta_l=\eta_l(\xi)$; one in $(-\infty,-c_l)$, denoted by $-\hat\eta_l=-\hat\eta_l(\xi)$; and one in $(-c_l,0)$. Moreover, as $t\to+\infty$,
\begin{align}
\eta_l&=c_l+\frac{\xi_{\mathrm{I}}^{+}-\xi}{6(\alpha+4\beta c_l)}+\mathcal{O}(t^{-4/3}), \label{eq:etal}\\
\hat\eta_l&=\hat\eta_l^{\,0}+\mathcal{O}(t^{-2/3}),\qquad
\hat\eta_l^{\,0}:=\frac{\alpha+3\beta c_l+\sqrt{\alpha^{2}+9\beta^{2}c_l^{2}}}{6\beta}>c_l. \label{eq:hatetal}
\end{align}
\end{lemma}

\begin{proof}
By \eqref{eq:gI-prime},
\begin{equation*}
n_{\mathrm{I}}(c_l)=c_l\big(\xi-\xi_{\mathrm{I}}^{+}\big)<0,\qquad
n_{\mathrm{I}}(-c_l)=c_l\big(4\alpha c_l-(\xi-\xi_{\mathrm{I}}^{+})\big)>0,\qquad
n_{\mathrm{I}}(0)=-2\alpha c_l^{2}<0,
\end{equation*}
and $n_{\mathrm{I}}(k)\to\pm\infty$ as $k\to\pm\infty$. Hence $n_{\mathrm{I}}$ changes sign on each of the intervals $(-\infty,-c_l)$, $(-c_l,0)$ and $(c_l,+\infty)$, which gives the three real zeros; since $n_{\mathrm{I}}$ is cubic, they are simple. At $\xi=\xi_{\mathrm{I}}^{+}$ we have the factorization
\begin{equation*}
n_{\mathrm{I}}(k)\big|_{\xi=\xi_{\mathrm{I}}^{+}}=2(k-c_l)\big(6\beta k^{2}+2(\alpha+3\beta c_l)k+\alpha c_l\big),
\end{equation*}
whose zeros are $c_l$, $-\hat\eta_l^{\,0}$ and $\big(-\alpha-3\beta c_l+\sqrt{\alpha^{2}+9\beta^{2}c_l^{2}}\big)/(6\beta)\in(-c_l,0)$; the inequality $\hat\eta_l^{\,0}>c_l$ is equivalent to $\sqrt{\alpha^{2}+9\beta^{2}c_l^{2}}>3\beta c_l-\alpha$, which holds since $\alpha\beta c_l>0$. Since $\partial_\xi n_{\mathrm{I}}(k)=k$ and $\partial_k n_{\mathrm{I}}(c_l)\big|_{\xi=\xi_{\mathrm{I}}^{+}}=6c_l(\alpha+4\beta c_l)\neq0$, the implicit function theorem together with $|\xi-\xi_{\mathrm{I}}^{+}|<Ct^{-2/3}$ yields \eqref{eq:etal} and \eqref{eq:hatetal}.
\end{proof}

Consequently, only the saddle point $\eta_l$ tends to the branch point $c_l$, at the rate $\mathcal{O}(t^{-2/3})$, while the saddle point $-\hat\eta_l$ stays at the positive distance $\hat\eta_l^{\,0}-c_l+\mathcal{O}(t^{-2/3})$ from $-c_l$. The third zero of $n_{\mathrm{I}}$ lies on the branch cut of $g_{\mathrm{I}}$ and plays no role in the subsequent analysis.

\begin{figure}[htbp]
\centering
\includegraphics[width=0.78\textwidth]{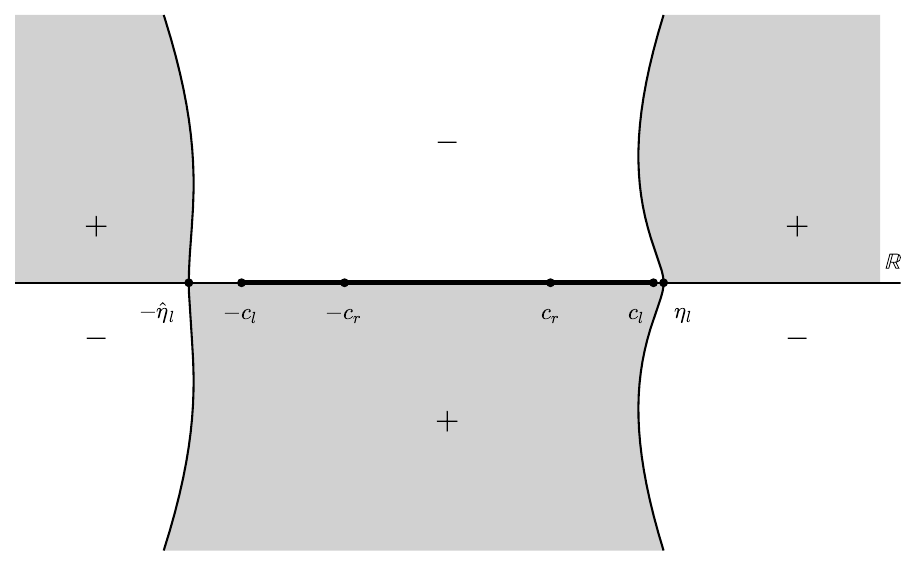}
\caption{The signature table of $\operatorname{Im}g_{\mathrm{I}}$ for $\xi$ satisfying \eqref{eq:TIp-half}. In the shaded regions $\operatorname{Im}g_{\mathrm{I}}>0$, in the white regions $\operatorname{Im}g_{\mathrm{I}}<0$.}
\label{fig:sign-TIp}
\end{figure}

We next describe the sign of $\operatorname{Im}g_{\mathrm{I}}$. By item (a) of Proposition~\ref{prop:gI}, $\operatorname{Im}g_{\mathrm{I}}(\bar k)=-\operatorname{Im}g_{\mathrm{I}}(k)$. Under \eqref{eq:TIp-half}, \eqref{eq:pI-endpoints} gives $p_{\mathrm{I}}(\pm c_l)<0$; since $p_{\mathrm{I}}$ is convex, $p_{\mathrm{I}}(k)<0$ for $k\in[-c_l,c_l]$, and \eqref{eq:gI-cut} implies
\begin{equation}\label{eq:Im-gI-cut}
\operatorname{Im}g_{\mathrm{I},+}(k)=p_{\mathrm{I}}(k)\sqrt{c_l^{2}-k^{2}}<0,\qquad
\big|e^{-2itg_{\mathrm{I},+}(k)}\big|=e^{2tp_{\mathrm{I}}(k)\sqrt{c_l^{2}-k^{2}}}<1,\qquad k\in(-c_l,c_l).
\end{equation}
On $\mathbb{R}\setminus[-c_l,c_l]$ the function $g_{\mathrm{I}}$ is real, so that for $k\in\mathbb{C}^{+}$ close to $\mathbb{R}\setminus[-c_l,c_l]$ the sign of $\operatorname{Im}g_{\mathrm{I}}(k)$ is that of $g_{\mathrm{I}}'(\operatorname{Re}k)$. By \eqref{eq:gI-prime} and Lemma~\ref{lem:TIp-saddle}, $g_{\mathrm{I}}'>0$ on $(-\infty,-\hat\eta_l)\cup(\eta_l,+\infty)$ and $g_{\mathrm{I}}'<0$ on $(-\hat\eta_l,-c_l)\cup(c_l,\eta_l)$. The signature table of $\operatorname{Im}g_{\mathrm{I}}$ is illustrated in Figure~\ref{fig:sign-TIp}.

\subsubsection{RH problem for $M^{(1)}$}\label{subsubsec:TIp-RHP1}

By \eqref{eq:gI-infty}, $e^{it(g_{\mathrm{I}}(k)-\theta(k))\sigma_3}\to e^{-i\alpha c_l^{2}t\sigma_3}$ as $k\to\infty$. Therefore, with the help of $g_{\mathrm{I}}$, we define
\begin{equation}\label{eq:M1-TIp}
M^{(1)}(k)=M^{(1)}(k;x,t):=e^{i\alpha c_l^{2}t\sigma_3}M(k)\,e^{it\left(g_{\mathrm{I}}(k)-\theta(k)\right)\sigma_3}.
\end{equation}
Then $M^{(1)}$ satisfies the following RH problem.

\begin{rhp}\label{rhp:M1-TIp}
Find $M^{(1)}(k)=M^{(1)}(k;x,t)$ with the properties
\begin{itemize}
\item[\rm(a)] $M^{(1)}(k)$ is holomorphic for $k\in\mathbb{C}\setminus\mathbb{R}$.
\item[\rm(b)] $M^{(1)}(k)$ has continuous boundary values $M^{(1)}_{\pm}(k)$ on $\mathbb{R}\setminus\{\pm c_l,\pm c_r\}$ and
\begin{equation*}
M^{(1)}_{+}(k)=M^{(1)}_{-}(k)V^{(1)}(k),\qquad k\in\mathbb{R},
\end{equation*}
where
\begin{equation}\label{eq:V1-TIp}
V^{(1)}(k)=\begin{cases}
\begin{pmatrix}1-r(k)r^{*}(k)&-r^{*}(k)e^{-2itg_{\mathrm{I}}(k)}\\ r(k)e^{2itg_{\mathrm{I}}(k)}&1\end{pmatrix}, & k\in(-\infty,-c_l)\cup(c_l,+\infty),\\[4mm]
\begin{pmatrix}0&-r^{*}_{-}(k)\\ r_{+}(k)&e^{-2itg_{\mathrm{I},+}(k)}\end{pmatrix}, & k\in(-c_l,-c_r)\cup(c_r,c_l),\\[4mm]
\begin{pmatrix}0&-1\\ 1&0\end{pmatrix}, & k\in(-c_r,c_r).
\end{cases}
\end{equation}
\item[\rm(c)] $M^{(1)}(k)=I+\mathcal{O}(k^{-1})$ as $k\to\infty$ in $\mathbb{C}\setminus\mathbb{R}$.
\item[\rm(d)] $M^{(1)}(k)$ admits the same singular behavior as $M(k)$ at the branch points $\pm c_l$ and $\pm c_r$; see RH problem~\ref{rhp:M}.
\end{itemize}
\end{rhp}

Indeed, the constant factor $e^{i\alpha c_l^{2}t\sigma_3}$ does not affect the jump, so that
\begin{equation*}
V^{(1)}=e^{-it(g_{\mathrm{I},-}-\theta)\sigma_3}\,V\,e^{it(g_{\mathrm{I},+}-\theta)\sigma_3},\qquad k\in\mathbb{R}.
\end{equation*} On $\mathbb{R}\setminus[-c_l,c_l]$ we have $g_{\mathrm{I},+}=g_{\mathrm{I},-}=g_{\mathrm{I}}$, and $\theta$ is simply replaced by $g_{\mathrm{I}}$ in the off-diagonal entries of \eqref{eq:V}. On $(-c_l,c_l)$, item (c) of Proposition~\ref{prop:gI} gives $e^{\mp it(g_{\mathrm{I},+}+g_{\mathrm{I},-})}=1$ for the off-diagonal entries and $e^{it(g_{\mathrm{I},-}-g_{\mathrm{I},+})}=e^{-2itg_{\mathrm{I},+}}$ for the $(2,2)$-entry, which yields \eqref{eq:V1-TIp}. Item (c) follows from \eqref{eq:gI-infty}, and item (d) from the boundedness of $g_{\mathrm{I}}$ and $\theta$ near $\pm c_l$ and $\pm c_r$. By \eqref{eq:Im-gI-cut}, the $(2,2)$-entry of $V^{(1)}$ on $(-c_l,-c_r)\cup(c_r,c_l)$ is bounded by $1$, and it decays exponentially as $t\to+\infty$, uniformly for $k$ bounded away from $\pm c_l$.

Moreover, it follows from \eqref{eq:symM}, item (a) of Proposition~\ref{prop:gI} and $\overline{\theta(\bar k)}=\theta(k)$ that
\begin{equation}\label{eq:sym-M1-TIp}
M^{(1)}(k)=\sigma_1\overline{M^{(1)}(\bar k)}\sigma_1 ,
\end{equation}
and, by \eqref{eq:recon}, \eqref{eq:M1-TIp} and \eqref{eq:gI-infty}, the solution of \eqref{eq:Hirota} is recovered from
\begin{equation}\label{eq:q-M1-TIp}
q(x,t)=2i\,e^{-2i\alpha c_l^{2}t}\lim_{k\to\infty}\big(kM^{(1)}(k)\big)_{12}.
\end{equation}

\subsection{Second transformation: $M^{(1)}\to M^{(2)}$}\label{subsec:TIp-M2}

In this subsection, we introduce a scalar function $D_{\mathrm{I}}$ which removes the reflection coefficient from the off-diagonal entries of $V^{(1)}$ on $(-c_l,-c_r)\cup(c_r,c_l)$; this paves the way for the subsequent contour deformation along $(-\infty,-\hat\eta_l)$ and $(\eta_l,+\infty)$. The segment $[c_l,\eta_l]$, whose length is $\mathcal{O}(t^{-2/3})$ by Lemma~\ref{lem:TIp-saddle}, is kept on the real line. The segment $[-\hat\eta_l,-c_l]$ has length $\hat\eta_l^{\,0}-c_l+\mathcal{O}(t^{-2/3})$, which does not shrink as $t\to+\infty$; it will be handled in Section~\ref{subsec:TIp-M3} with the help of an additional scalar function.

\subsubsection{The function $D_{\mathrm{I}}$}\label{subsubsec:TIp-DI}

We first fix the branch of $\log r_+$ on $(-c_l,-c_r)\cup(c_r,c_l)$. Recall from Proposition~\ref{prop:scattering} that $|r_+(k)|=1$ for $k\in(-c_l,-c_r)\cup(c_r,c_l)$, and from \eqref{eq:sym-r} that $r_+(-k)=\overline{r_+(k)}$ there. Let $\log r_+$ denote the branch which is continuous on $[c_r,c_l]$ and satisfies
\begin{equation}\label{eq:arg-br0}
\arg b_{r,0}:=\operatorname{Im}\log r_+(c_r)\in(-\pi,\pi),
\end{equation}
which is possible by \eqref{eq:r-ne-1}, and extend it to $[-c_l,-c_r]$ by $\log r_+(k):=\overline{\log r_+(-k)}$. Then $\log r_+$ is purely imaginary and odd on $(-c_l,-c_r)\cup(c_r,c_l)$. We further set
\begin{equation}\label{eq:arg-bl0}
\arg b_{l,0}:=\operatorname{Im}\log r_+(c_l),\qquad \nu_r:=\frac{\arg b_{r,0}}{2\pi}\in\Big(-\frac12,\frac12\Big);
\end{equation}
note that $\arg b_{l,0}$ is determined by continuity and need not lie in $(-\pi,\pi]$. Now define
\begin{equation}\label{eq:DI}
D_{\mathrm{I}}(k):=\exp\bigg\{\frac{X_l(k)}{2\pi i}\Big(\int_{-c_l}^{-c_r}+\int_{c_r}^{c_l}\Big)\frac{\log r_+(z)}{X_{l,+}(z)(z-k)}\,\mathrm{d}z\bigg\},\qquad k\in\mathbb{C}\setminus[-c_l,c_l].
\end{equation}
The function $D_{\mathrm{I}}$ depends only on $r$, $c_l$ and $c_r$. In particular, it is independent of $(x,t)$ and of the coefficients $\alpha,\beta$, and it coincides with the function introduced in \cite[(3.6)]{FWZ}. Its properties are collected in the following proposition.

\begin{proposition}\label{prop:DI}
The function $D_{\mathrm{I}}$ defined in \eqref{eq:DI} has the following properties.
\begin{itemize}
\item[\rm(a)] $D_{\mathrm{I}}(k)$ is holomorphic for $k\in\mathbb{C}\setminus[-c_l,c_l]$.
\item[\rm(b)] $D_{\mathrm{I}}$ satisfies the jump relations
\begin{equation}\label{eq:DI-jump}
D_{\mathrm{I},+}(k)D_{\mathrm{I},-}(k)=\begin{cases} r_+(k), & k\in(-c_l,-c_r)\cup(c_r,c_l),\\[1mm] 1, & k\in(-c_r,c_r).\end{cases}
\end{equation}
\item[\rm(c)] $D_{\mathrm{I}}$ admits the symmetries
\begin{equation}\label{eq:DI-sym}
D_{\mathrm{I}}(-k)=D_{\mathrm{I}}(k)^{-1},\qquad \overline{D_{\mathrm{I}}(\bar k)}=D_{\mathrm{I}}(k)^{-1},\qquad k\in\mathbb{C}\setminus[-c_l,c_l].
\end{equation}
In particular, $|D_{\mathrm{I}}(k)|=1$ for $k\in\mathbb{R}\setminus[-c_l,c_l]$, and $D_{\mathrm{I},+}(k)/D_{\mathrm{I},-}(k)=|D_{\mathrm{I},+}(k)|^{2}>0$ for $k\in(-c_l,c_l)$.
\item[\rm(d)] $D_{\mathrm{I}}(k)=1+\mathcal{O}(k^{-1})$ as $k\to\infty$.
\item[\rm(e)] At the branch points $\pm c_l$,
\begin{equation}\label{eq:DI-cl}
D_{\mathrm{I}}(k)^{\pm1}=e^{\frac{i}{2}\arg b_{l,0}}\Big(1+\frac{b_{l,1}}{2\pi i\,b_{l,0}}\,(\pm k-c_l)^{\frac12}\log(\pm k-c_l)+\mathcal{O}\big(|k\mp c_l|^{\frac12}\big)\Big),\qquad k\to\pm c_l .
\end{equation}
At the points $\pm c_r$, for $\varepsilon\in\{+,-\}$ and $k\in\mathbb{C}^{\varepsilon}$,
\begin{equation}\label{eq:DI-cr}
\begin{aligned}
D_{\mathrm{I}}(k)&=e^{d^{\varepsilon}_{\mathrm{I},r}}\,(k-c_r)^{-\varepsilon\nu_r}\big(1+\mathcal{O}(|k-c_r|^{\frac12})\big), && k\to c_r,\\
D_{\mathrm{I}}(k)&=e^{-d^{-\varepsilon}_{\mathrm{I},r}}\,(-k-c_r)^{-\varepsilon\nu_r}\big(1+\mathcal{O}(|k+c_r|^{\frac12})\big), && k\to -c_r,
\end{aligned}
\end{equation}
where $d^{-}_{\mathrm{I},r}:=-\overline{d^{+}_{\mathrm{I},r}}$ and
\begin{equation}\label{eq:dIr}
d^{+}_{\mathrm{I},r}:=\frac{\sqrt{c_l^{2}-c_r^{2}}}{2\pi}\bigg(\int_{-c_l}^{-c_r}\frac{\log r_+(z)}{X_{l,+}(z)(z-c_r)}\,\mathrm{d}z+\int_{c_r}^{c_l}\Big(\frac{\log r_+(z)}{X_{l,+}(z)}-\frac{i\arg b_{r,0}}{X_{l,+}(c_r)}\Big)\frac{\mathrm{d}z}{z-c_r}\bigg)+\nu_r\log(c_r-c_l).
\end{equation}
In \eqref{eq:DI-cl}--\eqref{eq:dIr}, the principal branches are taken for the power functions and for the logarithms other than $\log r_+$; in particular, $\log(c_r-c_l)=\log(c_l-c_r)+i\pi$.
\item[\rm(f)] $D_{\mathrm{I}}$ is analytic in a neighbourhood of $\mathbb{R}\setminus[-c_l,c_l]$, and
\begin{equation}\label{eq:DI-eta}
D_{\mathrm{I}}(\eta_l)=e^{id_{\mathrm{I},\eta}},\qquad D_{\mathrm{I}}(-\hat\eta_l)=e^{-id_{\mathrm{I},\hat\eta}},
\end{equation}
with the real numbers
\begin{equation}\label{eq:dIeta}
\begin{aligned}
d_{\mathrm{I},\eta}&:=-\frac{\sqrt{\eta_l^{2}-c_l^{2}}}{2\pi}\Big(\int_{-c_l}^{-c_r}+\int_{c_r}^{c_l}\Big)\frac{\log r_+(z)}{X_{l,+}(z)(z-\eta_l)}\,\mathrm{d}z,\\
d_{\mathrm{I},\hat\eta}&:=-\frac{\sqrt{\hat\eta_l^{\,2}-c_l^{2}}}{2\pi}\Big(\int_{-c_l}^{-c_r}+\int_{c_r}^{c_l}\Big)\frac{\log r_+(z)}{X_{l,+}(z)(z-\hat\eta_l)}\,\mathrm{d}z.
\end{aligned}
\end{equation}
\end{itemize}
\end{proposition}

\begin{proof}
Write $S:=(-c_l,-c_r)\cup(c_r,c_l)$, $h(z):=\log r_+(z)/X_{l,+}(z)$ for $z\in S$, and
\begin{equation*}
F(k):=\frac{1}{2\pi i}\int_S\frac{h(z)}{z-k}\,\mathrm{d}z,\qquad\text{so that}\qquad D_{\mathrm{I}}(k)=e^{X_l(k)F(k)}.
\end{equation*}
Since $X_{l,+}(z)=i\sqrt{c_l^{2}-z^{2}}$ and $\log r_+$ is purely imaginary and odd on $S$, the density $h$ is real-valued and odd on $S$; moreover, $h$ has integrable singularities $\mathcal{O}(|z\mp c_l|^{-1/2})$ at $\pm c_l$.

(a) $F$ is holomorphic in $\mathbb{C}\setminus\overline{S}$ and $X_l$ is holomorphic in $\mathbb{C}\setminus[-c_l,c_l]$, which gives (a).

(b) By the Sokhotski--Plemelj formula, $F_+-F_-=h$ on $S$ and $F_+=F_-$ on $(-c_r,c_r)$. Since $X_{l,-}=-X_{l,+}$ on $(-c_l,c_l)$, we obtain $(X_lF)_++(X_lF)_-=X_{l,+}(F_+-F_-)$, which equals $\log r_+$ on $S$ and vanishes on $(-c_r,c_r)$. Exponentiating gives \eqref{eq:DI-jump}. We note that $D_{\mathrm{I},+}\neq D_{\mathrm{I},-}$ on $(-c_r,c_r)$ in general, so that the domain of holomorphy in (a) cannot be enlarged.

(c) Changing variables $z\to-z$ and using that $h$ is odd, we get $F(-k)=F(k)$; since $h$ is real-valued, $\overline{F(\bar k)}=-F(k)$. Together with $X_l(-k)=-X_l(k)$ and $\overline{X_l(\bar k)}=X_l(k)$, this yields \eqref{eq:DI-sym}. For $k\in\mathbb{R}\setminus[-c_l,c_l]$, $X_l(k)$ is real and $F(k)$ is purely imaginary, whence $|D_{\mathrm{I}}(k)|=1$. For $k\in(-c_l,c_l)$, the second relation in \eqref{eq:DI-sym} gives $D_{\mathrm{I},-}(k)=\overline{D_{\mathrm{I},+}(k)}^{\,-1}$, and hence $D_{\mathrm{I},+}/D_{\mathrm{I},-}=|D_{\mathrm{I},+}|^{2}$.

(d) As $k\to\infty$, $X_l(k)F(k)=-\frac{1}{2\pi i}\int_S h(z)\,\mathrm{d}z+\mathcal{O}(k^{-1})=\mathcal{O}(k^{-1})$, since $h$ is odd.

(e) By \eqref{eq:r-endpoints} and \eqref{eq:b-cond}, $\log r_+(z)=\log b_{l,0}+i\frac{b_{l,1}}{b_{l,0}}(c_l-z)^{\frac12}+\mathcal{O}(c_l-z)$ as $z\nearrow c_l$, where $\log b_{l,0}=i\arg b_{l,0}$ and $b_{l,1}/b_{l,0}\in\mathbb{R}\setminus\{0\}$. Hence
\begin{equation*}
h(z)=\frac{\log b_{l,0}}{i(2c_l)^{\frac12}}(c_l-z)^{-\frac12}+\frac{b_{l,1}}{(2c_l)^{\frac12}b_{l,0}}+\mathcal{O}\big((c_l-z)^{\frac12}\big),\qquad z\nearrow c_l.
\end{equation*}
Using $\int_{c_r}^{c_l}\frac{(c_l-z)^{-1/2}}{z-k}\mathrm{d}z=-\pi(k-c_l)^{-\frac12}+\mathcal{O}(1)$ and $\int_{c_r}^{c_l}\frac{\mathrm{d}z}{z-k}=\log(k-c_l)+\mathcal{O}(1)$ as $k\to c_l$, and the boundedness of the Cauchy integral of the remainder, we find
\begin{equation*}
F(k)=\frac{\log b_{l,0}}{2(2c_l)^{\frac12}}(k-c_l)^{-\frac12}+\frac{b_{l,1}}{2\pi i\,(2c_l)^{\frac12}b_{l,0}}\log(k-c_l)+\mathcal{O}(1),\qquad k\to c_l.
\end{equation*}
Multiplying by $X_l(k)=(2c_l)^{\frac12}(k-c_l)^{\frac12}\big(1+\mathcal{O}(k-c_l)\big)$ and exponentiating, we arrive at \eqref{eq:DI-cl} at $c_l$; the expansion at $-c_l$ follows from $D_{\mathrm{I}}(k)^{-1}=D_{\mathrm{I}}(-k)$.
Near $c_r$, the density $h$ is bounded, $h(z)\to h_r:=\frac{i\arg b_{r,0}}{X_{l,+}(c_r)}=\frac{\arg b_{r,0}}{\sqrt{c_l^{2}-c_r^{2}}}$ as $z\searrow c_r$ with $h(z)-h_r=\mathcal{O}(|z-c_r|^{1/2})$, while $h$ is not defined on $(-c_r,c_r)$. Writing $\int_{c_r}^{c_l}\frac{h(z)}{z-k}\mathrm{d}z=\int_{c_r}^{c_l}\frac{h(z)-h_r}{z-k}\mathrm{d}z+h_r\big(\log(k-c_l)-\log(k-c_r)\big)$ for $k\in\mathbb{C}^{+}$, and using $X_l(k)=i\sqrt{c_l^{2}-c_r^{2}}+\mathcal{O}(k-c_r)$ for $k\in\mathbb{C}^{+}$, we obtain
$X_l(k)F(k)=-\nu_r\log(k-c_r)+d^{+}_{\mathrm{I},r}+\mathcal{O}(|k-c_r|^{\frac12})$ as $k\to c_r$ in $\mathbb{C}^{+}$, which is \eqref{eq:DI-cr} at $c_r$ for $\varepsilon=+$. The case $\varepsilon=-$ follows from the second relation in \eqref{eq:DI-sym}, and the expansions at $-c_r$ from the first one.

(f) For real $k\notin[-c_l,c_l]$, $D_{\mathrm{I}}(k)=e^{id_{\mathrm{I}}(k)}$ with $d_{\mathrm{I}}(k):=-\frac{X_l(k)}{2\pi}\int_S\frac{h(z)}{z-k}\mathrm{d}z\in\mathbb{R}$. Taking $k=\eta_l$ and $k=\hat\eta_l$, and using $D_{\mathrm{I}}(-\hat\eta_l)=D_{\mathrm{I}}(\hat\eta_l)^{-1}$, we obtain \eqref{eq:DI-eta}.
\end{proof}

\subsubsection{RH problem for $M^{(2)}$}\label{subsubsec:TIp-RHP2}

With the help of $D_{\mathrm{I}}$, we define
\begin{equation}\label{eq:M2-TIp}
M^{(2)}(k)=M^{(2)}(k;x,t):=M^{(1)}(k)\,D_{\mathrm{I}}(k)^{-\sigma_3}.
\end{equation}
Then $M^{(2)}$ satisfies the following RH problem.

\begin{rhp}\label{rhp:M2-TIp}
Find $M^{(2)}(k)=M^{(2)}(k;x,t)$ with the properties
\begin{itemize}
\item[\rm(a)] $M^{(2)}(k)$ is holomorphic for $k\in\mathbb{C}\setminus\mathbb{R}$.
\item[\rm(b)] $M^{(2)}(k)$ has continuous boundary values $M^{(2)}_{\pm}(k)$ on $\mathbb{R}\setminus\{\pm c_l,\pm c_r\}$ and
\begin{equation*}
M^{(2)}_{+}(k)=M^{(2)}_{-}(k)V^{(2)}(k),\qquad k\in\mathbb{R},
\end{equation*}
where
\begin{equation}\label{eq:V2-TIp}
V^{(2)}(k)=\begin{cases}
\begin{pmatrix}1-r(k)r^{*}(k)&-D_{\mathrm{I}}^{2}(k)r^{*}(k)e^{-2itg_{\mathrm{I}}(k)}\\ D_{\mathrm{I}}^{-2}(k)r(k)e^{2itg_{\mathrm{I}}(k)}&1\end{pmatrix}, & k\in(-\infty,-c_l)\cup(c_l,+\infty),\\[4mm]
\begin{pmatrix}0&-1\\ 1&\dfrac{D_{\mathrm{I},+}(k)}{D_{\mathrm{I},-}(k)}e^{-2itg_{\mathrm{I},+}(k)}\end{pmatrix}, & k\in(-c_l,-c_r)\cup(c_r,c_l),\\[5mm]
\begin{pmatrix}0&-1\\ 1&0\end{pmatrix}, & k\in(-c_r,c_r).
\end{cases}
\end{equation}
\item[\rm(c)] $M^{(2)}(k)=I+\mathcal{O}(k^{-1})$ as $k\to\infty$ in $\mathbb{C}\setminus\mathbb{R}$.
\item[\rm(d)] $M^{(2)}(k)=\mathcal{O}(1)$ as $k\to\pm c_l$, and, as $k\to\pm c_r$,
\begin{equation*}
M^{(2)}(k)=\begin{cases}
\mathcal{O}\begin{pmatrix}|k\mp c_r|^{\frac14+\nu_r}&|k\mp c_r|^{-\frac14-\nu_r}\\ |k\mp c_r|^{\frac14+\nu_r}&|k\mp c_r|^{-\frac14-\nu_r}\end{pmatrix}, & k\in\mathbb{C}^{+},\\[5mm]
\mathcal{O}\begin{pmatrix}|k\mp c_r|^{-\frac14-\nu_r}&|k\mp c_r|^{\frac14+\nu_r}\\ |k\mp c_r|^{-\frac14-\nu_r}&|k\mp c_r|^{\frac14+\nu_r}\end{pmatrix}, & k\in\mathbb{C}^{-}.
\end{cases}
\end{equation*}
\end{itemize}
\end{rhp}

Indeed, $V^{(2)}=D_{\mathrm{I},-}^{\sigma_3}V^{(1)}D_{\mathrm{I},+}^{-\sigma_3}$ on $\mathbb{R}$. On $\mathbb{R}\setminus[-c_l,c_l]$ we have $D_{\mathrm{I},\pm}=D_{\mathrm{I}}$. On $(-c_l,-c_r)\cup(c_r,c_l)$, \eqref{eq:DI-jump} and the relation $r_+=(r^{*}_{-})^{-1}$ (see Proposition~\ref{prop:scattering}) give $-r^{*}_{-}D_{\mathrm{I},+}D_{\mathrm{I},-}=-1$ and $r_+D_{\mathrm{I},+}^{-1}D_{\mathrm{I},-}^{-1}=1$ for the off-diagonal entries, while the $(2,2)$-entry becomes $D_{\mathrm{I},+}D_{\mathrm{I},-}^{-1}e^{-2itg_{\mathrm{I},+}}$. On $(-c_r,c_r)$ we use $D_{\mathrm{I},+}D_{\mathrm{I},-}=1$. This proves \eqref{eq:V2-TIp}. Item (c) follows from item (d) of Proposition~\ref{prop:DI}, and item (d) from item (d) of RH problem~\ref{rhp:M1-TIp} (i.e., the behaviour of $M$ at $\pm c_l$ and $\pm c_r$ in RH problem~\ref{rhp:M}) together with \eqref{eq:DI-cl} and \eqref{eq:DI-cr}.

Moreover, by \eqref{eq:sym-M1-TIp}, \eqref{eq:DI-sym} and \eqref{eq:q-M1-TIp}, we have
\begin{equation}\label{eq:sym-q-M2-TIp}
M^{(2)}(k)=\sigma_1\overline{M^{(2)}(\bar k)}\sigma_1,\qquad
q(x,t)=2i\,e^{-2i\alpha c_l^{2}t}\lim_{k\to\infty}\big(kM^{(2)}(k)\big)_{12}.
\end{equation}

\begin{remark}\label{rem:V2-22}
By item (c) of Proposition~\ref{prop:DI} and \eqref{eq:gI-cut}, the $(2,2)$-entry of $V^{(2)}$ on $(-c_l,-c_r)\cup(c_r,c_l)$ equals $|D_{\mathrm{I},+}(k)|^{2}e^{2tp_{\mathrm{I}}(k)\sqrt{c_l^{2}-k^{2}}}>0$. By \eqref{eq:Im-gI-cut}, it decays exponentially as $t\to+\infty$, uniformly on compact subsets of $(-c_l,-c_r)\cup(c_r,c_l)$. However, by \eqref{eq:DI-cl} and $|e^{\frac{i}{2}\arg b_{l,0}}|=1$, it tends to $1$ as $k\to\pm c_l$, so that it cannot be neglected near $\pm c_l$; this will be taken into account in the subsequent sections.
\end{remark}

\subsection{Third transformation: $M^{(2)}\to M^{(3)}$}\label{subsec:TIp-M3}

In this subsection we open lenses. By the signature table in Figure~\ref{fig:sign-TIp}, on $(-\infty,-\hat\eta_l)\cup(\eta_l,+\infty)$ the jump matrix $V^{(2)}$ is factorized into an upper triangular factor, which is continued to $\mathbb{C}^{-}$, and a lower triangular factor, which is continued to $\mathbb{C}^{+}$. On $(-\hat\eta_l,-c_l)$ the signs of $\operatorname{Im}g_{\mathrm{I}}$ are reversed, and the factorization of $V$ on $\mathbb{R}\setminus[-c_l,c_l]$ involving $r_2=r^{*}/(1-rr^{*})$ in \eqref{eq:fac1} has to be used; it contains the diagonal factor $(1-rr^{*})^{\sigma_3}$. Since the length of $(-\hat\eta_l,-c_l)$ does not shrink as $t\to+\infty$ (Lemma~\ref{lem:TIp-saddle}), this factor is removed with the help of a scalar function $\delta_{\mathrm{I}}$, introduced in Section~\ref{subsubsec:TIp-delta}. The short interval $[c_l,\eta_l]$ is kept on the real line. Finally, since $-c_l$ is a hard edge of $g_{\mathrm{I}}$, the lenses emanating from $-\hat\eta_l$ to the right are not closed at $-c_l$; in the spirit of \cite{XZ}, they will be extended across $-c_l$ into the band $(-c_l,-c_r)$, which avoids a local parametrix at $-c_l$ (see Remark~\ref{rem:delta-cl}).

\subsubsection{The function $\delta_{\mathrm{I}}$}\label{subsubsec:TIp-delta}

Define
\begin{equation}\label{eq:deltaI}
\delta_{\mathrm{I}}(k)=\delta_{\mathrm{I}}(k;\xi):=\exp\bigg\{\frac{X_l(k)}{2\pi i}\int_{-\hat\eta_l}^{-c_l}\frac{\log\big(1-|r(s)|^{2}\big)}{X_l(s)(s-k)}\,\mathrm{d}s\bigg\},\qquad k\in\mathbb{C}\setminus[-\hat\eta_l,c_l],
\end{equation}
where $X_l(s)=-\sqrt{s^{2}-c_l^{2}}<0$ for $s\in(-\hat\eta_l,-c_l)$. By Proposition~\ref{prop:scattering}, $0<1-|r(s)|^{2}<1$ for $s\in(-\hat\eta_l,-c_l)$. Moreover, by \eqref{eq:r-endpoints} and \eqref{eq:b-cond}, $|r(s)|^{2}=1+2\frac{b_{l,1}}{b_{l,0}}(-c_l-s)^{1/2}+\mathcal{O}(-c_l-s)$ as $s\nearrow-c_l$; since $|r|<1$ there and $b_{l,1}/b_{l,0}\in\mathbb{R}\setminus\{0\}$, we have $b_{l,1}/b_{l,0}=-|b_{l,1}|$ and
\begin{equation}\label{eq:1-r2-cl}
1-|r(s)|^{2}=2|b_{l,1}|\,(-c_l-s)^{\frac12}\big(1+\mathcal{O}((-c_l-s)^{\frac12})\big),\qquad s\nearrow-c_l .
\end{equation}
Hence the integrand in \eqref{eq:deltaI} has an integrable singularity of order $|s+c_l|^{-1/2}\log|s+c_l|$ at $-c_l$. The factor $X_l(k)$ in \eqref{eq:deltaI}, which is absent in the classical $\delta$-functions, guarantees that the jumps of the form $\left(\begin{smallmatrix}0&-1\\1&\ast\end{smallmatrix}\right)$ on $(-c_l,c_l)$ are preserved; see item (b) below. We also introduce
\begin{equation}\label{eq:nu-phiI}
\nu=\nu(\xi):=-\frac{1}{2\pi}\log\big(1-|r(-\hat\eta_l)|^{2}\big)>0,\qquad
\varphi_{\mathrm{I}}=\varphi_{\mathrm{I}}(\xi):=\frac{1}{2\pi}\int_{-\hat\eta_l}^{-c_l}\frac{\log\big(1-|r(s)|^{2}\big)}{X_l(s)}\,\mathrm{d}s>0,
\end{equation}
where $\varphi_{\mathrm{I}}>0$ because both $\log(1-|r(s)|^{2})$ and $X_l(s)$ are negative on $(-\hat\eta_l,-c_l)$.

\begin{proposition}\label{prop:deltaI}
Let $\xi$ satisfy \eqref{eq:TIp-half} and let $t$ be large enough. The function $\delta_{\mathrm{I}}$ defined in \eqref{eq:deltaI} has the following properties.
\begin{itemize}
\item[\rm(a)] $\delta_{\mathrm{I}}(k)$ is holomorphic for $k\in\mathbb{C}\setminus[-\hat\eta_l,c_l]$.
\item[\rm(b)] $\delta_{\mathrm{I}}$ satisfies the jump relations
\begin{equation}\label{eq:deltaI-jump}
\delta_{\mathrm{I},+}(k)=\delta_{\mathrm{I},-}(k)\big(1-|r(k)|^{2}\big),\quad k\in(-\hat\eta_l,-c_l),\qquad
\delta_{\mathrm{I},+}(k)\,\delta_{\mathrm{I},-}(k)=1,\quad k\in(-c_l,c_l).
\end{equation}
\item[\rm(c)] $\overline{\delta_{\mathrm{I}}(\bar k)}=\delta_{\mathrm{I}}(k)^{-1}$ for $k\in\mathbb{C}\setminus[-\hat\eta_l,c_l]$. In particular, $|\delta_{\mathrm{I}}(k)|=1$ for $k\in\mathbb{R}\setminus[-\hat\eta_l,c_l]$; $|\delta_{\mathrm{I},\pm}(k)|^{2}=(1-|r(k)|^{2})^{\pm1}$ for $k\in(-\hat\eta_l,-c_l)$; and $\delta_{\mathrm{I},+}(k)/\delta_{\mathrm{I},-}(k)=|\delta_{\mathrm{I},+}(k)|^{2}>0$ for $k\in(-c_l,c_l)$.
\item[\rm(d)] $\delta_{\mathrm{I}}(k)=e^{i\varphi_{\mathrm{I}}}\big(1+\mathcal{O}(k^{-1})\big)$ as $k\to\infty$.
\item[\rm(e)] As $k\to-\hat\eta_l$,
\begin{equation}\label{eq:deltaI-PC}
\delta_{\mathrm{I}}(k)=(-k-\hat\eta_l)^{-i\nu}\,\delta_{b}(k),\qquad
\big|\delta_b(k)-e^{i\phi_{\hat\eta}}\big|\lesssim|k+\hat\eta_l|\,\big|\log|k+\hat\eta_l|\big|,
\end{equation}
where
\begin{equation}\label{eq:phi-hateta}
\phi_{\hat\eta}:=\nu\log(\hat\eta_l-c_l)-\frac{1}{2\pi}\int_{-\hat\eta_l}^{-c_l}\Big(\frac{X_l(-\hat\eta_l)}{X_l(s)}\log\big(1-|r(s)|^{2}\big)-\log\big(1-|r(-\hat\eta_l)|^{2}\big)\Big)\frac{\mathrm{d}s}{s+\hat\eta_l}\in\mathbb{R}.
\end{equation}
\item[\rm(f)] At the branch points $\pm c_l$,
\begin{align}
\delta_{\mathrm{I}}(k)^{\pm1}&=\big(2|b_{l,1}|\big)^{\frac12}(k+c_l)^{\frac14}\Big(1+\mathcal{O}\big(|k+c_l|^{\frac12}\big|\log|k+c_l|\big|\big)\Big),\qquad k\to-c_l,\ k\in\mathbb{C}^{\pm},\label{eq:deltaI-mcl}\\
\delta_{\mathrm{I}}(k)&=1+\mathcal{O}\big(|k-c_l|^{\frac12}\big),\qquad k\to c_l.\label{eq:deltaI-cl}
\end{align}
\item[\rm(g)] $\delta_{\mathrm{I}}$ and $\delta_{\mathrm{I}}^{-1}$ are bounded in $\mathbb{C}\setminus D_{\varepsilon}(-c_l)$ for every $\varepsilon>0$, uniformly with respect to $\xi$ satisfying \eqref{eq:TIp-half}.
\end{itemize}
In \eqref{eq:deltaI-PC}--\eqref{eq:deltaI-cl}, the principal branches of the power functions and of the logarithms are taken.
\end{proposition}

\begin{proof}
Write $h_\delta(s):=\log(1-|r(s)|^{2})/X_l(s)$ for $s\in(-\hat\eta_l,-c_l)$ and $F_\delta(k):=\frac{1}{2\pi i}\int_{-\hat\eta_l}^{-c_l}\frac{h_\delta(s)}{s-k}\mathrm{d}s$, so that $\delta_{\mathrm{I}}=e^{X_lF_\delta}$. The density $h_\delta$ is real-valued.

(a) is immediate. (b) By the Sokhotski--Plemelj formula, $F_{\delta,+}-F_{\delta,-}=h_\delta$ on $(-\hat\eta_l,-c_l)$, where $X_l$ is continuous; this gives the first relation. On $(-c_l,c_l)$, $F_\delta$ is continuous while $X_{l,-}=-X_{l,+}$, so that $(X_lF_\delta)_++(X_lF_\delta)_-=0$.

(c) Since $h_\delta$ is real, $\overline{F_\delta(\bar k)}=-F_\delta(k)$, and $\overline{X_l(\bar k)}=X_l(k)$. The remaining statements follow from this symmetry and from (b).

(d) As $k\to\infty$, $X_l(k)F_\delta(k)=-\frac{1}{2\pi i}\int_{-\hat\eta_l}^{-c_l}h_\delta(s)\,\mathrm{d}s+\mathcal{O}(k^{-1})=i\varphi_{\mathrm{I}}+\mathcal{O}(k^{-1})$.

(e) Write $L(s):=\log(1-|r(s)|^{2})$ and $L_0:=L(-\hat\eta_l)=-2\pi\nu$. Then
\begin{equation*}
X_l(k)F_\delta(k)=\frac{L_0}{2\pi i}\int_{-\hat\eta_l}^{-c_l}\frac{\mathrm{d}s}{s-k}+\frac{1}{2\pi i}\int_{-\hat\eta_l}^{-c_l}\Big(\frac{X_l(k)}{X_l(s)}L(s)-L_0\Big)\frac{\mathrm{d}s}{s-k}.
\end{equation*}
For $k\in\mathbb{C}\setminus\mathbb{R}$, $\int_{-\hat\eta_l}^{-c_l}\frac{\mathrm{d}s}{s-k}=\log(-k-c_l)-\log(-k-\hat\eta_l)$, so the first term equals $-i\nu\log(-k-\hat\eta_l)+i\nu\log(-k-c_l)$. The density of the second term is Lipschitz in $s$ and vanishes at $s=-\hat\eta_l$ when $k=-\hat\eta_l$; hence the second term is continuous at $-\hat\eta_l$ with a modulus of continuity $\mathcal{O}(|k+\hat\eta_l|\,|\log|k+\hat\eta_l||)$. Evaluating at $k=-\hat\eta_l$ gives \eqref{eq:deltaI-PC} with \eqref{eq:phi-hateta}.

(f) By \eqref{eq:1-r2-cl}, $h_\delta(s)=-\frac{\log(2|b_{l,1}|)+\frac12\log(-c_l-s)}{(2c_l)^{1/2}(-c_l-s)^{1/2}}+\mathcal{O}(1)$ as $s\nearrow-c_l$. Using
$\int_0^{\infty}\frac{u^{-1/2}}{u+w}\mathrm{d}u=\pi w^{-1/2}$ and $\int_0^{\infty}\frac{u^{-1/2}\log u}{u+w}\mathrm{d}u=\pi w^{-1/2}\log w$ with $w=k+c_l$, we obtain
$F_\delta(k)=\frac{1}{2i(2c_l)^{1/2}}(k+c_l)^{-1/2}\big(\log(2|b_{l,1}|)+\tfrac12\log(k+c_l)\big)+\mathcal{O}(\log|k+c_l|)$ as $k\to-c_l$. Since $X_l(k)=\pm i(2c_l)^{1/2}(k+c_l)^{1/2}\big(1+\mathcal{O}(k+c_l)\big)$ for $k\in\mathbb{C}^{\pm}$, it follows that $X_l(k)F_\delta(k)=\pm\big(\tfrac12\log(2|b_{l,1}|)+\tfrac14\log(k+c_l)\big)+\mathcal{O}(|k+c_l|^{1/2}|\log|k+c_l||)$, which yields \eqref{eq:deltaI-mcl}. Near $c_l$, $F_\delta$ is holomorphic and $X_l(k)=\mathcal{O}(|k-c_l|^{1/2})$, which gives \eqref{eq:deltaI-cl}.

(g) Away from $-\hat\eta_l$ and $-c_l$, the boundedness follows from (a)--(d). Near $-\hat\eta_l$ it follows from (e) and $|(-k-\hat\eta_l)^{-i\nu}|=e^{\nu\arg(-k-\hat\eta_l)}\in[e^{-\pi\nu},e^{\pi\nu}]$. The uniformity in $\xi$ follows from Lemma~\ref{lem:TIp-saddle}, since $\hat\eta_l$ stays in a compact subset of $(c_l,+\infty)$.
\end{proof}

Let
\begin{equation}\label{eq:w-def}
w(k):=D_{\mathrm{I}}(k)\,\delta_{\mathrm{I}}(k),\qquad r_2(k)=\frac{r^{*}(k)}{1-r(k)r^{*}(k)},\quad k\in\mathbb{R}\setminus[-c_l,c_l],
\end{equation}
where $r_2$ is as in \eqref{eq:fac1}. By \eqref{eq:V2-TIp}, \eqref{eq:fac1} and Propositions~\ref{prop:DI} and \ref{prop:deltaI}, the conjugation of $V^{(2)}$ by $\delta_{\mathrm{I}}$ takes the form
\begin{equation}\label{eq:V2-delta}
\delta_{\mathrm{I},-}^{\sigma_3}V^{(2)}\delta_{\mathrm{I},+}^{-\sigma_3}=\begin{cases}
\begin{pmatrix}1&-w^{2}r^{*}e^{-2itg_{\mathrm{I}}}\\0&1\end{pmatrix}\begin{pmatrix}1&0\\w^{-2}re^{2itg_{\mathrm{I}}}&1\end{pmatrix}, & k\in(-\infty,-\hat\eta_l)\cup(\eta_l,+\infty),\\[5mm]
\begin{pmatrix}1&0\\w_-^{-2}r_2^{*}e^{2itg_{\mathrm{I}}}&1\end{pmatrix}\begin{pmatrix}1&-w_+^{2}r_2e^{-2itg_{\mathrm{I}}}\\0&1\end{pmatrix}, & k\in(-\hat\eta_l,-c_l),\\[5mm]
\begin{pmatrix}1-rr^{*}&-w^{2}r^{*}e^{-2itg_{\mathrm{I}}}\\w^{-2}re^{2itg_{\mathrm{I}}}&1\end{pmatrix}, & k\in(c_l,\eta_l),\\[5mm]
\begin{pmatrix}0&-1\\1&\dfrac{w_+}{w_-}e^{-2itg_{\mathrm{I},+}}\end{pmatrix}, & k\in(-c_l,-c_r)\cup(c_r,c_l),\\[5mm]
\begin{pmatrix}0&-1\\1&0\end{pmatrix}, & k\in(-c_r,c_r).
\end{cases}
\end{equation}
Indeed, on $(-\hat\eta_l,-c_l)$ the diagonal factor $(1-rr^{*})^{\sigma_3}$ in the factorization of \eqref{eq:fac1} involving $r_2$ is cancelled by the first relation in \eqref{eq:deltaI-jump}, while on $(-c_l,c_l)$ the second relation in \eqref{eq:deltaI-jump} leaves the off-diagonal entries unchanged; on the remaining intervals $\delta_{\mathrm{I}}$ has no jump.

\subsubsection{Analytic approximation}\label{subsubsec:TIp-approx}

Let
\begin{equation}\label{eq:cm}
c_m:=\frac{c_l+c_r}{2},\qquad\text{so that}\qquad -c_m\in(-c_l,-c_r).
\end{equation}
We define the lens boundaries
\begin{equation}\label{eq:Gamma12-TIp}
\Gamma_1^{(3)}:=\eta_l+e^{\frac{\pi i}{3}}\mathbb{R}^{+},\qquad
\Gamma_2^{(3)}:=-\hat\eta_l+e^{\frac{3\pi i}{4}}\mathbb{R}^{+},
\end{equation}
and let $\Gamma_3^{(3)}$ be a smooth arc in $\mathbb{C}^{+}$ from $-\hat\eta_l$ to $-c_m$ which leaves $-\hat\eta_l$ in the direction $e^{\frac{\pi i}{4}}$. For $j=1,2,3$, $\Gamma_j^{(3)*}$ denotes the complex conjugate of $\Gamma_j^{(3)}$, and all these contours are oriented from left to right. Let $U_1^{(3)}$ be the open region between $(\eta_l,+\infty)$ and $\Gamma_1^{(3)}$, $U_2^{(3)}$ the open region between $(-\infty,-\hat\eta_l)$ and $\Gamma_2^{(3)}$, and $U_3^{(3)}$ the open region bounded by $[-\hat\eta_l,-c_m]$ and $\Gamma_3^{(3)}$; let $U_j^{(3)*}$ be the complex conjugate of $U_j^{(3)}$, $j=1,2,3$. These regions are illustrated in Figure~\ref{fig:contour-M3-TIp}. The arc $\Gamma_3^{(3)}$ is chosen such that
\begin{equation}\label{eq:U-sign-TIp}
U_1^{(3)}\cup U_2^{(3)}\subset\big\{k\in\mathbb{C}^{+}:\operatorname{Im}g_{\mathrm{I}}(k)>0\big\},\qquad
U_3^{(3)}\subset\big\{k\in\mathbb{C}^{+}:\operatorname{Im}g_{\mathrm{I}}(k)<0\big\},
\end{equation}
which is possible by the signature table in Figure~\ref{fig:sign-TIp}; by item (a) of Proposition~\ref{prop:gI}, the reversed inequalities hold on $U_j^{(3)*}$. Note that the hard edge $-c_l$ lies on the common boundary $[-\hat\eta_l,-c_m]$ of $U_3^{(3)}$ and $U_3^{(3)*}$, at a positive distance from $\Gamma_3^{(3)}\cup\Gamma_3^{(3)*}$, whereas the segment $[c_l,\eta_l]$ is kept on the real line. The directions of the contours at $\eta_l$ and at $-\hat\eta_l$ are adapted to the local analysis near $c_l$ and to the parabolic cylinder parametrix at $-\hat\eta_l$, respectively. The following proposition establishes the analytic approximations of $r$ on $(-\infty,-\hat\eta_l)\cup(\eta_l,+\infty)$ and of $r_2$ on $(-\hat\eta_l,-c_l)$ in the regions shown in Figure~\ref{fig:contour-M3-TIp}. Throughout, let
\begin{equation}\label{eq:varsigma}
\varsigma(k):=(-c_l-k)^{\frac12},\qquad k\in\mathbb{C}\setminus[-c_l,+\infty),
\end{equation}
with the principal branch. It maps $U_3^{(3)}$ (resp. $U_3^{(3)*}$) into the fourth (resp. first) quadrant, and its boundary values on $(-c_l,+\infty)$ are $\varsigma_\pm(k)=\mp i(k+c_l)^{1/2}$.

\begin{proposition}\label{prop:ra-TIp}
Let $\xi$ satisfy \eqref{eq:TIp-half}. There exist continuous functions
\begin{gather*}
r_a:\overline{U_1^{(3)}}\cup\overline{U_2^{(3)}}\to\mathbb{C},\qquad
r_r:(-\infty,-\hat\eta_l)\cup(\eta_l,+\infty)\to\mathbb{C},\\
r_{2,a}:\overline{U_3^{(3)}}\setminus\{-c_l\}\to\mathbb{C},\qquad
r_{2,r}:(-\hat\eta_l,-c_l)\to\mathbb{C},
\end{gather*}
with the following properties.
\begin{itemize}
\item[\rm(a)] $r=r_a+r_r$ on $(-\infty,-\hat\eta_l)\cup(\eta_l,+\infty)$, and $r_2=r_{2,a}+r_{2,r}$ on $(-\hat\eta_l,-c_l)$.
\item[\rm(b)] $r_a$ is holomorphic in $U_1^{(3)}\cup U_2^{(3)}$, and
\begin{align*}
&|r_a(k)-r(\eta_l)|\lesssim|k-\eta_l|^{\frac12}+|k-\eta_l|\,e^{\frac t4|\operatorname{Im}g_{\mathrm{I}}(k)|}, && k\in\overline{U_1^{(3)}},\\
&|r_a(k)-r(-\hat\eta_l)|\lesssim|k+\hat\eta_l|\,e^{\frac t4|\operatorname{Im}g_{\mathrm{I}}(k)|}, && k\in\overline{U_2^{(3)}},\\
&|r_a(k)|\lesssim\frac{e^{\frac t4|\operatorname{Im}g_{\mathrm{I}}(k)|}}{1+|k|^{2}}, && k\in\overline{U_1^{(3)}}\cup\overline{U_2^{(3)}}.
\end{align*}
\item[\rm(c)] $r_{2,a}$ is holomorphic in $U_3^{(3)}$, and
\begin{align*}
&|r_{2,a}(k)-r_2(-\hat\eta_l)|\lesssim|k+\hat\eta_l|\,e^{\frac t4|\operatorname{Im}g_{\mathrm{I}}(k)|},&& k\in\overline{U_3^{(3)}},\ |k+\hat\eta_l|\le\tfrac12(\hat\eta_l-c_l),\\
&|\varsigma(k)\,r_{2,a}(k)|\lesssim e^{\frac t4|\operatorname{Im}g_{\mathrm{I}}(k)|},&& k\in\overline{U_3^{(3)}}\setminus\{-c_l\}.
\end{align*}
\item[\rm(d)] Let $r^{*}_{2,a}(k):=\overline{r_{2,a}(\bar k)}$ for $k\in\overline{U_3^{(3)*}}\setminus\{-c_l\}$, and let $r_{2,a,+}$, $r^{*}_{2,a,-}$ denote the boundary values on $(-c_l,-c_m)$ from $U_3^{(3)}$ and $U_3^{(3)*}$, respectively. Then the function
\begin{equation}\label{eq:Ba}
B_a(k):=\frac{1}{r_+(k)}+\frac{r^{*}_{2,a,-}(k)}{r_+(k)^{2}}+r_{2,a,+}(k),\qquad k\in(-c_l,-c_m),
\end{equation}
satisfies $|B_a(k)|\lesssim|k+c_l|\,e^{\frac t4|\operatorname{Im}g_{\mathrm{I},+}(k)|}$.
\item[\rm(e)] As $t\to+\infty$, $\|r_r\|_{L^{p}((-\infty,-\hat\eta_l)\cup(\eta_l,+\infty))}=\mathcal{O}(t^{-1})$ for $p\in[1,+\infty]$, and $|r_{2,r}(k)|\lesssim t^{-1}|k+c_l|$ for $k\in(-\hat\eta_l,-c_l)$.
\end{itemize}
\end{proposition}

\begin{proof}
The proofs are analogous to those of Lemma 5.2 in \cite{S-J}.
\end{proof}

\subsubsection{RH problem for $M^{(3)}$}\label{subsubsec:TIp-RHP3}

With the functions of Proposition~\ref{prop:ra-TIp} and $w=D_{\mathrm{I}}\delta_{\mathrm{I}}$ from \eqref{eq:w-def}, we define, with $r_a^{*}(k):=\overline{r_a(\bar k)}$,
\begin{equation}\label{eq:G-TIp}
G(k):=\begin{cases}
\begin{pmatrix}1&0\\-w^{-2}r_ae^{2itg_{\mathrm{I}}}&1\end{pmatrix}, & k\in U_1^{(3)}\cup U_2^{(3)},\\[4mm]
\begin{pmatrix}1&-w^{2}r_a^{*}e^{-2itg_{\mathrm{I}}}\\0&1\end{pmatrix}, & k\in U_1^{(3)*}\cup U_2^{(3)*},\\[4mm]
\begin{pmatrix}1&w^{2}r_{2,a}e^{-2itg_{\mathrm{I}}}\\0&1\end{pmatrix}, & k\in U_3^{(3)},\\[4mm]
\begin{pmatrix}1&0\\w^{-2}r^{*}_{2,a}e^{2itg_{\mathrm{I}}}&1\end{pmatrix}, & k\in U_3^{(3)*},\\[3mm]
I, & \text{elsewhere},
\end{cases}
\end{equation}
and
\begin{equation}\label{eq:M3-TIp}
M^{(3)}(k)=M^{(3)}(k;x,t):=e^{i\varphi_{\mathrm{I}}\sigma_3}\,M^{(2)}(k)\,\delta_{\mathrm{I}}(k)^{-\sigma_3}\,G(k).
\end{equation}
Then $M^{(3)}$ satisfies the following RH problem.

\begin{figure}[htbp]
\centering
\includegraphics[width=0.98\textwidth]{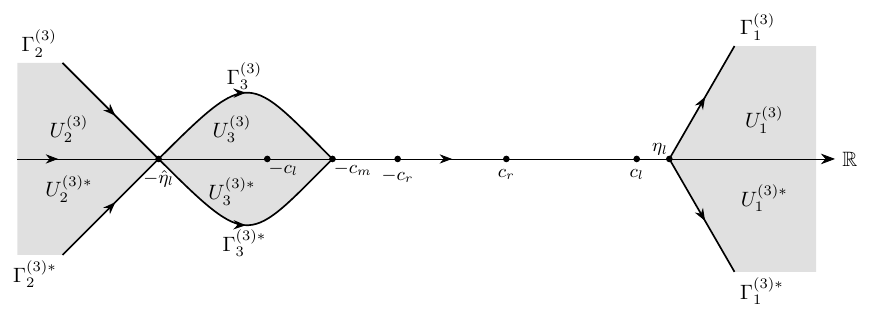}
\caption{The jump contour $\Gamma^{(3)}:=\bigcup_{j=1}^{3}\big(\Gamma_j^{(3)}\cup\Gamma_j^{(3)*}\big)\cup\mathbb{R}$ and the regions $U_j^{(3)}$, $U_j^{(3)*}$ (shaded), $j=1,2,3$.}
\label{fig:contour-M3-TIp}
\end{figure}

\begin{rhp}\label{rhp:M3-TIp}
Find $M^{(3)}(k)=M^{(3)}(k;x,t)$ with the properties
\begin{itemize}
\item[\rm(a)] $M^{(3)}(k)$ is holomorphic for $k\in\mathbb{C}\setminus\Gamma^{(3)}$, where $\Gamma^{(3)}=\bigcup_{j=1}^{3}\big(\Gamma_j^{(3)}\cup\Gamma_j^{(3)*}\big)\cup\mathbb{R}$; see Figure~\ref{fig:contour-M3-TIp}.
\item[\rm(b)] $M^{(3)}(k)$ has continuous boundary values $M^{(3)}_{\pm}(k)$ on $\Gamma^{(3)}$, except at the points $-\hat\eta_l$, $\pm c_l$, $\pm c_r$, $-c_m$ and $\eta_l$, and $M^{(3)}_+(k)=M^{(3)}_-(k)V^{(3)}(k)$, where
\begin{equation}\label{eq:V3-TIp}
V^{(3)}(k)=\begin{cases}
\begin{pmatrix}1&0\\w^{-2}r_ae^{2itg_{\mathrm{I}}}&1\end{pmatrix}, & k\in\Gamma_1^{(3)}\cup\Gamma_2^{(3)},\\[4mm]
\begin{pmatrix}1&-w^{2}r_a^{*}e^{-2itg_{\mathrm{I}}}\\0&1\end{pmatrix}, & k\in\Gamma_1^{(3)*}\cup\Gamma_2^{(3)*},\\[4mm]
\begin{pmatrix}1&-w^{2}r_{2,a}e^{-2itg_{\mathrm{I}}}\\0&1\end{pmatrix}, & k\in\Gamma_3^{(3)},\\[4mm]
\begin{pmatrix}1&0\\w^{-2}r^{*}_{2,a}e^{2itg_{\mathrm{I}}}&1\end{pmatrix}, & k\in\Gamma_3^{(3)*},\\[4mm]
\begin{pmatrix}1-r_rr_r^{*}&-w^{2}r_r^{*}e^{-2itg_{\mathrm{I}}}\\w^{-2}r_re^{2itg_{\mathrm{I}}}&1\end{pmatrix}, & k\in(-\infty,-\hat\eta_l)\cup(\eta_l,+\infty),\\[4mm]
\begin{pmatrix}1&0\\w_-^{-2}r^{*}_{2,r}e^{2itg_{\mathrm{I}}}&1\end{pmatrix}\begin{pmatrix}1&-w_+^{2}r_{2,r}e^{-2itg_{\mathrm{I}}}\\0&1\end{pmatrix}, & k\in(-\hat\eta_l,-c_l),\\[4mm]
\begin{pmatrix}1-rr^{*}&-w^{2}r^{*}e^{-2itg_{\mathrm{I}}}\\w^{-2}re^{2itg_{\mathrm{I}}}&1\end{pmatrix}, & k\in(c_l,\eta_l),\\[4mm]
\begin{pmatrix}0&-1\\1&w_+^{2}e^{-2itg_{\mathrm{I},+}}B_a\end{pmatrix}, & k\in(-c_l,-c_m),\\[4mm]
\begin{pmatrix}0&-1\\1&\dfrac{w_+}{w_-}e^{-2itg_{\mathrm{I},+}}\end{pmatrix}, & k\in(-c_m,-c_r)\cup(c_r,c_l),\\[5mm]
\begin{pmatrix}0&-1\\1&0\end{pmatrix}, & k\in(-c_r,c_r),
\end{cases}
\end{equation}
with $r_r^{*}:=\overline{r_r}$ and $r^{*}_{2,r}:=\overline{r_{2,r}}$ on the real line, and $B_a$ given in \eqref{eq:Ba}.
\item[\rm(c)] $M^{(3)}(k)=I+\mathcal{O}(k^{-1})$ as $k\to\infty$ in $\mathbb{C}\setminus\Gamma^{(3)}$.
\item[\rm(d)] $M^{(3)}(k)=\mathcal{O}(|k+c_l|^{-1/4})$ as $k\to-c_l$, $M^{(3)}(k)=\mathcal{O}(1)$ as $k\to c_l$, and $M^{(3)}$ has the same behaviour as $M^{(2)}$ at $\pm c_r$; see RH problem~\ref{rhp:M2-TIp}.
\end{itemize}
\end{rhp}

In addition, since $G=I$ outside the lenses and $\delta_{\mathrm{I}}(k)=e^{i\varphi_{\mathrm{I}}}(1+\mathcal{O}(k^{-1}))$,
\begin{equation}\label{eq:q-M3-TIp}
q(x,t)=2i\,e^{-2i\alpha c_l^{2}t-2i\varphi_{\mathrm{I}}}\lim_{k\to\infty}\big(kM^{(3)}(k)\big)_{12},
\end{equation}
where the limit is taken in $\mathbb{C}\setminus\big(\overline{U_1^{(3)}}\cup\overline{U_2^{(3)}}\cup\overline{U_1^{(3)*}}\cup\overline{U_2^{(3)*}}\big)$.

\subsection{Analysis of RH problem for $M^{(3)}$}\label{subsec:TIp-M3-analysis}

It is readily seen from RH problem~\ref{rhp:M3-TIp} and \eqref{eq:U-sign-TIp} that $V^{(3)}\to I$ exponentially as $t\to+\infty$ on $\Gamma_j^{(3)}\cup\Gamma_j^{(3)*}$, $j=1,2,3$, except around the points $c_l$ and $-\hat\eta_l$. Moreover, Proposition~\ref{prop:ra-TIp} ensures that $V^{(3)}$ decays to the identity matrix on $(-\infty,-\hat\eta_l)\cup(-\hat\eta_l,-c_l)\cup(\eta_l,+\infty)$, and that on $(-c_l,c_l)$ the matrix $V^{(3)}$ tends to $\left(\begin{smallmatrix}0&-1\\1&0\end{smallmatrix}\right)$, except near $c_l$, as $t\to+\infty$. Therefore, we construct a global parametrix together with local parametrices near the soft edge $c_l$ and near the stationary point $-\hat\eta_l$.

\subsubsection{Global parametrix}\label{subsubsec:TIp-global}

For $\xi$ satisfying \eqref{eq:TIp-half}, the global parametrix $M^{(\infty)}(k)$ satisfies the following RH problem.

\begin{rhp}\label{rhp:Minf-TIp}
Find $M^{(\infty)}(k)$ with the properties
\begin{itemize}
\item[\rm(a)] $M^{(\infty)}(k)$ is holomorphic for $k\in\mathbb{C}\setminus[-c_l,c_l]$.
\item[\rm(b)] $M^{(\infty)}(k)$ has continuous boundary values $M^{(\infty)}_{\pm}(k)$ on $(-c_l,c_l)$ satisfying the jump condition
\begin{equation*}
M^{(\infty)}_{+}(k)=M^{(\infty)}_{-}(k)V^{(\infty)},\qquad V^{(\infty)}=\begin{pmatrix}0&-1\\1&0\end{pmatrix}.
\end{equation*}
\item[\rm(c)] $M^{(\infty)}(k)=I+\mathcal{O}(k^{-1})$ as $k\to\infty$ in $\mathbb{C}\setminus[-c_l,c_l]$.
\item[\rm(d)] $M^{(\infty)}(k)=\mathcal{O}\big(|k\mp c_l|^{-1/4}\big)$ as $k\to\pm c_l$.
\end{itemize}
\end{rhp}

It follows from a direct calculation that the unique solution of RH problem~\ref{rhp:Minf-TIp} is given by
\begin{equation}\label{eq:Minf-TIp}
M^{(\infty)}(k)=\Delta_l(k)=\frac12\begin{pmatrix}\chi_l(k)+\chi_l(k)^{-1}&i\big(\chi_l(k)-\chi_l(k)^{-1}\big)\\[1mm]-i\big(\chi_l(k)-\chi_l(k)^{-1}\big)&\chi_l(k)+\chi_l(k)^{-1}\end{pmatrix},\qquad
\chi_l(k)=\Big(\frac{k-c_l}{k+c_l}\Big)^{\frac14},
\end{equation}
where $\Delta_l$ and $\chi_l$ are defined in \eqref{eq:Delta} with $j=l$. Indeed, $\chi_{l,+}=i\chi_{l,-}$ on $(-c_l,c_l)$, which gives (b); items (c) and (d) follow from $\chi_l(k)=1+\mathcal{O}(k^{-1})$ and $\chi_l(k)^{\pm1}=\mathcal{O}(|k\mp c_l|^{\pm1/4})$; and $\det\Delta_l\equiv1$. If $\widetilde M$ is another solution, then $\widetilde M(M^{(\infty)})^{-1}$ has no jump, tends to $I$ at infinity and is $\mathcal{O}(|k\mp c_l|^{-1/2})$ near $\pm c_l$, so that these singularities are removable and $\widetilde M=M^{(\infty)}$ by Liouville's theorem. Moreover, $M^{(\infty)}(k)=\sigma_1\overline{M^{(\infty)}(\bar k)}\sigma_1$, and
\begin{equation}\label{eq:Minf-exp-TIp}
M^{(\infty)}(k)=I+\frac{c_l}{2k}\,\sigma_2+\mathcal{O}(k^{-2}),\qquad k\to\infty,\qquad \sigma_2=\begin{pmatrix}0&-i\\ i&0\end{pmatrix}.
\end{equation}

\subsubsection{Local parametrices near $c_l$ and $-\hat\eta_l$}\label{subsubsec:TIp-local}

Fix $\epsilon\in(\frac12,\frac23)$ and let
\begin{equation}\label{eq:disks-TIp}
\begin{aligned}
D^{(r)}&:=D_\varrho(c_l)=\{k:|k-c_l|<\varrho\}, && \varrho:=t^{-\frac23+\epsilon},\\
D^{(l)}&:=D_{\varrho_0}(-\hat\eta_l)=\{k:|k+\hat\eta_l|<\varrho_0\}, && \varrho_0:=\tfrac13\big(\hat\eta_l^{\,0}-c_l\big),
\end{aligned}
\end{equation}
where $\hat\eta_l^{\,0}$ is given in \eqref{eq:hatetal}. Set
\begin{equation*}
\Gamma^{(j)}:=D^{(j)}\cap\Gamma^{(3)},\qquad j\in\{r,l\}.
\end{equation*}

For $j\in\{r,l\}$, we intend to solve the following local RH problem for $M^{(j)}$.

\begin{rhp}\label{rhp:Mloc-TIp}
Find $M^{(j)}(k)$, $j\in\{r,l\}$, with the properties
\begin{itemize}
\item[\rm(a)] $M^{(j)}(k)$ is holomorphic for $k\in D^{(j)}\setminus\Gamma^{(j)}$.
\item[\rm(b)] $M^{(j)}(k)$ has continuous boundary values $M^{(j)}_{\pm}(k)$ on $\Gamma^{(j)}\setminus\{c_l,\eta_l,-\hat\eta_l\}$ satisfying
\begin{equation*}
M^{(j)}_{+}(k)=M^{(j)}_{-}(k)\,V^{(3)}(k)\big|_{\Gamma^{(j)}},
\end{equation*}
where $V^{(3)}$ is given in \eqref{eq:V3-TIp}.
\item[\rm(c)] $M^{(j)}(k)M^{(\infty)}(k)^{-1}\to I$ as $t\to+\infty$, uniformly for $k\in\partial D^{(j)}$, where $M^{(\infty)}$ is given in \eqref{eq:Minf-TIp}.
\end{itemize}
\end{rhp}

The solution $M^{(l)}$ of RH problem~\ref{rhp:Mloc-TIp} is constructed in a standard manner with the help of the parabolic cylinder model problem; since $-\hat\eta_l$ stays at a positive distance from $-c_l$, this is the usual construction at a single stationary point (cf. \cite{XZ}), and it will be given in Section~\ref{subsubsec:TIp-local-PC}. The construction of $M^{(r)}$ near $c_l$, based on the Painlev\'e XXXIV model problem, is the main new ingredient. In the rest of this subsection we focus on it.

First, we introduce the following change of variables near $c_l$. Recall from \eqref{eq:gI-exp}--\eqref{eq:gI-coef} that
\begin{equation}\label{eq:gI-cl-exp}
g_{\mathrm{I}}(k)=g^{(1)}_{\mathrm{I},+}(k-c_l)^{\frac12}+g^{(2)}_{\mathrm{I},+}(k-c_l)^{\frac32}+\mathcal{O}\big(|k-c_l|^{\frac52}\big),\qquad k\to c_l,
\end{equation}
where
\begin{equation}\label{eq:g12-TI}
g^{(1)}_{\mathrm{I},+}=(2c_l)^{\frac12}\big(\xi-\xi_{\mathrm{I}}^{+}\big),\qquad
g^{(2)}_{\mathrm{I},+}=(2c_l)^{\frac12}\Big(8\beta c_l+2\alpha+\frac{\xi-\xi_{\mathrm{I}}^{+}}{4c_l}\Big).
\end{equation}
In fact, \eqref{eq:gI-cl-exp} can be written in closed form. Let
\begin{equation*}
F_{\mathrm{I}}(k):=p_{\mathrm{I}}(k)(k+c_l)^{\frac12},\qquad H_{\mathrm{I}}(k):=\frac{F_{\mathrm{I}}(k)-F_{\mathrm{I}}(c_l)}{k-c_l},
\end{equation*}
which are holomorphic in $\{|k-c_l|<2c_l\}$. Since $X_l(k)=(k-c_l)^{1/2}(k+c_l)^{1/2}$, we have
\begin{equation}\label{eq:gI-H}
g_{\mathrm{I}}(k)=g^{(1)}_{\mathrm{I},+}(k-c_l)^{\frac12}+H_{\mathrm{I}}(k)(k-c_l)^{\frac32},\qquad H_{\mathrm{I}}(c_l)=g^{(2)}_{\mathrm{I},+}.
\end{equation}
For $\xi$ satisfying \eqref{eq:TIp-half}, \eqref{eq:g12-TI} gives $g^{(1)}_{\mathrm{I},+}=\mathcal{O}(t^{-2/3})$ and $g^{(1)}_{\mathrm{I},+}<0$, while $g^{(2)}_{\mathrm{I},+}=2(2c_l)^{1/2}(4\beta c_l+\alpha)+\mathcal{O}(t^{-2/3})$ is positive and bounded away from zero. Hence $H_{\mathrm{I}}$ does not vanish in $D_\varrho(c_l)$ for $t$ large, and
\begin{equation}\label{eq:gI-first-order}
\big|g_{\mathrm{I}}(k)-H_{\mathrm{I}}(k)(k-c_l)^{\frac32}\big|=\big|g^{(1)}_{\mathrm{I},+}\big|\,|k-c_l|^{\frac12}\lesssim t^{-\frac23}|k-c_l|^{\frac12},\qquad k\in D_\varrho(c_l).
\end{equation}

For $k\in D_\varrho(c_l)$, we define
\begin{equation}\label{eq:zeta-r-TIp}
\zeta_r(k):=-\Big(\frac32\Big)^{\frac23}t^{\frac23}(k-c_l)\,H_{\mathrm{I}}(k)^{\frac23},
\end{equation}
with the principal branch of $H_{\mathrm{I}}^{2/3}$. Then $\zeta_r$ is holomorphic and univalent in $D_\varrho(c_l)$ for $t$ large, and
\begin{equation*}
\zeta_r(c_l)=0,\qquad \zeta_r'(c_l)=-\Big(\frac32\Big)^{\frac23}t^{\frac23}\big(g^{(2)}_{\mathrm{I},+}\big)^{\frac23}<0,\qquad \zeta_r(\bar k)=\overline{\zeta_r(k)}.
\end{equation*}
In particular, $\zeta_r$ maps $D_\varrho(c_l)\cap\mathbb{C}^{\pm}$ into $\mathbb{C}^{\mp}$, $(c_l-\varrho,c_l)$ into $\mathbb{R}^{+}$ and $(c_l,c_l+\varrho)$ into $\mathbb{R}^{-}$. Moreover, with principal branches,
\begin{align}
\zeta_r(k)^{\frac12}&=\mp i\Big(\frac32\Big)^{\frac13}t^{\frac13}H_{\mathrm{I}}(k)^{\frac13}(k-c_l)^{\frac12},&& k\in D_\varrho(c_l)\cap\mathbb{C}^{\pm},\label{eq:zeta-r-half}\\
\zeta_r(k)^{\frac32}&=\pm\frac{3i}{2}\,t\,H_{\mathrm{I}}(k)(k-c_l)^{\frac32},&& k\in D_\varrho(c_l)\cap\mathbb{C}^{\pm}.\label{eq:zeta-r-threehalf}
\end{align}
We next define
\begin{equation*}
S(k)=S(k;\xi):=\pm\,it\,\frac{g^{(1)}_{\mathrm{I},+}(k-c_l)^{\frac12}}{\zeta_r(k)^{\frac12}},\qquad k\in D_\varrho(c_l)\cap\mathbb{C}^{\pm}.
\end{equation*}
By \eqref{eq:zeta-r-half}, $S(k)=-(\frac23)^{1/3}t^{2/3}g^{(1)}_{\mathrm{I},+}H_{\mathrm{I}}(k)^{-1/3}$, so that $S$ extends to a holomorphic function in $D_\varrho(c_l)$, and
\begin{equation}\label{eq:s-TIp}
s:=S(c_l)=-\Big(\frac23\Big)^{\frac13}t^{\frac23}\frac{g^{(1)}_{\mathrm{I},+}}{\big(g^{(2)}_{\mathrm{I},+}\big)^{\frac13}}
=\kappa\,t^{\frac23}\big(\xi_{\mathrm{I}}^{+}-\xi\big)\big(1+\mathcal{O}(t^{-\frac23})\big),\qquad
\kappa:=\Big(\frac{2c_l}{3(4\beta c_l+\alpha)}\Big)^{\frac13}.
\end{equation}
Thus $0<s\le\kappa C(1+o(1))$ by \eqref{eq:TIp-half}, and $S(k)=s\,\big(g^{(2)}_{\mathrm{I},+}/H_{\mathrm{I}}(k)\big)^{1/3}=s\big(1+\mathcal{O}(k-c_l)\big)$. It follows from \eqref{eq:gI-H}, \eqref{eq:zeta-r-half} and \eqref{eq:zeta-r-threehalf} that
\begin{equation}\label{eq:theta-hat-identity}
\frac43\,\zeta_r(k)^{\frac32}+2S(k)\,\zeta_r(k)^{\frac12}=\begin{cases}2itg_{\mathrm{I}}(k),& k\in D_\varrho(c_l)\cap\mathbb{C}^{+},\\ -2itg_{\mathrm{I}}(k),& k\in D_\varrho(c_l)\cap\mathbb{C}^{-}.\end{cases}
\end{equation}
Furthermore, by \eqref{eq:etal}, \eqref{eq:zeta-r-TIp} and \eqref{eq:s-TIp}, the image of the saddle point $\eta_l$ satisfies
\begin{equation}\label{eq:zeta-r-eta}
\zeta_r(\eta_l)=-\frac s2+\mathcal{O}(t^{-\frac23}),
\end{equation}
which is the critical point $\zeta=-s/2$ of the function $\frac23\zeta^{3/2}+s\zeta^{1/2}$, up to $\mathcal{O}(t^{-2/3})$.

Similarly, near the stationary point $-\hat\eta_l$ we introduce a local variable adapted to the quadratic behaviour of $g_{\mathrm{I}}$. By \eqref{eq:gI-prime} and $n_{\mathrm{I}}(-\hat\eta_l)=0$,
\begin{equation}\label{eq:gI-second-heta}
g_{\mathrm{I}}''(-\hat\eta_l)=\frac{n_{\mathrm{I}}'(-\hat\eta_l)}{X_l(-\hat\eta_l)}=-\frac{36\beta\hat\eta_l^{\,2}-8\alpha\hat\eta_l+\xi-6\beta c_l^{2}}{\sqrt{\hat\eta_l^{\,2}-c_l^{2}}}<0,
\end{equation}
since $-\hat\eta_l$ is the smallest zero of the cubic $n_{\mathrm{I}}$, so that $n_{\mathrm{I}}'(-\hat\eta_l)>0$. For $k\in D^{(l)}$, we define
\begin{equation}\label{eq:zeta-l-TIp}
\zeta_l(k):=\big(2t\,|g_{\mathrm{I}}''(-\hat\eta_l)|\big)^{\frac12}(k+\hat\eta_l)\,\psi_l(k),\qquad
\psi_l(k):=\bigg(\frac{2\big(g_{\mathrm{I}}(-\hat\eta_l)-g_{\mathrm{I}}(k)\big)}{|g_{\mathrm{I}}''(-\hat\eta_l)|(k+\hat\eta_l)^{2}}\bigg)^{\frac12},
\end{equation}
where, decreasing $\varrho_0$ in \eqref{eq:disks-TIp} if necessary, $\psi_l$ is holomorphic and non-vanishing in $D^{(l)}$ with $\psi_l(-\hat\eta_l)=1$, and $\zeta_l$ is univalent in $D^{(l)}$. Then
\begin{equation*}
\zeta_l(-\hat\eta_l)=0,\qquad \zeta_l'(-\hat\eta_l)=\big(2t|g_{\mathrm{I}}''(-\hat\eta_l)|\big)^{\frac12}>0,\qquad \zeta_l(\bar k)=\overline{\zeta_l(k)},
\end{equation*}
and
\begin{equation}\label{eq:zeta-l-identity}
2itg_{\mathrm{I}}(k)=2itg_{\mathrm{I}}(-\hat\eta_l)-\frac i2\,\zeta_l(k)^{2},\qquad k\in D^{(l)}.
\end{equation}
In particular, $\zeta_l$ maps $(-\hat\eta_l-\varrho_0,-\hat\eta_l)$ and $(-\hat\eta_l,-\hat\eta_l+\varrho_0)$ into $\mathbb{R}^{-}$ and $\mathbb{R}^{+}$, respectively. The four segments of $\Gamma^{(l)}$ are mapped onto curves tangent at $\zeta_l=0$ to the rays $\arg\zeta_l=\pm\frac\pi4,\pm\frac{3\pi}4$.

Since $\zeta_r$ is univalent in $D^{(r)}$, we denote by $k=k(\zeta_r)$ its inverse and set
\begin{equation*}
\widehat D^{(r)}:=\zeta_r\big(D^{(r)}\big),\qquad
\widehat\Gamma^{(r)}_1:=\zeta_r\big(\Gamma_1^{(3)*}\cap D^{(r)}\big)\subset\mathbb{C}^{+},\qquad
\widehat\Gamma^{(r)*}_1:=\zeta_r\big(\Gamma_1^{(3)}\cap D^{(r)}\big)\subset\mathbb{C}^{-},
\end{equation*}
and $\widehat\Gamma^{(r)}:=\zeta_r(\Gamma^{(r)})=\widehat\Gamma^{(r)}_1\cup\widehat\Gamma^{(r)*}_1\cup\big(\zeta_r(c_l+\varrho),\zeta_r(c_l-\varrho)\big)$. Since $\zeta_r'(c_l)<0$, the curves $\widehat\Gamma^{(r)}_1$ and $\widehat\Gamma^{(r)*}_1$ leave $\zeta_r(\eta_l)$ in the directions $e^{\frac{2\pi i}{3}}$ and $e^{-\frac{2\pi i}{3}}$, respectively. We orient the real segment of $\widehat\Gamma^{(r)}$ from left to right, and $\widehat\Gamma^{(r)}_1$, $\widehat\Gamma^{(r)*}_1$ towards $\zeta_r(\eta_l)$. This orientation is opposite to the one induced by $\zeta_r$ from $\Gamma^{(r)}$.

\begin{figure}[htbp]
\centering
\includegraphics[width=0.9\textwidth]{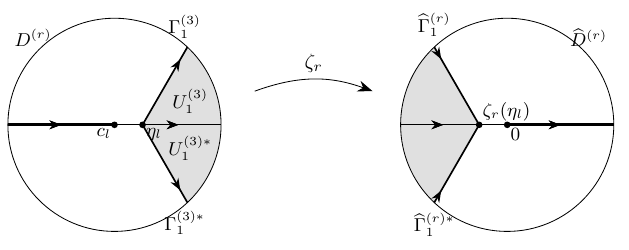}
\caption{The disk $D^{(r)}=D_\varrho(c_l)$ with the contour $\Gamma^{(r)}$ (left), and its image $\widehat D^{(r)}=\zeta_r(D^{(r)})$ with the contour $\widehat\Gamma^{(r)}$ (right).}
\label{fig:map-zeta-r-TIp}
\end{figure}

Let $\widehat M^{(r)}(\zeta_r):=M^{(r)}(k(\zeta_r))$. By RH problem~\ref{rhp:Mloc-TIp}, $\widehat M^{(r)}$ is holomorphic in $\widehat D^{(r)}\setminus\widehat\Gamma^{(r)}$ and satisfies $\widehat M^{(r)}_{+}(\zeta_r)=\widehat M^{(r)}_{-}(\zeta_r)\widehat V^{(3)}(\zeta_r)$ on $\widehat\Gamma^{(r)}$. Because of the reversed orientation, $\widehat V^{(3)}(\zeta_r)=V^{(3)}(k(\zeta_r))^{-1}$, that is,
\begin{equation}\label{eq:Vhat3-TIp}
\widehat V^{(3)}(\zeta_r)=\begin{cases}
\begin{pmatrix}1&w^{2}r_a^{*}e^{-2itg_{\mathrm{I}}}\\0&1\end{pmatrix}, & \zeta_r\in\widehat\Gamma^{(r)}_1,\\[4mm]
\begin{pmatrix}1&0\\-w^{-2}r_ae^{2itg_{\mathrm{I}}}&1\end{pmatrix}, & \zeta_r\in\widehat\Gamma^{(r)*}_1,\\[4mm]
\begin{pmatrix}1&w^{2}r_r^{*}e^{-2itg_{\mathrm{I}}}\\-w^{-2}r_re^{2itg_{\mathrm{I}}}&1-r_rr_r^{*}\end{pmatrix}, & \zeta_r\in\big(\zeta_r(c_l+\varrho),\zeta_r(\eta_l)\big),\\[4mm]
\begin{pmatrix}1&w^{2}r^{*}e^{-2itg_{\mathrm{I}}}\\-w^{-2}re^{2itg_{\mathrm{I}}}&1-rr^{*}\end{pmatrix}, & \zeta_r\in\big(\zeta_r(\eta_l),0\big),\\[4mm]
\begin{pmatrix}\dfrac{w_+}{w_-}e^{-2itg_{\mathrm{I},+}}&1\\-1&0\end{pmatrix}, & \zeta_r\in\big(0,\zeta_r(c_l-\varrho)\big),
\end{cases}
\end{equation}
where all functions on the right-hand side are evaluated at $k=k(\zeta_r)$, and the subscript $\pm$ refers to boundary values in the $k$-plane. The contours $\Gamma^{(r)}$, $\widehat\Gamma^{(r)}$ are illustrated in Figure~\ref{fig:map-zeta-r-TIp}.

It follows from the definition of $\varrho$ in \eqref{eq:disks-TIp} and from \eqref{eq:etal} that, as $t\to+\infty$,
\begin{equation}\label{eq:k-cl-TIp}
k-c_l=\mathcal{O}\big(t^{-\frac23+\epsilon}\big),\qquad k-\eta_l=\mathcal{O}\big(t^{-\frac23+\epsilon}\big),\qquad k\in D^{(r)}.
\end{equation}
With respect to item (e) of Proposition~\ref{prop:DI}, which contains a logarithmic term, to \eqref{eq:deltaI-cl}, and to item (c) of Propositions~\ref{prop:DI} and \ref{prop:deltaI}, we have, for large $t$,
\begin{equation}\label{eq:w-cl-TIp}
\begin{aligned}
&w(k)=e^{\frac i2\arg b_{l,0}}+\mathcal{O}\big(t^{-\frac13+\frac\epsilon2}\log t\big), && k\in D^{(r)},\\
&\frac{w_+(k)}{w_-(k)}=|w_+(k)|^{2}=1+\mathcal{O}\big(t^{-\frac13+\frac\epsilon2}\log t\big), && k\in(c_l-\varrho,c_l).
\end{aligned}
\end{equation}
Moreover, by item (b) of Proposition~\ref{prop:ra-TIp} and by \eqref{eq:r-endpoints}, which gives $r(\eta_l)-r(c_l)=\mathcal{O}(|\eta_l-c_l|^{1/2})=\mathcal{O}(t^{-1/3})$ with $r(c_l):=b_{l,0}$,
\begin{equation}\label{eq:ra-cl-TIp}
\begin{aligned}
|r_a(k)-r(c_l)|&\le|r_a(k)-r(\eta_l)|+|r(\eta_l)-r(c_l)|\lesssim|k-\eta_l|^{\frac12}+|k-\eta_l|\,e^{\frac t4|\operatorname{Im}g_{\mathrm{I}}(k)|}+|\eta_l-c_l|^{\frac12}\\
&\lesssim t^{-\frac13+\frac\epsilon2}+t^{-\frac23+\epsilon}e^{\frac t4|\operatorname{Im}g_{\mathrm{I}}(k)|},\qquad k\in D^{(r)}\cap\overline{U_1^{(3)}}.
\end{aligned}
\end{equation}
Since $w(c_l)^{-2}r(c_l)=e^{-i\arg b_{l,0}}b_{l,0}=1$, \eqref{eq:w-cl-TIp} and \eqref{eq:ra-cl-TIp} imply
\begin{equation}\label{eq:wra-TIp}
\begin{aligned}
\big|w(k)^{-2}r_a(k)-w(c_l)^{-2}r(c_l)\big|&=\big|w(k)^{-2}r_a(k)-1\big|\\
&\lesssim t^{-\frac13+\frac\epsilon2}\log t+t^{-\frac23+\epsilon}e^{\frac t4|\operatorname{Im}g_{\mathrm{I}}(k)|},\qquad k\in D^{(r)}\cap\overline{U_1^{(3)}}.
\end{aligned}
\end{equation}
By item (c) of Propositions~\ref{prop:DI} and \ref{prop:deltaI}, $\overline{w(\bar k)}=w(k)^{-1}$; together with $r_a^{*}(k)=\overline{r_a(\bar k)}$ this gives
\begin{equation*}
w(k)^{2}r_a^{*}(k)=\overline{w(\bar k)^{-2}r_a(\bar k)},\qquad k\in D^{(r)}\cap\overline{U_1^{(3)*}},
\end{equation*}
so that the same bound holds for $|w(k)^{2}r_a^{*}(k)-1|$ on $D^{(r)}\cap\overline{U_1^{(3)*}}$. Similarly, on the real segment $(c_l,\eta_l)$, where $r_a$ is replaced by $r$ in \eqref{eq:Vhat3-TIp},
\begin{equation}\label{eq:wr-real-TIp}
\big|w(k)^{-2}r(k)-1\big|+\big|w(k)^{2}r^{*}(k)-1\big|+\big|r(k)r^{*}(k)-1\big|\lesssim t^{-\frac13}\log t,\qquad k\in(c_l,\eta_l).
\end{equation}
In view of \eqref{eq:Vhat3-TIp} and the estimates \eqref{eq:w-cl-TIp}--\eqref{eq:wr-real-TIp}, the RH problem for $\widehat M^{(r)}$ is approximated by the following model RH problem. It is equivalent to the Painlev\'e XXXIV model problem with parameters $(b,\omega)=(0,1)$; see Remark~\ref{rem:omega-TIp} below.

\begin{rhp}\label{rhp:Nhat-TIp}
Find $\widehat N^{(r)}(\zeta_r)=\widehat N^{(r)}(\zeta_r;s)$ with the properties
\begin{itemize}
\item[\rm(a)] $\widehat N^{(r)}(\zeta_r)$ is holomorphic for $\zeta_r\in\mathbb{C}\setminus\widehat\Sigma^{(r)}$.
\item[\rm(b)] $\widehat N^{(r)}$ has continuous boundary values on $\widehat\Sigma^{(r)}\setminus\{\zeta_r(\eta_l),0\}$ and $\widehat N^{(r)}_+(\zeta_r)=\widehat N^{(r)}_-(\zeta_r)\widehat V_N^{(r)}(\zeta_r)$, where
\begin{equation}\label{eq:VN-TIp}
\widehat V_N^{(r)}(\zeta_r)=\begin{cases}
\begin{pmatrix}1&e^{2\hat\theta(\zeta_r)}\\0&1\end{pmatrix}, & \zeta_r\in\widehat\Sigma_1^{(r)},\\[4mm]
\begin{pmatrix}1&0\\-e^{2\hat\theta(\zeta_r)}&1\end{pmatrix}, & \zeta_r\in\widehat\Sigma_1^{(r)*},\\[4mm]
\begin{pmatrix}1&0\\-e^{2\hat\theta_-(\zeta_r)}&1\end{pmatrix}\begin{pmatrix}1&e^{2\hat\theta_+(\zeta_r)}\\0&1\end{pmatrix}, & \zeta_r\in(\zeta_r(\eta_l),0),\\[4mm]
\begin{pmatrix}e^{-2\hat\theta(\zeta_r)}&1\\-1&0\end{pmatrix}, & \zeta_r\in(0,+\infty),
\end{cases}
\end{equation}
with $\hat\theta(\zeta_r)=\frac23\zeta_r^{\frac32}+s\,\zeta_r^{\frac12}$.
\item[\rm(c)] As $\zeta_r\to\infty$ in $\mathbb{C}\setminus\widehat\Sigma^{(r)}$,
\begin{equation}\label{eq:Nhat-infty-TIp}
\widehat N^{(r)}(\zeta_r)=\big(I+\mathcal{O}(\zeta_r^{-1})\big)\,\zeta_r^{-\frac{\sigma_3}{4}}\,\frac{1}{\sqrt2}\begin{pmatrix}1&i\\ i&1\end{pmatrix}Q^{(r)}(\zeta_r),\qquad
Q^{(r)}(\zeta_r):=\begin{cases}\sigma_1,&\zeta_r\in\mathbb{C}^{+},\\ \sigma_3,&\zeta_r\in\mathbb{C}^{-}.\end{cases}
\end{equation}
\item[\rm(d)] $\widehat N^{(r)}(\zeta_r)$ is bounded near $\zeta_r(\eta_l)$ and near $0$.
\end{itemize}
\end{rhp}

\begin{figure}[htbp]
\centering
\includegraphics[width=0.62\textwidth]{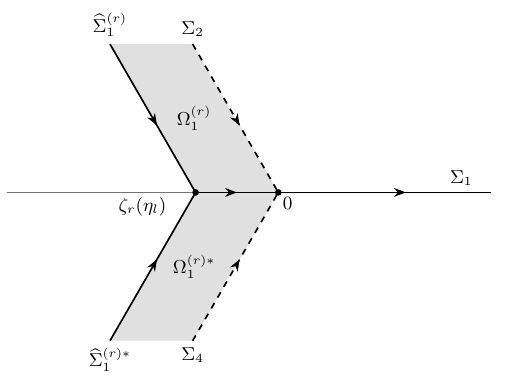}
\caption{The jump contour of $\widehat N^{(r)}$ (solid) and the auxiliary rays $\Sigma_2$, $\Sigma_4$ (dashed).}
\label{fig:Shat-TIp}
\end{figure}

Next, as in \cite{FWZ}, we remove the jumps of $\widehat N^{(r)}$ on $\widehat\Sigma_1^{(r)}$, $\widehat\Sigma_1^{(r)*}$ and $(\zeta_r(\eta_l),0)$ by moving the vertex of the two rays from $\zeta_r(\eta_l)$ to $0$. Let
\begin{equation}\label{eq:Sigma-j-TIp}
\Sigma_1:=(0,+\infty),\qquad \Sigma_2:=e^{\frac{2\pi i}{3}}\mathbb{R}^{+},\qquad \Sigma_3:=(-\infty,0),\qquad \Sigma_4:=e^{-\frac{2\pi i}{3}}\mathbb{R}^{+},
\end{equation}
where $\Sigma_1$ is oriented from $0$ to $+\infty$, and $\Sigma_2$, $\Sigma_3$, $\Sigma_4$ are oriented towards $0$. This is the jump contour of the Painlev\'e XXXIV model problem; see Appendix~\ref{app:P34} and \cite[Appendix~A]{FWZ}. Denote by $\Omega_1^{(r)}$ the half-strip in $\mathbb{C}^{+}$ bounded by $\widehat\Sigma_1^{(r)}$, $[\zeta_r(\eta_l),0]$ and $\Sigma_2$, and let $\Omega_1^{(r)*}:=\{\bar\zeta_r:\zeta_r\in\Omega_1^{(r)}\}$; see Figure~\ref{fig:Shat-TIp}. Define
\begin{equation}\label{eq:Gr-TIp}
G^{(r)}(\zeta_r):=\begin{cases}
\begin{pmatrix}1&-e^{2\hat\theta(\zeta_r)}\\0&1\end{pmatrix}, & \zeta_r\in\Omega_1^{(r)},\\[4mm]
\begin{pmatrix}1&0\\-e^{2\hat\theta(\zeta_r)}&1\end{pmatrix}, & \zeta_r\in\Omega_1^{(r)*},\\[3mm]
I, & \text{elsewhere},
\end{cases}
\end{equation}
and
\begin{equation}\label{eq:Shat-TIp}
\widehat S^{(r)}(\zeta_r)=\widehat S^{(r)}(\zeta_r;s):=\widehat N^{(r)}(\zeta_r)\,G^{(r)}(\zeta_r).
\end{equation}

\begin{rhp}\label{rhp:Shat-TIp}
Find $\widehat S^{(r)}(\zeta_r)=\widehat S^{(r)}(\zeta_r;s)$ with the properties
\begin{itemize}
\item[\rm(a)] $\widehat S^{(r)}(\zeta_r)$ is holomorphic for $\zeta_r\in\mathbb{C}\setminus(\Sigma_1\cup\Sigma_2\cup\Sigma_4)$.
\item[\rm(b)] $\widehat S^{(r)}$ has continuous boundary values on $(\Sigma_1\cup\Sigma_2\cup\Sigma_4)\setminus\{0\}$ and $\widehat S^{(r)}_+(\zeta_r)=\widehat S^{(r)}_-(\zeta_r)\widehat V_S^{(r)}(\zeta_r)$, where
\begin{equation}\label{eq:VS-TIp}
\widehat V_S^{(r)}(\zeta_r)=\begin{cases}
\begin{pmatrix}e^{-2\hat\theta(\zeta_r)}&1\\-1&0\end{pmatrix}, & \zeta_r\in\Sigma_1,\\[4mm]
\begin{pmatrix}1&e^{2\hat\theta(\zeta_r)}\\0&1\end{pmatrix}, & \zeta_r\in\Sigma_2,\\[4mm]
\begin{pmatrix}1&0\\-e^{2\hat\theta(\zeta_r)}&1\end{pmatrix}, & \zeta_r\in\Sigma_4.
\end{cases}
\end{equation}
\item[\rm(c)] As $\zeta_r\to\infty$ in $\mathbb{C}\setminus(\Sigma_1\cup\Sigma_2\cup\Sigma_4)$,
\begin{equation*}
\widehat S^{(r)}(\zeta_r)=\big(I+\mathcal{O}(\zeta_r^{-1})\big)\,\zeta_r^{-\frac{\sigma_3}{4}}\,\frac{1}{\sqrt2}\begin{pmatrix}1&i\\ i&1\end{pmatrix}Q^{(r)}(\zeta_r),
\end{equation*}
with $Q^{(r)}$ given in \eqref{eq:Nhat-infty-TIp}.
\item[\rm(d)] $\widehat S^{(r)}(\zeta_r)$ is bounded near $0$.
\end{itemize}
\end{rhp}

Consequently, $\widehat S^{(r)}$ can be expressed through the Painlev\'e XXXIV parametrix with $(b,\omega)=(0,1)$:
\begin{equation}\label{eq:Shat-P34-TIp}
\widehat S^{(r)}(\zeta_r)=\begin{pmatrix}1&0\\ ia(s)&1\end{pmatrix}M^{P34}(\zeta_r;0,1,s)\,e^{\hat\theta(\zeta_r)\sigma_3}\,Q^{(r)}(\zeta_r),\qquad a(s)=\frac{s^{2}}{4}.
\end{equation}
Finally, as $\zeta_r\to\infty$,
\begin{equation}\label{eq:Shat-infty-TIp}
\begin{aligned}
&\widehat S^{(r)}(\zeta_r)=\Big(I+\frac{M_1^{(r)}(s)}{\zeta_r}+\mathcal{O}(\zeta_r^{-2})\Big)\zeta_r^{-\frac{\sigma_3}{4}}\frac{1}{\sqrt2}\begin{pmatrix}1&i\\ i&1\end{pmatrix}Q^{(r)}(\zeta_r),\\
&M_1^{(r)}(s)=\begin{pmatrix}\dfrac{s^{4}}{32}-\dfrac{s}{4}&\dfrac{is^{2}}{4}\\[3mm] \dfrac{i(s^{6}-16s^{3}+28)}{192}&-\dfrac{s^{4}}{32}+\dfrac{s}{4}\end{pmatrix}.
\end{aligned}
\end{equation}
Since $\widehat S^{(r)}=\widehat N^{(r)}$ outside $\Omega_1^{(r)}\cup\Omega_1^{(r)*}$, and $G^{(r)}-I$ is exponentially small as $\zeta_r\to\infty$ in $\Omega_1^{(r)}\cup\Omega_1^{(r)*}$, the expansion \eqref{eq:Shat-infty-TIp} also holds for $\widehat N^{(r)}$.

We now compare $\widehat M^{(r)}$ with $\widehat N^{(r)}$. Since $\widehat M^{(r)}$ has not been constructed yet, we define the error function by its own RH problem and set $\widehat M^{(r)}:=E^{(r)}\widehat N^{(r)}$ afterwards. Hence the mismatch between $\widehat V^{(3)}$ and $\widehat V_N^{(r)}$, which is of order $t^{-1/3}\log t$, cannot be moved into the error problem in the $k$-plane, where it would be enlarged by a factor of order $t^{1/3}$. Let
\begin{equation}\label{eq:GammaE-TIp}
\widehat\Gamma_E^{(r)}:=\Big(\widehat\Gamma_1^{(r)}\cup\widehat\Gamma_1^{(r)*}\cup\big(\zeta_r(\eta_l),0\big)\cup\big(0,\zeta_r(c_l-\tfrac{\varrho}{2})\big)\Big)\cap\zeta_r\big(D_{\varrho/2}(c_l)\big),
\end{equation}
with the orientation of $\widehat\Gamma^{(r)}$; see Figure~\ref{fig:map-zeta-r-TIp}. Compared with $\widehat\Gamma^{(r)}$, two parts are removed. The first is the segment $(\zeta_r(c_l+\varrho),\zeta_r(\eta_l))$, on which $\widehat N^{(r)}$ has no jump and $\widehat V^{(3)}-I=\mathcal{O}(r_r)$. The second is the part of $\widehat\Gamma^{(r)}$ outside $\zeta_r(D_{\varrho/2}(c_l))$, on which the jump below is $\mathcal{O}(e^{-ct^{3\epsilon/2}})$. Both are kept in the error analysis of Section~\ref{subsec:TIp-err}. On $\widehat\Gamma_E^{(r)}$ we define
\begin{equation}\label{eq:VE-TIp}
V_E^{(r)}(\zeta_r):=\widehat N^{(r)}_-(\zeta_r)\,\widehat V^{(3)}(\zeta_r)\,\widehat V_N^{(r)}(\zeta_r)^{-1}\,\widehat N^{(r)}_-(\zeta_r)^{-1},
\end{equation}
where $\widehat V^{(3)}$ and $\widehat V_N^{(r)}$ are given in \eqref{eq:Vhat3-TIp} and \eqref{eq:VN-TIp}.

\begin{rhp}\label{rhp:E-TIp}
Find $E^{(r)}(\zeta_r)=E^{(r)}(\zeta_r;x,t)$ with the properties
\begin{itemize}
\item[\rm(a)] $E^{(r)}(\zeta_r)$ is holomorphic for $\zeta_r\in\mathbb{C}\setminus\widehat\Gamma_E^{(r)}$.
\item[\rm(b)] $E^{(r)}$ has continuous boundary values on $\widehat\Gamma_E^{(r)}$ except at its endpoints and at $\zeta_r(\eta_l)$, $0$, and $E^{(r)}_+(\zeta_r)=E^{(r)}_-(\zeta_r)V_E^{(r)}(\zeta_r)$ for $\zeta_r\in\widehat\Gamma_E^{(r)}$.
\item[\rm(c)] $E^{(r)}(\zeta_r)=I+\mathcal{O}(\zeta_r^{-1})$ as $\zeta_r\to\infty$.
\end{itemize}
\end{rhp}

If $E^{(r)}$ solves RH problem~\ref{rhp:E-TIp}, then $E^{(r)}\widehat N^{(r)}$ has the jump $\widehat V^{(3)}$ on $\widehat\Gamma_E^{(r)}$, since $\widehat N^{(r)}_+=\widehat N^{(r)}_-\widehat V_N^{(r)}$ there.

\begin{proposition}\label{prop:E-TIp}
Let $\xi$ satisfy \eqref{eq:TIp-half} and $\epsilon\in(\frac12,\frac23)$. There exist constants $c>0$ and $t_0>0$ such that the following statements hold for $t\ge t_0$, uniformly in $\xi$.
\begin{itemize}
\item[\rm(a)] The jump matrix satisfies
\begin{equation}\label{eq:VE-est-TIp}
\big|V_E^{(r)}(\zeta_r)-I\big|\lesssim t^{-\frac13}\log t\;e^{-c|\zeta_r|^{3/2}},\qquad \zeta_r\in\widehat\Gamma_E^{(r)}.
\end{equation}
In particular, $\big\|(1+|\zeta_r|)^{m}(V_E^{(r)}-I)\big\|_{L^{p}(\widehat\Gamma_E^{(r)})}=\mathcal{O}(t^{-1/3}\log t)$ for every $m\ge0$ and $p\in[1,\infty]$.
\item[\rm(b)] RH problem~\ref{rhp:E-TIp} has a unique solution, and
\begin{equation}\label{eq:E-L2-TIp}
\big\|E^{(r)}_\pm-I\big\|_{L^{2}(\widehat\Gamma_E^{(r)})}=\mathcal{O}\big(t^{-\frac13}\log t\big).
\end{equation}
\item[\rm(c)] Let $E_1^{(r)}:=-\frac{1}{2\pi i}\int_{\widehat\Gamma_E^{(r)}}E^{(r)}_-(\zeta)\big(V_E^{(r)}(\zeta)-I\big)\,\mathrm{d}\zeta$. Then $E_1^{(r)}=\mathcal{O}(t^{-1/3}\log t)$, and
\begin{equation}\label{eq:E-exp-TIp}
E^{(r)}(\zeta_r)=I+\frac{E_1^{(r)}}{\zeta_r}+\mathcal{O}\big(t^{-\frac13}\log t\,|\zeta_r|^{-2}\big),
\end{equation}
uniformly for $\zeta_r$ with $\operatorname{dist}(\zeta_r,\widehat\Gamma_E^{(r)})\ge\frac13|\zeta_r|$. In particular, $E^{(r)}(\zeta_r)-I=\mathcal{O}(t^{-1/3-\epsilon}\log t)$ uniformly for $\zeta_r\in\partial\widehat D^{(r)}$.
\end{itemize}
\end{proposition}

\begin{proof}
Throughout, $k=k(\zeta_r)$,
\begin{equation*}
X:=\widehat V^{(3)}\big(\widehat V_N^{(r)}\big)^{-1},\qquad
L(a):=\begin{pmatrix}1&0\\a&1\end{pmatrix},\qquad U(b):=\begin{pmatrix}1&b\\0&1\end{pmatrix},
\end{equation*}
so that $V_E^{(r)}-I=\widehat N^{(r)}_-(X-I)\widehat N^{(r)\,-1}_-$. By \eqref{eq:zeta-r-TIp},
\begin{equation*}
|\zeta_r|=\Big(\frac32\Big)^{\frac23}t^{\frac23}\,|H_{\mathrm{I}}(k)|^{\frac23}\,|k-c_l|,\qquad
|\zeta_r'(k)|=\Big(\frac32\Big)^{\frac23}t^{\frac23}\big(g^{(2)}_{\mathrm{I},+}\big)^{\frac23}\big(1+\mathcal{O}(k-c_l)\big),
\end{equation*}
where $H_{\mathrm{I}}(k)=g^{(2)}_{\mathrm{I},+}(1+\mathcal{O}(k-c_l))$ and $g^{(2)}_{\mathrm{I},+}$ is bounded and bounded away from zero. Together with \eqref{eq:etal}, there is a constant $C>0$ such that, for $k\in D^{(r)}$,
\begin{equation}\label{eq:k-zeta-TIp}
\begin{gathered}
C^{-1}t^{-\frac23}|\zeta_r|\le|k-c_l|\le C\,t^{-\frac23}|\zeta_r|,\qquad |k-\eta_l|\le C\,t^{-\frac23}(1+|\zeta_r|),\\
C^{-1}t^{\frac23}|\mathrm{d}k|\le|\mathrm{d}\zeta_r|\le C\,t^{\frac23}|\mathrm{d}k| .
\end{gathered}
\end{equation}
All $L^{p}$ norms are taken with respect to $|\mathrm{d}\zeta_r|$ on $\widehat\Gamma_E^{(r)}$, and $u:=|k-c_l|$.

We first compute $X$. By \eqref{eq:Vhat3-TIp} and \eqref{eq:VN-TIp},
\begin{equation}\label{eq:X-TIp}
X=\begin{cases}
U\big(w^{2}r_a^{*}e^{2\hat\theta_S}-e^{2\hat\theta}\big), & \zeta_r\in\widehat\Gamma_1^{(r)},\\[1mm]
L\big(e^{2\hat\theta}-w^{-2}r_ae^{2\hat\theta_S}\big), & \zeta_r\in\widehat\Gamma_1^{(r)*},\\[1mm]
L(a)\,U(b)\,U\big(-e^{2\hat\theta_+}\big)\,L\big(e^{2\hat\theta_-}\big), & \zeta_r\in(\zeta_r(\eta_l),0),\\[1mm]
U\big(e^{-2\hat\theta}-|w_+|^{2}e^{-2\hat\theta_S}\big), & \zeta_r\in\big(0,\zeta_r(c_l-\tfrac\varrho2)\big),
\end{cases}
\end{equation}
where $a:=-w^{-2}r\,e^{2\hat\theta_{S,-}}$ and $b:=w^{2}r^{*}e^{2\hat\theta_{S,+}}$.

On $\widehat\Gamma_1^{(r)}$ we split $X_{12}$ into $I_{1}$ and $I_{2}$. Since $e^{2\hat\theta_S}=e^{2\hat\theta}e^{2(S(k)-s)\zeta_r^{1/2}}$,
\begin{equation}\label{eq:X12-split-TIp}
|X_{12}|\le I_1+I_2,\qquad
I_1:=\big|\big(w^{2}r_a^{*}-1\big)e^{2\hat\theta}\big|,\qquad
I_2:=\big|w^{2}r_a^{*}e^{2\hat\theta}\big(e^{2(S(k)-s)\zeta_r^{1/2}}-1\big)\big| .
\end{equation}
On the rays $\zeta_r(\eta_l)+e^{\pm\frac{2\pi i}3}\mathbb{R}^{+}$ we have $\operatorname{Re}\hat\theta\le-\frac12|\zeta_r|^{3/2}+C$ and $|\zeta_r|\ge|\zeta_r(\eta_l)|$; moreover $|e^{2\hat\theta_S}|=e^{-2t|\operatorname{Im}g_{\mathrm{I}}|}$ on $\widehat\Gamma_1^{(r)}$ and $|\hat\theta_S-\hat\theta|=|S-s||\zeta_r|^{\frac12}\lesssim t^{-\frac23}|\zeta_r|^{\frac32}$. Hence, by \eqref{eq:k-zeta-TIp},
\begin{equation}\label{eq:phase-absorb-TIp}
e^{\frac t4|\operatorname{Im}g_{\mathrm{I}}(k)|}\,\big|e^{2\hat\theta}\big|=\big|e^{2\hat\theta_S}\big|^{-\frac18}\big|e^{2\hat\theta}\big|\lesssim e^{-c|\zeta_r|^{3/2}}\le e^{-c'tu^{3/2}},\qquad \zeta_r\in\widehat\Gamma_1^{(r)} .
\end{equation}

By item (e) of Proposition~\ref{prop:DI}, \eqref{eq:deltaI-cl}, item (b) of Proposition~\ref{prop:ra-TIp}, $|r(\eta_l)-r(c_l)|\lesssim|\eta_l-c_l|^{1/2}$, $w(c_l)^{-2}r(c_l)=1$, and $|\eta_l-c_l|\lesssim u$, $|k-\eta_l|\lesssim u$ on $\widehat\Gamma_1^{(r)}$,
\begin{equation*}
\big|w^{2}r_a^{*}-1\big|\lesssim u^{\frac12}\big(1+|\log u|\big)+u\,e^{\frac t4|\operatorname{Im}g_{\mathrm{I}}(k)|}.
\end{equation*}
Together with \eqref{eq:phase-absorb-TIp},
\begin{equation}\label{eq:I1-pointwise-TIp}
I_1\lesssim\big(u^{\frac12}(1+|\log u|)+u\big)e^{-ctu^{3/2}}\lesssim t^{-\frac13}\log t\;e^{-c|\zeta_r|^{3/2}} .
\end{equation}
With $u=t^{-2/3}x$ and $|\mathrm{d}\zeta_r|\le C\,t^{2/3}\mathrm{d}u$,
\begin{equation}\label{eq:I1-norms-TIp}
\begin{aligned}
\|I_1\|_{L^{1}(\widehat\Gamma_1^{(r)})}&\lesssim t^{\frac23}\!\int_0^{\varrho}\!u^{\frac12}(1+|\log u|)e^{-ctu^{3/2}}\mathrm{d}u\\
&\le t^{-\frac13}\!\int_0^{\infty}\!x^{\frac12}\big(1+\tfrac23\log t+|\log x|\big)e^{-cx^{3/2}}\mathrm{d}x=\mathcal{O}\big(t^{-\frac13}\log t\big),\\
\|I_1\|^{2}_{L^{2}(\widehat\Gamma_1^{(r)})}&\lesssim t^{\frac23}\!\int_0^{\varrho}\!u(1+|\log u|)^{2}e^{-2ctu^{3/2}}\mathrm{d}u=\mathcal{O}\big(t^{-\frac23}\log^{2}t\big),\\
\|I_1\|_{L^{\infty}(\widehat\Gamma_1^{(r)})}&\lesssim\sup_{u>0}u^{\frac12}(1+|\log u|)e^{-ctu^{3/2}}=\mathcal{O}\big(t^{-\frac13}\log t\big),
\end{aligned}
\end{equation}
where the terms with $u$ instead of $u^{1/2}(1+|\log u|)$ give $\mathcal{O}(t^{-2/3})$.

Since $S$ is holomorphic in $D^{(r)}$ with $S(c_l)=s$, we have $|S(k)-s|\lesssim u$, and therefore, by \eqref{eq:k-zeta-TIp},
\begin{equation*}
\big|e^{2(S(k)-s)\zeta_r^{1/2}}-1\big|\le2|S-s||\zeta_r|^{\frac12}e^{2|S-s||\zeta_r|^{1/2}}\le C\,t^{-\frac23}|\zeta_r|^{\frac32}e^{C t^{-2/3}|\zeta_r|^{3/2}} .
\end{equation*}
Using $|w^{2}r_a^{*}|\le1+|w^{2}r_a^{*}-1|$, the bound for $|w^{2}r_a^{*}-1|$ above and \eqref{eq:phase-absorb-TIp},
\begin{equation}\label{eq:I2-TIp}
I_2\lesssim t^{-\frac23}|\zeta_r|^{\frac32}e^{-c|\zeta_r|^{3/2}},\qquad
\|I_2\|_{L^{p}(\widehat\Gamma_1^{(r)})}=\mathcal{O}\big(t^{-\frac23}\big),\quad p=1,2,\infty .
\end{equation}
By \eqref{eq:X12-split-TIp}, \eqref{eq:I1-pointwise-TIp} and \eqref{eq:I2-TIp},
\begin{equation}\label{eq:X-Gamma1-TIp}
|X-I|\lesssim t^{-\frac13}\log t\;e^{-c|\zeta_r|^{3/2}},\qquad \zeta_r\in\widehat\Gamma_1^{(r)} .
\end{equation}

We now conjugate by $\widehat N^{(r)}$. By item (d) of RH problem~\ref{rhp:Nhat-TIp}, \eqref{eq:Nhat-infty-TIp} and $\det\widehat N^{(r)}=-1$, so that $\widehat N^{(r)\,-1}=-\operatorname{adj}\widehat N^{(r)}$,
\begin{equation}\label{eq:Nhat-growth-TIp}
\big|\widehat N^{(r)}_\pm(\zeta_r)\big|+\big|\widehat N^{(r)}_\pm(\zeta_r)^{-1}\big|\lesssim(1+|\zeta_r|)^{\frac14},
\qquad
\big|V_E^{(r)}-I\big|\le\big|\widehat N^{(r)}_-\big|\big|\widehat N^{(r)\,-1}_-\big||X-I|\lesssim(1+|\zeta_r|)^{\frac12}|X-I|,
\end{equation}
uniformly for $s\in(0,\kappa C]$. Thus, by \eqref{eq:X-Gamma1-TIp}, $|V_E^{(r)}-I|\lesssim t^{-1/3}\log t\,e^{-c|\zeta_r|^{3/2}}$ on $\widehat\Gamma_1^{(r)}$.

On $\widehat\Gamma_1^{(r)*}$, by $\overline{w(\bar k)}=w(k)^{-1}$, $r_a^{*}(k)=\overline{r_a(\bar k)}$, $\overline{\hat\theta(\bar\zeta_r)}=\hat\theta(\zeta_r)$ and $\overline{S(\bar k)}=S(k)$,
\begin{equation*}
X_{21}(\zeta_r)=-\overline{X_{12}(\bar\zeta_r)},\qquad \zeta_r\in\widehat\Gamma_1^{(r)*},
\end{equation*}
so that the estimates \eqref{eq:X12-split-TIp}--\eqref{eq:Nhat-growth-TIp} hold verbatim.

Consider next the segment $(\zeta_r(c_l+\varrho),\zeta_r(\eta_l))$. Here $\widehat V_N^{(r)}=I$, $|w|=|e^{2\hat\theta_{S,\pm}}|=1$, and by item (e) of Proposition~\ref{prop:ra-TIp},
\begin{equation*}
X-I=\begin{pmatrix}0&w^{2}r_r^{*}e^{2\hat\theta_{S,+}}\\-w^{-2}r_re^{2\hat\theta_{S,-}}&-r_rr_r^{*}\end{pmatrix},\qquad |X-I|\lesssim|r_r|=\mathcal{O}(t^{-1}).
\end{equation*}
The segment has length at most $C|\zeta_r'(c_l)|\varrho\le C\,t^{\epsilon}$, so \eqref{eq:Nhat-growth-TIp} only gives
\begin{equation*}
\|V_E^{(r)}-I\|_{L^{\infty}}\lesssim t^{-1+\frac\epsilon2},\qquad \|V_E^{(r)}-I\|_{L^{2}}\lesssim t^{-1+\epsilon},\qquad \|V_E^{(r)}-I\|_{L^{1}}\lesssim t^{-1+\frac{3\epsilon}2},
\end{equation*}
and $t^{-1+3\epsilon/2}/(t^{-1/3}\log t)\to+\infty$ since $\epsilon>\frac12>\frac49$. Therefore this segment is not part of $\widehat\Gamma_E^{(r)}$ in \eqref{eq:GammaE-TIp}. The same applies to $\widehat\Gamma^{(r)}\setminus\zeta_r(D_{\varrho/2}(c_l))$, where $|X-I|=\mathcal{O}(e^{-ct^{3\epsilon/2}})$. Both parts are treated in the $k$-plane in Section~\ref{subsec:TIp-err}; see item (c) of Lemma~\ref{lem:match-TIp}.

On $(\zeta_r(\eta_l),0)$ we consider
\begin{equation*}
X=L(a)\,U\big(b-e^{2\hat\theta_+}\big)\,L\big(e^{2\hat\theta_-}\big)
=L\big(a+e^{2\hat\theta_-}\big)+\big(b-e^{2\hat\theta_+}\big)\begin{pmatrix}e^{2\hat\theta_-}&1\\ a\,e^{2\hat\theta_-}&a\end{pmatrix}.
\end{equation*}
Let $A:=w^{2}r^{*}e^{2(\hat\theta_{S,+}-\hat\theta_+)}$ and $B:=w^{-2}r\,e^{2(\hat\theta_{S,-}-\hat\theta_-)}$. Since $\hat\theta_++\hat\theta_-=\hat\theta_{S,+}+\hat\theta_{S,-}=0$ on $\mathbb{R}^{-}$, we have $AB=rr^{*}$ and
\begin{equation}\label{eq:X-seg-TIp}
X-I=\begin{pmatrix}A-1&(A-1)e^{2\hat\theta_+}\\(1-rr^{*})e^{2\hat\theta_-}&(B-1)+(1-rr^{*})\end{pmatrix}.
\end{equation}
For $k\in(c_l,\eta_l)$ we have $u\le\eta_l-c_l\lesssim t^{-2/3}$, so \eqref{eq:wr-real-TIp}, $|\hat\theta_{S,\pm}-\hat\theta_\pm|\lesssim t^{-2/3}$ and $|e^{2\hat\theta_\pm}|=1$ give $|X-I|\lesssim t^{-1/3}\log t$. Since $|\zeta_r|\le|\zeta_r(\eta_l)|\lesssim1$, \eqref{eq:Nhat-growth-TIp} yields $|V_E^{(r)}-I|\lesssim t^{-1/3}\log t$ there.

On $(0,\zeta_r(c_l-\frac\varrho2))$ we have $X-I=\big(e^{-2\hat\theta}-a_b\big)E_{12}$ with $a_b:=|w_+|^{2}e^{-2\hat\theta_S}$, and
\begin{equation*}
\big|e^{-2\hat\theta}-a_b\big|\le\big|1-|w_+|^{2}\big|e^{-2\hat\theta_S}+\big|e^{-2\hat\theta}-e^{-2\hat\theta_S}\big|,\qquad
\hat\theta,\ \hat\theta_S\ge\tfrac23\zeta_r^{3/2}\ \ (s,S>0).
\end{equation*}
By $|1-|w_+|^{2}|\lesssim u^{1/2}(1+|\log u|)$ and $|\hat\theta-\hat\theta_S|\lesssim t^{-2/3}\zeta_r^{3/2}$, the same arguments as for $I_1$ and $I_2$ give
\begin{equation*}
|X-I|\lesssim t^{-\frac13}\log t\;e^{-c\zeta_r^{3/2}},\qquad \big|V_E^{(r)}-I\big|\lesssim t^{-\frac13}\log t\;e^{-c\zeta_r^{3/2}} .
\end{equation*}

The above estimates prove \eqref{eq:VE-est-TIp}, and the weighted $L^{p}$ bounds in (a) follow by integration.

On the other hand,
\begin{equation*}
\mu:=I+(I-C_{V_E})^{-1}C_{V_E}I,\qquad \|\mu-I\|_{L^{2}}\lesssim\|V_E^{(r)}-I\|_{L^{2}}=\mathcal{O}\big(t^{-\frac13}\log t\big),
\end{equation*}
and
\begin{equation*}
E^{(r)}(\zeta_r)=I+\frac{1}{2\pi i}\int_{\widehat\Gamma_E^{(r)}}\frac{\mu(\zeta)\big(V_E^{(r)}(\zeta)-I\big)}{\zeta-\zeta_r}\,\mathrm{d}\zeta,\qquad E^{(r)}_-=\mu,
\end{equation*}
solves RH problem~\ref{rhp:E-TIp}. Since $\det V_E^{(r)}=1$, $\det E^{(r)}\equiv1$, and the quotient of two solutions has no jump and tends to $I$; this gives uniqueness and \eqref{eq:E-L2-TIp}. For (c),
\begin{equation*}
|E_1^{(r)}|\le\|V_E^{(r)}-I\|_{L^{1}}+\|\mu-I\|_{L^{2}}\|V_E^{(r)}-I\|_{L^{2}}=\mathcal{O}\big(t^{-\frac13}\log t\big),
\end{equation*}
\begin{equation*}
E^{(r)}(\zeta_r)-I-\frac{E_1^{(r)}}{\zeta_r}=\frac{1}{2\pi i\,\zeta_r}\int_{\widehat\Gamma_E^{(r)}}\frac{\zeta\,\mu(\zeta)\big(V_E^{(r)}(\zeta)-I\big)}{\zeta-\zeta_r}\,\mathrm{d}\zeta
=\mathcal{O}\big(t^{-\frac13}\log t\,|\zeta_r|^{-2}\big),
\end{equation*}
provided $\operatorname{dist}(\zeta_r,\widehat\Gamma_E^{(r)})\ge\frac13|\zeta_r|$.
Finally, $\widehat\Gamma_E^{(r)}\subset\{|\zeta_r|\le(\frac12+o(1))|\zeta_r'(c_l)|\varrho\}$ and $\partial\widehat D^{(r)}\subset\{|\zeta_r|=(1+o(1))|\zeta_r'(c_l)|\varrho\}$, where $|\zeta_r'(c_l)|\varrho=(\frac32)^{2/3}(g^{(2)}_{\mathrm{I},+})^{2/3}t^{\epsilon}$. Hence
\begin{equation*}
\big\|E^{(r)}-I\big\|_{L^{\infty}(\partial\widehat D^{(r)})}=\mathcal{O}\big(t^{-\frac13-\epsilon}\log t\big).
\end{equation*}
\end{proof}

We now construct the local parametrix near $c_l$. Define
\begin{equation}\label{eq:Mtilde-r-TIp}
\widetilde M^{(r)}(k)=\widetilde M^{(r)}(k;x,t):=P^{(r)}(k)\,E^{(r)}\big(\zeta_r(k)\big)\,\widehat N^{(r)}\big(\zeta_r(k)\big),\qquad k\in D^{(r)},
\end{equation}
where $E^{(r)}$ is the solution of RH problem~\ref{rhp:E-TIp}, $\widehat N^{(r)}$ is the solution of RH problem~\ref{rhp:Nhat-TIp}, and
\begin{equation}\label{eq:P-r-TIp}
P^{(r)}(k)=P^{(r)}(k;x,t):=M^{(\infty)}(k)\,Q^{(r)}\big(\zeta_r(k)\big)^{-1}\,\frac{1}{\sqrt2}\begin{pmatrix}1&-i\\-i&1\end{pmatrix}\zeta_r(k)^{\frac{\sigma_3}{4}},\qquad k\in D^{(r)}\setminus\mathbb{R}.
\end{equation}
Here $M^{(\infty)}$ is given in \eqref{eq:Minf-TIp}, $Q^{(r)}$ in \eqref{eq:Nhat-infty-TIp}, and $\frac{1}{\sqrt2}\left(\begin{smallmatrix}1&-i\\-i&1\end{smallmatrix}\right)=\frac{1}{\sqrt2}(I-i\sigma_1)$ is the inverse of $U:=\frac{1}{\sqrt2}\left(\begin{smallmatrix}1&i\\ i&1\end{smallmatrix}\right)$. The factor $\zeta_r^{\sigma_3/4}$ is taken with the principal branch. By \eqref{eq:Shat-TIp} and \eqref{eq:Shat-P34-TIp}, we can also write
\begin{equation}\label{eq:Mtilde-P34-TIp}
\begin{aligned}
\widetilde M^{(r)}(k)&=P^{(r)}(k)\,E^{(r)}(\zeta_r)\begin{pmatrix}1&0\\ \frac{is^{2}}{4}&1\end{pmatrix}M^{P34}(\zeta_r;0,1,s)\\
&\qquad\times e^{\hat\theta(\zeta_r)\sigma_3}\,Q^{(r)}(\zeta_r)\,G^{(r)}(\zeta_r)^{-1},\qquad \zeta_r=\zeta_r(k).
\end{aligned}
\end{equation}

\begin{proposition}\label{prop:P-r-TIp}
The following statements hold.
\begin{itemize}
\item[\rm(a)] Let
\begin{equation}\label{eq:F-r-TIp}
F(k):=e^{-\frac{\pi i}{4}}\Big(\frac32\Big)^{\frac16}t^{\frac16}H_{\mathrm{I}}(k)^{\frac16}(k+c_l)^{\frac14},\qquad k\in D^{(r)},
\end{equation}
with principal branches, where $H_{\mathrm{I}}$ is given in \eqref{eq:gI-H}. Then $F$ is holomorphic and does not vanish in $D^{(r)}$, and
\begin{equation}\label{eq:P-closed-TIp}
P^{(r)}(k)=U\begin{pmatrix}F(k)&0\\0&-F(k)^{-1}\end{pmatrix},\qquad k\in D^{(r)}\setminus\mathbb{R}.
\end{equation}
Consequently, $P^{(r)}$ extends to a holomorphic function in $D^{(r)}$, and $\det P^{(r)}\equiv-1$.
\item[\rm(b)] As $k\to c_l$, $P^{(r)}(k)=P^{(r)}(c_l)+\mathcal{O}(t^{1/6}|k-c_l|)$, where
\begin{equation}\label{eq:P-cl-TIp}
P^{(r)}(c_l)=\begin{pmatrix}\frac{1-i}{2}\,K&\frac{1-i}{2}\,K^{-1}\\[1mm]\frac{1+i}{2}\,K&-\frac{1+i}{2}\,K^{-1}\end{pmatrix},\qquad K:=(2c_l)^{\frac14}\Big(\frac32\Big)^{\frac16}t^{\frac16}\big(g^{(2)}_{\mathrm{I},+}\big)^{\frac16}.
\end{equation}
In particular, $|P^{(r)}(k)|\asymp|P^{(r)}(k)^{-1}|\asymp t^{1/6}$ uniformly in $D^{(r)}$.
\item[\rm(c)] For every $2\times2$ matrix $X$,
\begin{equation}\label{eq:P-conj-TIp}
P^{(r)}XP^{(r)\,-1}=U\begin{pmatrix}X_{11}&-F^{2}X_{12}\\-F^{-2}X_{21}&X_{22}\end{pmatrix}U^{-1}.
\end{equation}
Moreover,
\begin{equation}\label{eq:F2-zeta-TIp}
\frac{F(k)^{2}}{\zeta_r(k)}=\frac{i\kappa_t\,t^{-1/3}}{k-c_l}\big(1+\mathcal{O}(k-c_l)\big),\qquad \kappa_t:=\Big(\frac23\Big)^{\frac13}(2c_l)^{\frac12}\big(g^{(2)}_{\mathrm{I},+}\big)^{-\frac13}=\kappa\big(1+\mathcal{O}(t^{-2/3})\big),
\end{equation}
with $\kappa$ given in \eqref{eq:s-TIp}.
\end{itemize}
\end{proposition}

\begin{proof}
(a) Since $H_{\mathrm{I}}(k)=g^{(2)}_{\mathrm{I},+}(1+\mathcal{O}(k-c_l))$ with $g^{(2)}_{\mathrm{I},+}>0$, and $\operatorname{Re}(k+c_l)>0$ in $D^{(r)}$, the function $F$ is holomorphic and nonvanishing there. By \eqref{eq:zeta-r-TIp} and the argument used for \eqref{eq:zeta-r-half}, $\zeta_r(k)^{1/4}=e^{\mp\frac{\pi i}{4}}(\frac32)^{1/6}t^{1/6}H_{\mathrm{I}}(k)^{1/6}(k-c_l)^{1/4}$ for $k\in D^{(r)}\cap\mathbb{C}^{\pm}$. Since $\chi_l(k)=(k-c_l)^{1/4}(k+c_l)^{-1/4}$, it follows that
\begin{equation*}
\zeta_r(k)^{\frac14}\chi_l(k)^{-1}=F(k),\quad k\in\mathbb{C}^{+},\qquad
\zeta_r(k)^{\frac14}\chi_l(k)^{-1}=iF(k),\quad k\in\mathbb{C}^{-}.
\end{equation*}
By \eqref{eq:Minf-TIp}, $M^{(\infty)}=U\chi_l^{-\sigma_3}U^{-1}$. Moreover, $U^{-1}\sigma_3U^{-1}=\sigma_3$ and $U^{-1}\sigma_1U^{-1}=-iI$. Recall that $\zeta_r$ maps $\mathbb{C}^{\pm}\cap D^{(r)}$ into $\mathbb{C}^{\mp}$, so that $Q^{(r)}(\zeta_r(k))=\sigma_3$ for $k\in\mathbb{C}^{+}$ and $\sigma_1$ for $k\in\mathbb{C}^{-}$. Hence, for $k\in\mathbb{C}^{+}$,
\begin{equation*}
P^{(r)}=U\chi_l^{-\sigma_3}\,U^{-1}\sigma_3U^{-1}\,\zeta_r^{\frac{\sigma_3}{4}}=U\chi_l^{-\sigma_3}\sigma_3\zeta_r^{\frac{\sigma_3}{4}}=U\big(\zeta_r^{1/4}\chi_l^{-1}\big)^{\sigma_3}\sigma_3=U F^{\sigma_3}\sigma_3,
\end{equation*}
and for $k\in\mathbb{C}^{-}$,
\begin{equation*}
P^{(r)}=-iU\chi_l^{-\sigma_3}\zeta_r^{\frac{\sigma_3}{4}}=-iU(iF)^{\sigma_3}=U\begin{pmatrix}F&0\\0&-F^{-1}\end{pmatrix}=UF^{\sigma_3}\sigma_3 .
\end{equation*}
This proves \eqref{eq:P-closed-TIp}. Its right-hand side is holomorphic in $D^{(r)}$ and has determinant $\det U\cdot(-1)=-1$.

(b) This follows from (a), since $F(c_l)=e^{-\pi i/4}K$, $H_{\mathrm{I}}(c_l)=g^{(2)}_{\mathrm{I},+}$ and $F'=\mathcal{O}(t^{1/6})$ in $D^{(r)}$. For instance, $P^{(r)}_{11}(c_l)=\frac{1}{\sqrt2}F(c_l)=\frac{1-i}{2}K$ and $P^{(r)}_{12}(c_l)=-\frac{i}{\sqrt2}F(c_l)^{-1}=\frac{1-i}{2}K^{-1}$.

(c) The identity \eqref{eq:P-conj-TIp} follows from \eqref{eq:P-closed-TIp}. By \eqref{eq:zeta-r-TIp} and \eqref{eq:F-r-TIp},
\begin{equation*}
\begin{split}
\frac{F(k)^{2}}{\zeta_r(k)}&=\frac{-i(\frac32)^{1/3}t^{1/3}H_{\mathrm{I}}^{1/3}(k+c_l)^{1/2}}{-(\frac32)^{2/3}t^{2/3}(k-c_l)H_{\mathrm{I}}^{2/3}}\\
&=\frac{i\,t^{-1/3}}{k-c_l}\Big(\frac23\Big)^{\frac13}\frac{(k+c_l)^{1/2}}{H_{\mathrm{I}}(k)^{1/3}},
\end{split}
\end{equation*}
which gives \eqref{eq:F2-zeta-TIp}. Finally, $g^{(2)}_{\mathrm{I},+}=2(2c_l)^{1/2}(4\beta c_l+\alpha)+\mathcal{O}(t^{-2/3})$, and
\begin{equation*}
\Big(\frac23\Big)^{1/3}(2c_l)^{1/2}\big(2(2c_l)^{1/2}(4\beta c_l+\alpha)\big)^{-1/3}=\kappa .
\end{equation*}
\end{proof}

\begin{proposition}\label{prop:Mtilde-r-TIp}
For $t$ large, the function $\widetilde M^{(r)}$ defined in \eqref{eq:Mtilde-r-TIp} is holomorphic in $D^{(r)}\setminus\Gamma^{(r)}$ and satisfies $\widetilde M^{(r)}_+=\widetilde M^{(r)}_-\widetilde V^{(r)}$ on $\Gamma^{(r)}$, where
\begin{equation}\label{eq:Vtilde-r-TIp}
\widetilde V^{(r)}(k)=\begin{cases}
V^{(3)}(k), & k\in\big(\Gamma^{(r)}\cap D_{\varrho/2}(c_l)\big)\setminus(\eta_l,c_l+\varrho),\\[1mm]
I, & k\in(\eta_l,c_l+\varrho),\\[1mm]
\widehat V_N^{(r)}\big(\zeta_r(k)\big)^{-1}, & k\in\big(\Gamma^{(r)}\setminus D_{\varrho/2}(c_l)\big)\setminus(\eta_l,c_l+\varrho).
\end{cases}
\end{equation}
Here $V^{(3)}$ is given in \eqref{eq:V3-TIp}, and $\widehat V_N^{(r)}$ in \eqref{eq:VN-TIp}. Moreover,
\begin{equation}\label{eq:Vtilde-diff-TIp}
\big\|V^{(3)}-\widetilde V^{(r)}\big\|_{L^{p}(\Gamma^{(r)}\setminus(\eta_l,c_l+\varrho))}=\mathcal{O}\big(e^{-ct^{3\epsilon/2}}\big),\qquad
V^{(3)}-\widetilde V^{(r)}=V^{(3)}-I\ \text{ on }(\eta_l,c_l+\varrho),
\end{equation}
for $p\in[1,\infty]$, and the second difference is $\mathcal{O}(|r_r|)$ by \eqref{eq:V3-TIp} and Proposition~\ref{prop:ra-TIp}(e).
\end{proposition}

\begin{proof}
By Proposition~\ref{prop:P-r-TIp}(a), $P^{(r)}$ has no jumps in $D^{(r)}$, so the jumps of $\widetilde M^{(r)}$ are those of $(E^{(r)}\widehat N^{(r)})\circ\zeta_r$. The orientation of $\widehat\Gamma^{(r)}$ is opposite to the one induced by $\zeta_r$; see the discussion before \eqref{eq:Vhat3-TIp}. Therefore a jump $J(\zeta_r)$ of a function of $\zeta_r$ becomes the jump $J(\zeta_r(k))^{-1}$ in the $k$-plane.

On $\zeta_r^{-1}(\widehat\Gamma_E^{(r)})$, the function $E^{(r)}\widehat N^{(r)}$ has the jump $\widehat V^{(3)}$ by RH problem~\ref{rhp:E-TIp}, and $\widehat V^{(3)}(\zeta_r(k))^{-1}=V^{(3)}(k)$ by \eqref{eq:Vhat3-TIp}. This gives the first line of \eqref{eq:Vtilde-r-TIp}.

On $(\eta_l,c_l+\varrho)$, the functions $E^{(r)}$ and $\widehat N^{(r)}$ have no jumps. This gives the second line.

On the remaining parts, $E^{(r)}$ is holomorphic, and $\widehat N^{(r)}$ has the jump $\widehat V_N^{(r)}$. This gives the third line.

On the third set, $|\zeta_r|\ge(\frac12-o(1))|\zeta_r'(c_l)|\varrho\gtrsim t^{\epsilon}$. The estimates in the proof of Proposition~\ref{prop:E-TIp}(a) show that the entries of $\widehat V^{(3)}(\widehat V_N^{(r)})^{-1}-I$ are bounded by $C\,t^{-1/3}\log t\,e^{-c|\zeta_r|^{3/2}}$. Moreover, $\widehat V_N^{(r)}$ is bounded there. This yields the first estimate in \eqref{eq:Vtilde-diff-TIp}.
\end{proof}

We now collect the properties of $\widetilde M^{(r)}$ that are needed in the error analysis. Throughout, $\partial D^{(r)}$ is oriented counterclockwise. We write $U=\frac{1}{\sqrt2}\left(\begin{smallmatrix}1&i\\ i&1\end{smallmatrix}\right)$ and $N:=U\left(\begin{smallmatrix}0&1\\0&0\end{smallmatrix}\right)U^{-1}=\frac12\left(\begin{smallmatrix}-i&1\\1&i\end{smallmatrix}\right)$, so that $N^{2}=0$. Moreover,
\begin{equation}\label{eq:Mtilde-M2-TIp}
\widehat N^{(r)}(\zeta_r)Q^{(r)}(\zeta_r)^{-1}U^{-1}\zeta_r^{\frac{\sigma_3}{4}}=I+\frac{M_1^{(r)}(s)}{\zeta_r}+\frac{M_2^{(r)}(s)}{\zeta_r^{2}}+\mathcal{O}(\zeta_r^{-3}),\qquad \zeta_r\to\infty,
\end{equation}
where $M_1^{(r)}$ is given in \eqref{eq:Shat-infty-TIp}, and $M_2^{(r)}(s)$ is a matrix whose entries are polynomials in $s$. This follows as in the proof of \eqref{eq:Shat-infty-TIp}.

\begin{lemma}\label{lem:match-TIp}
Let $\xi$ satisfy \eqref{eq:TIp-half} and $\epsilon\in(\frac12,\frac23)$. As $t\to+\infty$, the following statements hold uniformly in $\xi$.
\begin{itemize}
\item[\rm(a)] On $\partial D^{(r)}$,
\begin{equation}\label{eq:match-a-TIp}
\widetilde M^{(r)}(k)M^{(\infty)}(k)^{-1}-I=P^{(r)}(k)\,\frac{M_1^{(r)}(s)}{\zeta_r(k)}\,P^{(r)}(k)^{-1}+\mathcal{O}\big(t^{-\epsilon}\log t\big).
\end{equation}
The first term on the right-hand side equals $\frac{\kappa_t s^{2}t^{-1/3}}{4(k-c_l)}N+\mathcal{O}(t^{-1/3})$. In particular,
\begin{equation*}
\big\|\widetilde M^{(r)}M^{(\infty)\,-1}-I\big\|_{L^\infty(\partial D^{(r)})}=\mathcal{O}\big(t^{\frac13-\epsilon}\big).
\end{equation*}
\item[\rm(b)] We have
\begin{equation}\label{eq:match-b-TIp}
\frac{1}{2\pi i}\oint_{\partial D^{(r)}}\big(\widetilde M^{(r)}(k)M^{(\infty)}(k)^{-1}-I\big)\,\mathrm{d}k=\frac{\kappa_t s^{2}}{4}\,t^{-\frac13}N+\mathcal{O}\big(t^{-\frac23}\log t\big).
\end{equation}
\item[\rm(c)] For $p\in[1,\infty]$, $\|V^{(3)}-\widetilde V^{(r)}\|_{L^{p}(\Gamma^{(r)}\setminus(\eta_l,c_l+\varrho))}=\mathcal{O}(e^{-ct^{3\epsilon/2}})$, and $\|V^{(3)}-\widetilde V^{(r)}\|_{L^{p}((\eta_l,c_l+\varrho))}=\mathcal{O}(t^{-1})$.
\item[\rm(d)] $\det\widetilde M^{(r)}\equiv1$, and
\begin{equation}\label{eq:match-d-TIp}
\big|\widetilde M^{(r)}_\pm(k)\big|^{2}=\big|\widetilde M^{(r)}_\pm(k)^{-1}\big|^{2}\lesssim t^{\frac13}\big(1+|\zeta_r(k)|\big)^{-\frac12},\qquad k\in(\eta_l,c_l+\varrho).
\end{equation}
Consequently, on $(\eta_l,c_l+\varrho)$, the matrix $\widetilde M^{(r)}_-(V^{(3)}-\widetilde V^{(r)})\widetilde M^{(r)\,-1}_-$ is $\mathcal{O}(t^{-2/3})$ in $L^\infty$, $\mathcal{O}(t^{-4/3+\epsilon/2})$ in $L^{1}$, and $\mathcal{O}(t^{-1}(\log t)^{1/2})$ in $L^{2}$.
\end{itemize}
\end{lemma}

\begin{proof}
This follows from Propositions 4.6 and 4.7.
\end{proof}

The point $-c_l$ is a hard edge of $g_{\mathrm{I}}$, since $g^{(1)}_{\mathrm{I},-}=(2c_l)^{1/2}(4\alpha c_l-(\xi-\xi_{\mathrm{I}}^{+}))\neq0$ by \eqref{eq:gI-coef}. The following lemma shows that, owing to $\delta_{\mathrm{I}}$ and to the extension of the lenses across $-c_l$, no local parametrix is needed there: near $-c_l$, the global parametrix $M^{(\infty)}$ is already an admissible approximation of $M^{(3)}$.

\begin{lemma}\label{lem:hard-edge-TIp}
Fix $\varrho_1\in\big(0,\frac{c_l-c_r}{2}\big)$ so small that $D_{\varrho_1}(-c_l)$ does not intersect $D^{(l)}$, $\Gamma_3^{(3)}$ and $\Gamma_3^{(3)*}$. Then the following statements hold.
\begin{itemize}
\item[\rm(a)] As $t\to+\infty$, uniformly for $k\in(-c_l-\varrho_1,-c_l)$,
\begin{equation*}
\big|M^{(\infty)}(k)\big(V^{(3)}(k)-I\big)M^{(\infty)}(k)^{-1}\big|\lesssim t^{-1}|k+c_l|.
\end{equation*}
\item[\rm(b)] As $t\to+\infty$, uniformly for $k\in(-c_l,-c_l+\varrho_1)$,
\begin{equation*}
\big|M^{(\infty)}_-(k)\big(V^{(3)}(k)V^{(\infty)}(k)^{-1}-I\big)M^{(\infty)}_-(k)^{-1}\big|\lesssim t^{-2}.
\end{equation*}
\item[\rm(c)] For each fixed $t$, $M^{(3)}(k)M^{(\infty)}(k)^{-1}=\mathcal{O}\big(1+\big|\log|k+c_l|\big|\big)$ as $k\to-c_l$.
\end{itemize}
\end{lemma}

\begin{proof}
By \eqref{eq:Minf-TIp}, $M^{(\infty)}$ and $M^{(\infty)}{}^{-1}$ are $\mathcal{O}(|k+c_l|^{-1/4})$ near $-c_l$, so conjugation by $M^{(\infty)}$ amplifies a matrix at most by a factor $\mathcal{O}(|k+c_l|^{-1/2})$.

(a) On $(-\hat\eta_l,-c_l)$, the entries of $V^{(3)}-I$ in \eqref{eq:V3-TIp} are $-w_+^{2}r_{2,r}e^{-2itg_{\mathrm{I}}}$, $w_-^{-2}r^{*}_{2,r}e^{2itg_{\mathrm{I}}}$ and their product. Here $|e^{\pm2itg_{\mathrm{I}}}|=1$, and $|w_+^{2}|=|w_-^{-2}|=1-|r|^{2}\lesssim|k+c_l|^{1/2}$ by \eqref{eq:1-r2-cl}. Hence $|V^{(3)}-I|\lesssim t^{-1}|k+c_l|^{3/2}$ by Proposition~\ref{prop:ra-TIp}(e), which gives (a).

(b) On $(-c_l,-c_m)$, \eqref{eq:V3-TIp} gives $V^{(3)}(V^{(\infty)})^{-1}-I=\left(\begin{smallmatrix}0&0\\-\varepsilon&0\end{smallmatrix}\right)$ with $\varepsilon=w_+^{2}e^{-2itg_{\mathrm{I},+}}B_a$. We have $|\varepsilon|\lesssim|k+c_l|^{3/2}e^{-c_1t|k+c_l|^{1/2}}$, thus the left-hand side of (b) is bounded by a constant times $|k+c_l|e^{-c_1t|k+c_l|^{1/2}}\le\frac{4}{e^{2}c_1^{2}t^{2}}$.

(c) Let $k\in D_{\varrho_1}(-c_l)\cap\mathbb{C}^{+}$, which is contained in $U_3^{(3)}$. Then
\begin{equation*}
M^{(3)}=e^{i\varphi_{\mathrm{I}}\sigma_3}M^{(2)}\delta_{\mathrm{I}}^{-\sigma_3}\begin{pmatrix}1&X\\0&1\end{pmatrix},\qquad X:=w^{2}r_{2,a}e^{-2itg_{\mathrm{I}}},
\end{equation*}
and $M^{(2)}=\mathcal{O}(1)$ near $-c_l$ by item (d) of RH problem~\ref{rhp:M2-TIp}. We use the following facts:
\begin{itemize}
\item $D_{\mathrm{I}}(-c_l)^{2}=e^{-i\arg b_{l,0}}$ by \eqref{eq:DI-cl};
\item $\delta_{\mathrm{I}}^{2}=2|b_{l,1}|(k+c_l)^{1/2}\big(1+o(1)\big)$ by \eqref{eq:deltaI-mcl};
\item $\varsigma\,r_{2,a}\to\rho(0)=b_{l,0}/(2|b_{l,1}|)$ by the construction in the proof of Proposition~\ref{prop:ra-TIp};
\item $g_{\mathrm{I}}(-c_l)=0$.
\end{itemize}
Together with $(k+c_l)^{1/2}/\varsigma(k)=i$ for $k\in\mathbb{C}^{+}$, these give
\begin{equation*}
X=i+\mathcal{O}\big(|k+c_l|^{\frac12}\big(\big|\log|k+c_l|\big|+t\big)\big),\qquad k\to-c_l .
\end{equation*}
On the other hand, $\chi_l(k)=e^{\frac{\pi i}{4}}(2c_l)^{1/4}(k+c_l)^{-1/4}(1+\mathcal{O}(k+c_l))$ in $\mathbb{C}^{+}$. Hence
\begin{equation*}
M^{(\infty)}(k)^{-1}=\frac{\chi_l(k)}{2}\begin{pmatrix}1&-i\\ i&1\end{pmatrix}+\mathcal{O}\big(|k+c_l|^{\frac14}\big).
\end{equation*}
The case $k\in\mathbb{C}^{-}$ follows in the same way.
\end{proof}

By Lemma~\ref{lem:hard-edge-TIp}, the function $M^{(3)}(M^{(\infty)})^{-1}$ has at most a logarithmic singularity at $-c_l$, and its jump in $D_{\varrho_1}(-c_l)$ differs from $I$ by $\mathcal{O}(t^{-1})$ in $L^{1}\cap L^{\infty}$. Therefore no disk around $-c_l$ is needed in the small norm analysis of Section~\ref{subsec:TIp-err}.

\subsubsection{Parabolic cylinder parametrix near $-\hat\eta_l$}\label{subsubsec:TIp-local-PC}

We now construct the local parametrix in the disk $D^{(l)}=D_{\varrho_0}(-\hat\eta_l)$ from \eqref{eq:disks-TIp}. As for $D^{(r)}$, we may and do assume that $\Gamma^{(l)}\setminus\mathbb{R}$ is the $\zeta_l$-preimage of the four rays
\begin{equation}\label{eq:SigmaPC-TIp}
\Sigma^{\mathrm{PC}}:=e^{\frac{\pi i}{4}}\mathbb{R}^{+}\cup e^{-\frac{\pi i}{4}}\mathbb{R}^{+}\cup e^{\frac{3\pi i}{4}}\mathbb{R}^{+}\cup e^{-\frac{3\pi i}{4}}\mathbb{R}^{+}.
\end{equation}

\emph{The function $w$ near $-\hat\eta_l$.} By item (a) of Proposition~\ref{prop:DI}, $D_{\mathrm{I}}$ is holomorphic in $D^{(l)}$, and $D_{\mathrm{I}}(-\hat\eta_l)=e^{-id_{\mathrm{I},\hat\eta}}$ by \eqref{eq:DI-eta}. By \eqref{eq:deltaI-PC}, $\delta_{\mathrm{I}}(k)=(-k-\hat\eta_l)^{-i\nu}\delta_b(k)$ with $\delta_b(-\hat\eta_l)=e^{i\phi_{\hat\eta}}$. Moreover, $-k-\hat\eta_l=-\zeta_l(k)\big((2t|g_{\mathrm{I}}''(-\hat\eta_l)|)^{1/2}\psi_l(k)\big)^{-1}$ by \eqref{eq:zeta-l-TIp}. Hence
\begin{equation}\label{eq:w-PC-TIp}
\begin{gathered}
w(k)=w_0\,\big(-\zeta_l(k)\big)^{-i\nu}\,\omega(k),\qquad
w_0:=e^{-id_{\mathrm{I},\hat\eta}+i\phi_{\hat\eta}}\big(2t|g_{\mathrm{I}}''(-\hat\eta_l)|\big)^{\frac{i\nu}{2}},\\
\omega(k):=\frac{D_{\mathrm{I}}(k)\,\delta_b(k)\,\psi_l(k)^{i\nu}}{e^{-id_{\mathrm{I},\hat\eta}+i\phi_{\hat\eta}}},
\end{gathered}
\end{equation}
with principal branches. In particular, $(-\zeta_l)^{-i\nu}$ has its cut along $\zeta_l\in\mathbb{R}^{+}$, i.e.\ along $(-\hat\eta_l,-\hat\eta_l+\varrho_0)$, as $\delta_{\mathrm{I}}$. Moreover, $|w_0|=1$, and $\omega(k)=1+\mathcal{O}(|k+\hat\eta_l||\log|k+\hat\eta_l||)$ uniformly in $D^{(l)}$. We set
\begin{equation}\label{eq:E0-TIp}
r_0:=r(-\hat\eta_l),\qquad 1-|r_0|^{2}=e^{-2\pi\nu},\qquad
E_0:=\big(w_0\,e^{-itg_{\mathrm{I}}(-\hat\eta_l)}\big)^{\sigma_3},
\end{equation}
where $\nu$ is given in \eqref{eq:nu-phiI} and $g_{\mathrm{I}}(-\hat\eta_l)\in\mathbb{R}$. Then $E_0^{\pm1}$ are bounded. We arrive at the following model problem.
\begin{rhp}\label{rhp:PC-TIp}
Find $N^{\mathrm{PC}}(\zeta)=N^{\mathrm{PC}}(\zeta;r_0)$ with the properties
\begin{itemize}
\item[\rm(a)] $N^{\mathrm{PC}}(\zeta)$ is holomorphic for $\zeta\in\mathbb{C}\setminus\Sigma^{\mathrm{PC}}$.
\item[\rm(b)] $N^{\mathrm{PC}}_+(\zeta)=N^{\mathrm{PC}}_-(\zeta)V^{\mathrm{PC}}(\zeta)$ on $\Sigma^{\mathrm{PC}}\setminus\{0\}$, where
\begin{equation}\label{eq:VPC-TIp}
V^{\mathrm{PC}}(\zeta)=\begin{cases}
\begin{pmatrix}1&0\\ r_0(-\zeta)^{2i\nu}e^{-\frac i2\zeta^{2}}&1\end{pmatrix}, & \arg\zeta=\frac{3\pi}{4},\\[4mm]
\begin{pmatrix}1&-\bar r_0(-\zeta)^{-2i\nu}e^{\frac i2\zeta^{2}}\\0&1\end{pmatrix}, & \arg\zeta=-\frac{3\pi}{4},\\[4mm]
\begin{pmatrix}1&-\frac{\bar r_0}{1-|r_0|^{2}}(-\zeta)^{-2i\nu}e^{\frac i2\zeta^{2}}\\0&1\end{pmatrix}, & \arg\zeta=\frac{\pi}{4},\\[4mm]
\begin{pmatrix}1&0\\ \frac{r_0}{1-|r_0|^{2}}(-\zeta)^{2i\nu}e^{-\frac i2\zeta^{2}}&1\end{pmatrix}, & \arg\zeta=-\frac{\pi}{4},
\end{cases}
\end{equation}
with the principal branch of $(-\zeta)^{i\nu}$.
\item[\rm(c)] $N^{\mathrm{PC}}(\zeta)=I+\mathcal{O}(\zeta^{-1})$ as $\zeta\to\infty$.
\end{itemize}
\end{rhp}

\begin{proposition}\label{prop:PC-TIp}
RH problem~\ref{rhp:PC-TIp} has a unique solution, and $\det N^{\mathrm{PC}}\equiv1$. It is bounded on $\mathbb{C}$, uniformly for $\nu$ in compact subsets of $[0,\infty)$, and
\begin{equation}\label{eq:NPC-infty-TIp}
N^{\mathrm{PC}}(\zeta)=I+\frac{M_1^{\mathrm{PC}}}{\zeta}+\mathcal{O}(\zeta^{-2}),\qquad
M_1^{\mathrm{PC}}=\begin{pmatrix}0&i\beta_{12}\\-i\beta_{21}&0\end{pmatrix},
\end{equation}
where
\begin{equation}\label{eq:beta-PC-TIp}
\beta_{12}=\frac{\sqrt{2\pi}\,e^{\frac{3\pi i}{4}}e^{-\frac{\pi\nu}{2}}}{r_0\,\Gamma(i\nu)},\qquad \beta_{21}=\overline{\beta_{12}},\qquad |\beta_{12}|^{2}=\beta_{12}\beta_{21}=\nu .
\end{equation}
Explicitly, let $D_a(z)$ denote the parabolic cylinder function. For $0<\arg\zeta<\frac\pi4$, set
\begin{equation}\label{eq:PsiA-TIp}
\Psi(\zeta):=\begin{pmatrix}
e^{-\frac{3\pi\nu}{4}}D_{-i\nu}\big(e^{-\frac{\pi i}{4}}\zeta\big) & \frac{i\nu}{\beta_{21}}e^{\frac{5\pi\nu}{4}}e^{\frac{\pi i}{4}}D_{i\nu-1}\big(e^{\frac{\pi i}{4}}\zeta\big)\\[2mm]
-\frac{i\nu}{\beta_{12}}e^{-\frac{3\pi\nu}{4}}e^{-\frac{\pi i}{4}}D_{-i\nu-1}\big(e^{-\frac{\pi i}{4}}\zeta\big) & e^{\frac{5\pi\nu}{4}}D_{i\nu}\big(e^{\frac{\pi i}{4}}\zeta\big)
\end{pmatrix}.
\end{equation}
Extend $\Psi$ to $\mathbb{C}\setminus(\Sigma^{\mathrm{PC}}\cup\mathbb{R})$ by the constant jumps $J_{3\pi/4}=\left(\begin{smallmatrix}1&0\\ r_0&1\end{smallmatrix}\right)$, $J_{-3\pi/4}=\left(\begin{smallmatrix}1&-\bar r_0\\0&1\end{smallmatrix}\right)$, $J_{\pi/4}=\left(\begin{smallmatrix}1&-\bar r_0e^{2\pi\nu}\\0&1\end{smallmatrix}\right)$ and $J_{-\pi/4}=\left(\begin{smallmatrix}1&0\\ r_0e^{2\pi\nu}&1\end{smallmatrix}\right)$ on the corresponding rays, with $\Psi_+=\Psi_-J$, and by continuity across $\mathbb{R}^{-}$. Then
\begin{equation*}
N^{\mathrm{PC}}(\zeta)=\Psi(\zeta)\,(-\zeta)^{i\nu\sigma_3}e^{-\frac i4\zeta^{2}\sigma_3}.
\end{equation*}
\end{proposition}

\begin{proof}
Let $\Psi:=N^{\mathrm{PC}}(-\zeta)^{-i\nu\sigma_3}e^{\frac i4\zeta^{2}\sigma_3}$. Conjugating \eqref{eq:VPC-TIp} by $(-\zeta)^{-i\nu\sigma_3}e^{\frac i4\zeta^{2}\sigma_3}$, we see that $\Psi$ has the constant jumps $J_{\pm\pi/4}$, $J_{\pm3\pi/4}$ on $\Sigma^{\mathrm{PC}}$. Moreover, $\Psi_+=\Psi_-e^{-2\pi\nu\sigma_3}$ on $\mathbb{R}^{+}$, and $\Psi$ has no jump on $\mathbb{R}^{-}$. Hence $\Psi'\Psi^{-1}$ is entire. By (c),
\begin{equation*}
\Psi'(\zeta)\Psi(\zeta)^{-1}=\frac{i\zeta}{2}\sigma_3+\begin{pmatrix}0&\beta_{12}\\ \beta_{21}&0\end{pmatrix},\qquad \beta_{12}:=-i\big(M_1^{\mathrm{PC}}\big)_{12},\quad \beta_{21}:=i\big(M_1^{\mathrm{PC}}\big)_{21}.
\end{equation*}
Eliminating, the entries of the first column of $\Psi$ satisfy Weber's equation in the variable $e^{-\pi i/4}\zeta$. The entries of the second column satisfy Weber's equation in the variable $e^{\pi i/4}\zeta$. The parameters are $-i\beta_{12}\beta_{21}$ and $i\beta_{12}\beta_{21}$, respectively. Comparing with $\Psi\sim(-\zeta)^{-i\nu\sigma_3}e^{\frac i4\zeta^{2}\sigma_3}$ gives $\beta_{12}\beta_{21}=\nu$.

In the sector $0<\arg\zeta<\frac\pi4$, the asymptotic expansion $D_a(z)\sim z^{a}e^{-z^{2}/4}$ for $|\arg z|<\frac{3\pi}{4}$ applies to both columns. Together with the identity $D_a'(z)+\frac z2D_a(z)=aD_{a-1}(z)$, this yields \eqref{eq:PsiA-TIp}. The constants $e^{-3\pi\nu/4}$ and $e^{5\pi\nu/4}$ come from $(-\zeta)^{\mp i\nu}=\zeta^{\mp i\nu}e^{\mp\pi\nu}$ in this sector.

Continuing $\Psi$ by the jumps, the second column in the sector $\frac{3\pi}{4}<\arg\zeta<\pi$ equals $\Psi_{\cdot2}-\bar r_0e^{2\pi\nu}\Psi_{\cdot1}$, with $\Psi$ from \eqref{eq:PsiA-TIp}. It has the required asymptotics if and only if its exponentially growing part vanishes. By the connection formula
\begin{equation*}
D_a(z)=e^{\pi ia}D_a(-z)+\frac{\sqrt{2\pi}}{\Gamma(-a)}\,e^{\frac{\pi i}{2}(a+1)}D_{-a-1}(-iz)
\end{equation*}
this condition is equivalent to \eqref{eq:beta-PC-TIp}. The relation $\beta_{21}=\overline{\beta_{12}}$ follows from $\beta_{12}\beta_{21}=\nu$ and $|\Gamma(i\nu)|^{2}=\pi/(\nu\sinh\pi\nu)$, $1-|r_0|^{2}=e^{-2\pi\nu}$. Equivalently, it follows from the symmetry $N^{\mathrm{PC}}(\zeta)=\sigma_1\overline{N^{\mathrm{PC}}(\bar\zeta)}\sigma_1$.

\end{proof}

We now define the local parametrix near $-\hat\eta_l$:
\begin{equation}\label{eq:Mtilde-l-TIp}
\widetilde M^{(l)}(k)=\widetilde M^{(l)}(k;x,t):=M^{(\infty)}(k)\,E_0\,N^{\mathrm{PC}}\big(\zeta_l(k);r_0\big)\,E_0^{-1},\qquad k\in D^{(l)}.
\end{equation}
Since $D^{(l)}\cap[-c_l,c_l]=\emptyset$, the prefactor $M^{(\infty)}E_0$ is holomorphic in $D^{(l)}$, and it is bounded together with its inverse, uniformly in $t$. Hence the jumps of $\widetilde M^{(l)}$ in $D^{(l)}$ are
\begin{equation}\label{eq:Vtilde-l-TIp}
\widetilde V^{(l)}(k)=E_0\,V^{\mathrm{PC}}\big(\zeta_l(k)\big)\,E_0^{-1},\qquad k\in\Gamma^{(l)}\setminus\mathbb{R},
\end{equation}
and $\widetilde M^{(l)}$ has no jump on $\Gamma^{(l)}\cap\mathbb{R}$.

\begin{lemma}\label{lem:match-l-TIp}
Let $\xi$ satisfy \eqref{eq:TIp-half}. As $t\to+\infty$, the following statements hold uniformly in $\xi$.
\begin{itemize}
\item[\rm(a)] $\widetilde M^{(l)}$ and $\widetilde M^{(l)\,-1}$ are bounded in $D^{(l)}$, uniformly in $t$, and $\det\widetilde M^{(l)}\equiv1$.
\item[\rm(b)] On $\partial D^{(l)}$,
\begin{equation}\label{eq:match-l-a-TIp}
\widetilde M^{(l)}(k)M^{(\infty)}(k)^{-1}-I=M^{(\infty)}(k)E_0\frac{M_1^{\mathrm{PC}}}{\zeta_l(k)}E_0^{-1}M^{(\infty)}(k)^{-1}+\mathcal{O}(t^{-1})=\mathcal{O}(t^{-\frac12}).
\end{equation}
\item[\rm(c)] With $\partial D^{(l)}$ oriented counterclockwise,
\begin{equation}\label{eq:match-l-b-TIp}
\frac{1}{2\pi i}\oint_{\partial D^{(l)}}\big(\widetilde M^{(l)}M^{(\infty)\,-1}-I\big)\,\mathrm{d}k=\frac{M^{(\infty)}(-\hat\eta_l)E_0M_1^{\mathrm{PC}}E_0^{-1}M^{(\infty)}(-\hat\eta_l)^{-1}}{\big(2t|g_{\mathrm{I}}''(-\hat\eta_l)|\big)^{1/2}}+\mathcal{O}(t^{-1}),
\end{equation}
where
\begin{equation*}
E_0M_1^{\mathrm{PC}}E_0^{-1}=\begin{pmatrix}0&i\beta_{12}w_0^{2}e^{-2itg_{\mathrm{I}}(-\hat\eta_l)}\\-i\beta_{21}w_0^{-2}e^{2itg_{\mathrm{I}}(-\hat\eta_l)}&0\end{pmatrix}.
\end{equation*}
\item[\rm(d)] On $\Gamma^{(l)}\setminus\mathbb{R}$,
\begin{equation*}
\|V^{(3)}-\widetilde V^{(l)}\|_{L^{\infty}}=\mathcal{O}\big(t^{-\frac12}\log t\big),\quad
\begin{gathered}
\|V^{(3)}-\widetilde V^{(l)}\|_{L^{1}}=\mathcal{O}\big(t^{-1}\log t\big),\\
\|V^{(3)}-\widetilde V^{(l)}\|_{L^{2}}=\mathcal{O}\big(t^{-\frac34}\log t\big).
\end{gathered}
\end{equation*}
On $\Gamma^{(l)}\cap\mathbb{R}$, $\widetilde V^{(l)}=I$ and $\|V^{(3)}-I\|_{L^{p}}=\mathcal{O}(t^{-1})$ for $p\in[1,\infty]$. By (a), the same bounds hold for $\widetilde M^{(l)}_-(V^{(3)}-\widetilde V^{(l)})\widetilde M^{(l)\,-1}_-$.
\end{itemize}
\end{lemma}

\begin{proof}
According to \eqref{eq:NPC-infty-TIp}, \eqref{eq:match-l-b-TIp}, Propositions 4.6 and 4.7, Lemma \ref{lem:match-l-TIp} can be proved.
\end{proof}

\subsection{Small norm RH problem for $M^{(\mathrm{err})}$}\label{subsec:TIp-err}

\begin{figure}[htbp]
\centering
\includegraphics[width=\textwidth]{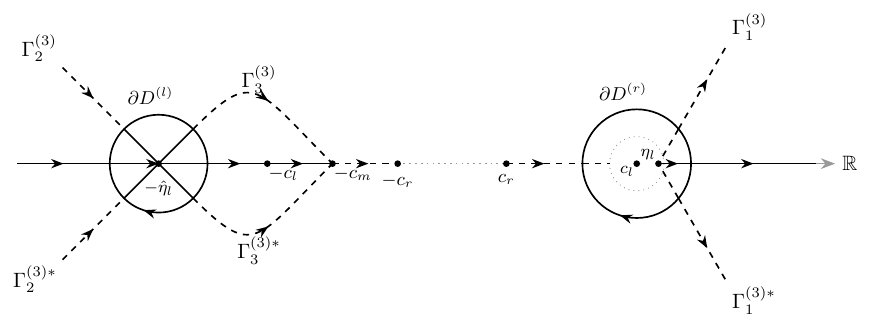}
\caption{The jump contour $\Gamma^{(\mathrm{err})}$.}
\label{fig:err-TIp}
\end{figure}

With the global parametrix $M^{(\infty)}$ from \eqref{eq:Minf-TIp} and the local parametrices $\widetilde M^{(r)}$ and $\widetilde M^{(l)}$ from \eqref{eq:Mtilde-r-TIp} and \eqref{eq:Mtilde-l-TIp}, we define
\begin{equation}\label{eq:Merr-TIp}
M^{(\mathrm{err})}(k)=M^{(\mathrm{err})}(k;x,t):=\begin{cases}
M^{(3)}(k)\,M^{(\infty)}(k)^{-1}, & k\in\mathbb{C}\setminus\big(\overline{D^{(r)}}\cup\overline{D^{(l)}}\big),\\[1mm]
M^{(3)}(k)\,\widetilde M^{(r)}(k)^{-1}, & k\in D^{(r)},\\[1mm]
M^{(3)}(k)\,\widetilde M^{(l)}(k)^{-1}, & k\in D^{(l)}.
\end{cases}
\end{equation}
Let $\Gamma^{(\mathrm{err})}:=\Gamma^{(3)}\cup\partial D^{(r)}\cup\partial D^{(l)}$, where the circles are oriented clockwise; see Figure~\ref{fig:err-TIp}. Then $M^{(\mathrm{err})}$ satisfies the following RH problem.

\begin{rhp}\label{rhp:Merr-TIp}
Find $M^{(\mathrm{err})}(k)=M^{(\mathrm{err})}(k;x,t)$ with the properties
\begin{itemize}
\item[\rm(a)] $M^{(\mathrm{err})}(k)$ is holomorphic for $k\in\mathbb{C}\setminus\Gamma^{(\mathrm{err})}$.
\item[\rm(b)] $M^{(\mathrm{err})}_+(k)=M^{(\mathrm{err})}_-(k)V^{(\mathrm{err})}(k)$ for $k\in\Gamma^{(\mathrm{err})}$, where
\begin{equation}\label{eq:Verr-TIp}
V^{(\mathrm{err})}(k)=\begin{cases}
M^{(\infty)}_-(k)\,V^{(3)}(k)\,M^{(\infty)}_+(k)^{-1}, & k\in\Gamma^{(3)}\setminus\big(\overline{D^{(r)}}\cup\overline{D^{(l)}}\big),\\[1mm]
\widetilde M^{(j)}(k)\,M^{(\infty)}(k)^{-1}, & k\in\partial D^{(j)},\ j\in\{r,l\},\\[1mm]
\widetilde M^{(j)}_-(k)\,V^{(3)}(k)\,\widetilde M^{(j)}_+(k)^{-1}, & k\in\Gamma^{(3)}\cap D^{(j)},\ j\in\{r,l\}.
\end{cases}
\end{equation}
\item[\rm(c)] $M^{(\mathrm{err})}(k)=I+\mathcal{O}(k^{-1})$ as $k\to\infty$.
\item[\rm(d)] $M^{(\mathrm{err})}(k)=\mathcal{O}\big(1+|\log|k+c_l||\big)$ as $k\to-c_l$, and $M^{(\mathrm{err})}(k)=\mathcal{O}\big(|k\mp c_r|^{-|\frac14+\nu_r|}\big)$ as $k\to\pm c_r$. Near all other points of self-intersection of $\Gamma^{(\mathrm{err})}$, $M^{(\mathrm{err})}$ is bounded.
\end{itemize}
\end{rhp}

\begin{proposition}\label{prop:Verr-TIp}
Let $\xi$ satisfy \eqref{eq:TIp-half}, $\epsilon\in(\frac12,\frac23)$ and $\nu_r\le0$. There is a constant $c>0$ such that, as $t\to+\infty$, uniformly in $\xi$, the following estimates hold for $p\in[1,\infty]$.
\begin{itemize}
\item[\rm(i)] $\|V^{(\mathrm{err})}-I\|_{L^{p}(\partial D^{(r)})}=\mathcal{O}\big(t^{(\frac13-\epsilon)+(\epsilon-\frac23)/p}\big)$. In particular, the $L^{\infty}$, $L^{1}$ and $L^{2}$ norms are $\mathcal{O}(t^{1/3-\epsilon})$, $\mathcal{O}(t^{-1/3})$ and $\mathcal{O}(t^{-\epsilon/2})$.
\item[\rm(ii)] $\|V^{(\mathrm{err})}-I\|_{L^{p}(\partial D^{(l)})}=\mathcal{O}(t^{-1/2})$.
\item[\rm(iii)] $\|V^{(\mathrm{err})}-I\|_{L^{p}(\Gamma^{(3)}\cap D^{(l)})}=\mathcal{O}\big(t^{-\frac12-\frac{1}{2p}}\log t\big)$.
\item[\rm(iv)] On $(\eta_l,c_l+\varrho)$, the $L^{\infty}$, $L^{1}$ and $L^{2}$ norms of $V^{(\mathrm{err})}-I$ are $\mathcal{O}(t^{-2/3})$, $\mathcal{O}(t^{-4/3+\epsilon/2})$ and $\mathcal{O}(t^{-1}(\log t)^{1/2})$.
\item[\rm(v)] On $\Sigma_{\mathbb{R}}:=\mathbb{R}\setminus\big([-c_l,c_l]\cup D^{(r)}\cup D^{(l)}\big)$, the $L^{\infty}$, $L^{1}$ and $L^{2}$ norms of $V^{(\mathrm{err})}-I$ are $\mathcal{O}(t^{-2/3-\epsilon/2})$, $\mathcal{O}(t^{-1})$ and $\mathcal{O}(t^{-1}(\log t)^{1/2})$.
\item[\rm(vi)] $\|V^{(\mathrm{err})}-I\|_{L^{p}((-c_l,-c_m))}=\mathcal{O}(t^{-2})$.
\item[\rm(vii)] On the remaining part $\Sigma_{\exp}$ of $\Gamma^{(\mathrm{err})}$, $\|(1+|k|)(V^{(\mathrm{err})}-I)\|_{L^{p}(\Sigma_{\exp})}=\mathcal{O}(e^{-ct^{3\epsilon/2}})$. Here $\Sigma_{\exp}$ consists of the lens boundaries outside $D^{(r)}\cup D^{(l)}$, of the band $(-c_m,-c_r)\cup(c_r,c_l-\varrho)$, and of $\Gamma^{(3)}\cap D^{(r)}\setminus\big(D_{\varrho/2}(c_l)\cup(\eta_l,c_l+\varrho)\big)$.
\end{itemize}
Consequently,
\begin{equation}\label{eq:Verr-total-TIp}
\big\|V^{(\mathrm{err})}-I\big\|_{L^{p}(\Gamma^{(\mathrm{err})})}=\mathcal{O}(t^{\gamma_p}),\qquad \gamma_1=-\frac13,\quad \gamma_2=-\frac\epsilon2,\quad \gamma_\infty=\frac13-\epsilon,
\end{equation}
and $\|V^{(\mathrm{err})}-I\|_{L^{1}(\Gamma^{(\mathrm{err})}\setminus(\partial D^{(r)}\cup\partial D^{(l)}))}=\mathcal{O}(t^{-1}\log t)$.
\end{proposition}

\begin{proof}
Items (i) and (ii) follow from Lemmas~\ref{lem:match-TIp}(a) and \ref{lem:match-l-TIp}(b). Items (iii) and (iv) are Lemma~\ref{lem:match-l-TIp}(d) and Lemma~\ref{lem:match-TIp}(d), respectively.

(v) On $\mathbb{R}\setminus[-\hat\eta_l,c_l]$ we have $|w|=1$ by item (c) of Propositions~\ref{prop:DI} and \ref{prop:deltaI}, and $g_{\mathrm{I}}$ is real. Hence $|V^{(3)}-I|\lesssim|r_r|$ there by \eqref{eq:V3-TIp}. By \eqref{eq:Minf-TIp}, $|M^{(\infty)}|^{2}\lesssim1+|k-c_l|^{-1/2}$ near $c_l$, and $M^{(\infty)}$ is bounded on the rest of this set. Hence $|V^{(\mathrm{err})}-I|\lesssim(1+|k-c_l|^{-1/2})|r_r(k)|$. Together with $\|r_r\|_{L^{p}}=\mathcal{O}(t^{-1})$ (Proposition~\ref{prop:ra-TIp}(e)) and $|k-c_l|\ge\varrho=t^{-2/3+\epsilon}$, this gives the bounds on $\mathbb{R}\setminus[-\hat\eta_l,c_l]$. On $(-\hat\eta_l+\varrho_0,-c_l)$, Lemma~\ref{lem:hard-edge-TIp}(a) gives $|V^{(\mathrm{err})}-I|\lesssim t^{-1}$.

(vi) This follows from Lemma~\ref{lem:hard-edge-TIp}(b).

(vii) This follows from \eqref{eq:V3-TIp} and item (b) of Proposition~\ref{prop:ra-TIp}.
\end{proof}

By Beals-Coifman theory \cite{BealsCoifman84}, RH problem~\ref{rhp:Merr-TIp} has a unique solution. It is given by
\begin{equation}\label{eq:Merr-int-TIp}
M^{(\mathrm{err})}(k)=I+\frac{1}{2\pi i}\int_{\Gamma^{(\mathrm{err})}}\frac{\mu(s)\big(V^{(\mathrm{err})}(s)-I\big)}{s-k}\,\mathrm{d}s,
\end{equation}
where $\mu=M^{(\mathrm{err})}_-$ satisfies $\mu-I\in L^{2}(\Gamma^{(\mathrm{err})})$ and
\begin{equation*}
\|\mu-I\|_{L^{2}(\Gamma^{(\mathrm{err})})}=\mathcal{O}(t^{-\epsilon/2}).
\end{equation*}
As $k\to\infty$ nontangentially to $\Gamma^{(\mathrm{err})}$,
\begin{equation*}
M^{(\mathrm{err})}(k)=I+\frac{M^{(\mathrm{err})}_1}{k}+o(k^{-1}),\qquad
M^{(\mathrm{err})}_1=-\frac{1}{2\pi i}\int_{\Gamma^{(\mathrm{err})}}\mu(s)\big(V^{(\mathrm{err})}(s)-I\big)\,\mathrm{d}s .
\end{equation*}

\begin{proposition}\label{prop:smallnorm-TIp}
Let $\xi$ satisfy \eqref{eq:TIp-half}, $\epsilon\in(\frac12,\frac23)$ and $\nu_r\le0$. Then the following statements hold for $t$ large enough,
\begin{equation}\label{eq:Merr1-TIp}
M^{(\mathrm{err})}_1=\sum_{j\in\{r,l\}}\frac{1}{2\pi i}\oint_{\partial D^{(j)}}\big(\widetilde M^{(j)}(k)M^{(\infty)}(k)^{-1}-I\big)\,\mathrm{d}k+\mathcal{O}(t^{-\epsilon}).
\end{equation}
Consequently, with $N=\frac12\left(\begin{smallmatrix}-i&1\\1&i\end{smallmatrix}\right)$ and $\kappa_t$ from \eqref{eq:F2-zeta-TIp},
\begin{equation}\label{eq:Merr1-explicit-TIp}
M^{(\mathrm{err})}_1=\frac{\kappa_t s^{2}}{4}\,t^{-\frac13}N+\frac{M^{(\infty)}(-\hat\eta_l)\,E_0M_1^{\mathrm{PC}}E_0^{-1}\,M^{(\infty)}(-\hat\eta_l)^{-1}}{\big(2t|g_{\mathrm{I}}''(-\hat\eta_l)|\big)^{1/2}}+\mathcal{O}(t^{-\epsilon}),
\end{equation}
where $E_0$ and $M_1^{\mathrm{PC}}$ are given in \eqref{eq:E0-TIp} and \eqref{eq:NPC-infty-TIp}.
\end{proposition}

\begin{proof}
We split $M^{(\mathrm{err})}_1=-\frac{1}{2\pi i}\int(V^{(\mathrm{err})}-I)-\frac{1}{2\pi i}\int(\mu-I)(V^{(\mathrm{err})}-I)$. The second integral is bounded by $\|\mu-I\|_{L^{2}}\|V^{(\mathrm{err})}-I\|_{L^{2}}=\mathcal{O}(t^{-\epsilon})$. In the first integral, the contribution of $\Gamma^{(\mathrm{err})}\setminus(\partial D^{(r)}\cup\partial D^{(l)})$ is $\mathcal{O}(t^{-1}\log t)$ by Proposition~\ref{prop:Verr-TIp}. On the circles, the clockwise orientation gives $-\frac{1}{2\pi i}\int_{\partial D^{(j)}}=\frac{1}{2\pi i}\oint_{\partial D^{(j)}}$ with the counterclockwise orientation, and $V^{(\mathrm{err})}-I=\widetilde M^{(j)}M^{(\infty)\,-1}-I$. This proves \eqref{eq:Merr1-TIp}, since $t^{-1}\log t=o(t^{-\epsilon})$. Finally, \eqref{eq:Merr1-explicit-TIp} follows from Lemma~\ref{lem:match-TIp}(b) and Lemma~\ref{lem:match-l-TIp}(c). Their error terms $\mathcal{O}(t^{-2/3}\log t)$ and $\mathcal{O}(t^{-1})$ are $o(t^{-\epsilon})$ because $\epsilon<\frac23$.
\end{proof}

\begin{remark}\label{rem:nur-TIp}
If $\nu_r\in(0,\frac12)$, then $a$ in the proof of Proposition~\ref{prop:Verr-TIp}(vii) satisfies $|a|\asymp e^{-ct}|k\mp c_r|^{-2\nu_r}$ near $\pm c_r$. This is unbounded, although exponentially small in $L^{1}$. In this case we modify $M^{(\mathrm{err})}$ in two small fixed disks $D_\delta(\pm c_r)$; we describe the construction at $c_r$.

Let $\chi\in C_c^\infty(c_r-\frac\delta2,c_r+\frac\delta2)$ with $\chi=1$ near $c_r$. Define
\begin{equation*}
\Phi(k):=\frac{1}{2\pi i}\int_{c_r}^{c_r+\delta}\frac{\chi(s)a(s)}{s-k}\,\mathrm{d}s,\qquad X(k):=M^{(\infty)}_{\uparrow}(k)E_{12}M^{(\infty)}_{\uparrow}(k)^{-1},
\end{equation*}
where $M^{(\infty)}_{\uparrow}$ is the analytic continuation of $M^{(\infty)}|_{\mathbb{C}^{+}}$ to $D_\delta(c_r)$. Then $X$ is holomorphic and $X^{2}=0$. Using $V^{(3)}=V^{(\infty)}(I+aE_{12})$ on the band, one checks that $K:=(I+\Phi X)M^{(\infty)}$ satisfies $K_+=K_-V^{(\infty)}(I+\chi aE_{12})$ on $(c_r-\delta,c_r+\delta)$.

We replace $M^{(\infty)}$ by $K$ in \eqref{eq:Merr-TIp} in $D_\delta(c_r)$. The new jumps are $I+\Phi X$ on $\partial D_\delta(c_r)$ and $I-(1-\chi)a\,K_-E_{21}K_-^{-1}$ on $(c_r,c_r+\delta)$; both are bounded and $\mathcal{O}(e^{-ct})$. Moreover, $M^{(3)}K^{-1}=M^{(3)}M^{(\infty)\,-1}-\Phi\,M^{(3)}E_{12}M^{(\infty)\,-1}$ in $\mathbb{C}^{+}$. By RH problem~\ref{rhp:M2-TIp}(d), the first column of $M^{(3)}$ is $\mathcal{O}(|k-c_r|^{1/4+\nu_r})$ there, so the new $M^{(\mathrm{err})}$ is $\mathcal{O}(|k-c_r|^{-1/4-\nu_r})$ near $c_r$; the same holds in $\mathbb{C}^{-}$. As $\frac14+\nu_r<1$ and there is no jump near $c_r$, the singularity is removable. With these changes, Propositions~\ref{prop:Verr-TIp} and \ref{prop:smallnorm-TIp} hold for all $\nu_r\in(-\frac12,\frac12)$.
\end{remark}

\subsection{Long-time asymptotics in $\mathcal{T}_{\mathrm{I}}^{+}$}\label{subsec:TIp-thm}

We now invert the transformations. For $k$ large and outside the lenses, \eqref{eq:Merr-TIp} gives $M^{(3)}(k)=M^{(\mathrm{err})}(k)M^{(\infty)}(k)$. By \eqref{eq:Minf-exp-TIp} and Proposition~\ref{prop:smallnorm-TIp}(b),
\begin{equation*}
\lim_{k\to\infty}k\big(M^{(3)}(k)\big)_{12}=\big(M^{(\mathrm{err})}_1\big)_{12}+\Big(\frac{c_l}{2}\sigma_2\Big)_{12}=\big(M^{(\mathrm{err})}_1\big)_{12}-\frac{ic_l}{2},
\end{equation*}
where the limit is taken along $i\mathbb{R}$, for instance. Together with \eqref{eq:q-M3-TIp}, this gives
\begin{equation}\label{eq:q-Merr-TIp}
q(x,t)=e^{-2i\alpha c_l^{2}t-2i\varphi_{\mathrm{I}}}\Big(c_l+2i\big(M^{(\mathrm{err})}_1\big)_{12}\Big).
\end{equation}
We evaluate the $(1,2)$-entries of the two terms in \eqref{eq:Merr1-explicit-TIp}.

\emph{The contribution of $c_l$.} The $(1,2)$-entry of $N=\frac12\left(\begin{smallmatrix}-i&1\\1&i\end{smallmatrix}\right)$ is $\frac12$. Hence the first term contributes $2i\cdot\frac{\kappa_ts^{2}}{8}t^{-1/3}=\frac{i\kappa_ts^{2}}{4}t^{-1/3}$ to the bracket in \eqref{eq:q-Merr-TIp}. By \eqref{eq:s-TIp} and the definition of $\kappa_t$ in \eqref{eq:F2-zeta-TIp}, we have exactly $s=\kappa_tt^{2/3}(\xi_{\mathrm{I}}^{+}-\xi)$. Moreover, $\kappa_t^{3}=\frac{2c_l}{3(4\beta c_l+\alpha)}(1+\mathcal{O}(\xi-\xi_{\mathrm{I}}^{+}))$. Therefore
\begin{equation}\label{eq:cl-term-TIp}
\frac{\kappa_ts^{2}}{4}t^{-\frac13}=\frac{\kappa_t^{3}}{4}\,t\,(\xi-\xi_{\mathrm{I}}^{+})^{2}=\frac{c_l\,(x-\xi_{\mathrm{I}}^{+}t)^{2}}{6(4\beta c_l+\alpha)\,t}\big(1+\mathcal{O}(t^{-2/3})\big).
\end{equation}

\emph{The contribution of $-\hat\eta_l$.} Let $R:=\sqrt{\hat\eta_l^{\,2}-c_l^{2}}$. Since $-\hat\eta_l<-c_l$, \eqref{eq:Minf-TIp} gives $\chi_l(-\hat\eta_l)=\big(\frac{\hat\eta_l+c_l}{\hat\eta_l-c_l}\big)^{1/4}>1$ and
\begin{equation*}
\left[M^{(\infty)}(-\hat\eta_l)\begin{pmatrix}0&y\\ \bar y&0\end{pmatrix}M^{(\infty)}(-\hat\eta_l)^{-1}\right]_{12}=a^{2}y+b^{2}\bar y,\qquad a^{2}=\frac{\hat\eta_l+R}{2R},\quad b^{2}=\frac{\hat\eta_l-R}{2R}.
\end{equation*}
Here $a=\frac12(\chi_l+\chi_l^{-1})$ and $b=\frac12(\chi_l-\chi_l^{-1})$ at $-\hat\eta_l$. By \eqref{eq:E0-TIp} and \eqref{eq:beta-PC-TIp}, $E_0M_1^{\mathrm{PC}}E_0^{-1}$ has this off-diagonal form with
\begin{equation*}
y=i\beta_{12}w_0^{2}e^{-2itg_{\mathrm{I}}(-\hat\eta_l)}=i\sqrt\nu\,e^{i\Theta},
\end{equation*}
where
\begin{equation}\label{eq:Theta-TIp}
\Theta=\Theta(x,t):=\frac{3\pi}{4}-\arg r(-\hat\eta_l)-\arg\Gamma(i\nu)-2d_{\mathrm{I},\hat\eta}+2\phi_{\hat\eta}+\nu\log\big(2t|g_{\mathrm{I}}''(-\hat\eta_l)|\big)-2tg_{\mathrm{I}}(-\hat\eta_l).
\end{equation}
Here we used $|\beta_{12}|^{2}=\nu$ and $w_0^{2}=e^{-2id_{\mathrm{I},\hat\eta}+2i\phi_{\hat\eta}}(2t|g''_{\mathrm{I}}(-\hat\eta_l)|)^{i\nu}$ from \eqref{eq:w-PC-TIp}. Since $X_l(-\hat\eta_l)=-R$, we have $g_{\mathrm{I}}(-\hat\eta_l)=-R\,p_{\mathrm{I}}(-\hat\eta_l)\in\mathbb{R}$. Equivalently, since $p_{\mathrm{I}}(k)=p_l(k)+\xi$ with $p_l(k)=4\beta k^{2}+2\alpha k+2\beta c_l^{2}$, we have $t\,g_{\mathrm{I}}(-\hat\eta_l)=-R\,\big(x+t\,p_l(-\hat\eta_l)\big)$.

Hence the second term in \eqref{eq:Merr1-explicit-TIp} contributes
\begin{equation}\label{eq:PC-term-TIp}
2i\,\frac{a^{2}y+b^{2}\bar y}{(2t|g''_{\mathrm{I}}(-\hat\eta_l)|)^{1/2}}=-\sqrt{\frac{2\nu}{t\,|g''_{\mathrm{I}}(-\hat\eta_l)|}}\;\Big(\frac{\hat\eta_l+R}{2R}e^{i\Theta}-\frac{\hat\eta_l-R}{2R}e^{-i\Theta}\Big).
\end{equation}

We have proved the main result of this section.

\appendix
\section{The Painlev\'e XXXIV parametrix}\label{app:P34}
The Painlev\'e XXXIV parametrix obeys the RH problem as below:
\begin{rhp}\label{rp:P34}
Find a $2\times2$ matrix-valued function $M^{P34}(\zeta)=M^{P34}(\zeta;b,\omega,s)$ with the
following properties.
\begin{itemize}
\item[\rm(a)] $M^{P34}$ is analytic in $\C\setminus\big(\Sigma_1\cup\Sigma_2\cup\Sigma_3\cup\Sigma_4\cup\{0\}\big)$, where
\begin{equation}\label{eq:P34-Sigma}
\Sigma_1=\R^{+},\qquad \Sigma_2=\e^{\frac{2\pi i}{3}}\R^{+},\qquad
\Sigma_3=\e^{\pi i}\R^{+},\qquad \Sigma_4=\e^{-\frac{2\pi i}{3}}\R^{+},
\end{equation}
with $\Sigma_1$ oriented away from $0$ and $\Sigma_2,\Sigma_3,\Sigma_4$ oriented towards $0$.
\item[\rm(b)] $M^{P34}$ has continuous boundary values on $\Sigma_j\setminus\{0\}$ and
$M^{P34}_+=M^{P34}_-V^{P34}_j$ on $\Sigma_j$, where
\begin{equation}\label{eq:P34-jumps}
V^{P34}_1=\begin{pmatrix}1&\omega\\0&1\end{pmatrix},\qquad
V^{P34}_2=\begin{pmatrix}1&0\\ \e^{2b\pi i}&1\end{pmatrix},\qquad
V^{P34}_3=\begin{pmatrix}0&1\\-1&0\end{pmatrix},\qquad
V^{P34}_4=\begin{pmatrix}1&0\\ \e^{-2b\pi i}&1\end{pmatrix}.
\end{equation}
\item[\rm(c)] As $\zeta\to\infty$, there exists a function $a(s)=a(s;b,\omega)$ such that
\begin{equation}\label{eq:A1}
M^{P34}(\zeta)=\begin{pmatrix}1&0\\-ia(s)&1\end{pmatrix}
\Big(I+\frac{M_1^{P34}(s)}{\zeta}+\cO(\zeta^{-2})\Big)\frac{\zeta^{-\frac14\sigma_3}}{\sqrt2}
\begin{pmatrix}1&i\\ i&1\end{pmatrix}\e^{-\varphi(\zeta;s)\sigma_3},
\end{equation}
where $\varphi(\zeta;s)=\tfrac23\zeta^{3/2}+s\zeta^{1/2}$.
with principal branches.
\item[\rm(d)] $M^{P34}$ is bounded near $0$ if $b\ge0$, and $M^{P34}(\zeta)=\cO(\zeta^{b})$ as
$\zeta\to0$ if $-\frac12<b<0$.
\end{itemize}
\end{rhp}

\begin{figure}[htbp]
\centering
\begin{tikzpicture}[scale=2.05,
    ray/.style={line width=1.0pt, draw=black!85},
    ax/.style={line width=0.5pt, draw=black!35, densely dotted},
    dom/.style={font=\small, text=black!70},
    lab/.style={font=\footnotesize, anchor=center, inner sep=2pt}]

  \fill[black!7] (0,0) -- (2.30,0) arc (0:120:2.30) -- cycle;

  \draw[ax] (-2.55,0) -- (2.55,0);
  \draw[ax] (0,-2.55) -- (0,2.55);

  \draw[ray, -{Stealth[length=2.2mm]}] (0,0) -- (2.30,0);                 
  \draw[ray, {Stealth[length=2.2mm]}-] (120:2.28) -- (0,0);               
  \draw[ray, {Stealth[length=2.2mm]}-] (-2.20,0) -- (0,0);                
  \draw[ray, {Stealth[length=2.2mm]}-] (-120:2.28) -- (0,0);              

  \node[lab, anchor=north west] at (0.55,-0.04) {$\Sigma_1$};
  \node[lab, anchor=north east] at (120:1.30) {$\Sigma_2$};
  \node[lab, anchor=north] at (-1.20,0.02) {$\Sigma_3$};
  \node[lab, anchor=south east] at (-120:1.30) {$\Sigma_4$};

  \node[lab, anchor=west] at (-0.10,0.10) {$0$};
  \node[lab, anchor=north] at (1.55,-0.12) {$\left(\begin{smallmatrix}1&\omega\\0&1\end{smallmatrix}\right)$};
  \node[lab, anchor=east] at (-0.50,1.34) {$\left(\begin{smallmatrix}1&0\\ \e^{2b\pi i}&1\end{smallmatrix}\right)$};
  \node[lab, anchor=north] at (-1.30,-0.12) {$\left(\begin{smallmatrix}0&1\\-1&0\end{smallmatrix}\right)$};
  \node[lab, anchor=east] at (-0.50,-1.34) {$\left(\begin{smallmatrix}1&0\\ \e^{-2b\pi i}&1\end{smallmatrix}\right)$};
  \node[dom] at (1.02,1.18) {$\Omega_1$};
  \node[dom] at (-1.30,0.86) {$\Omega_2$};
  \node[dom] at (-1.30,-0.86) {$\Omega_3$};
  \node[dom] at (1.02,-1.18) {$\Omega_4$};
\end{tikzpicture}
\caption{The jump contour of RH problem~\ref{rp:P34}.}
\label{fig:P34}
\end{figure}
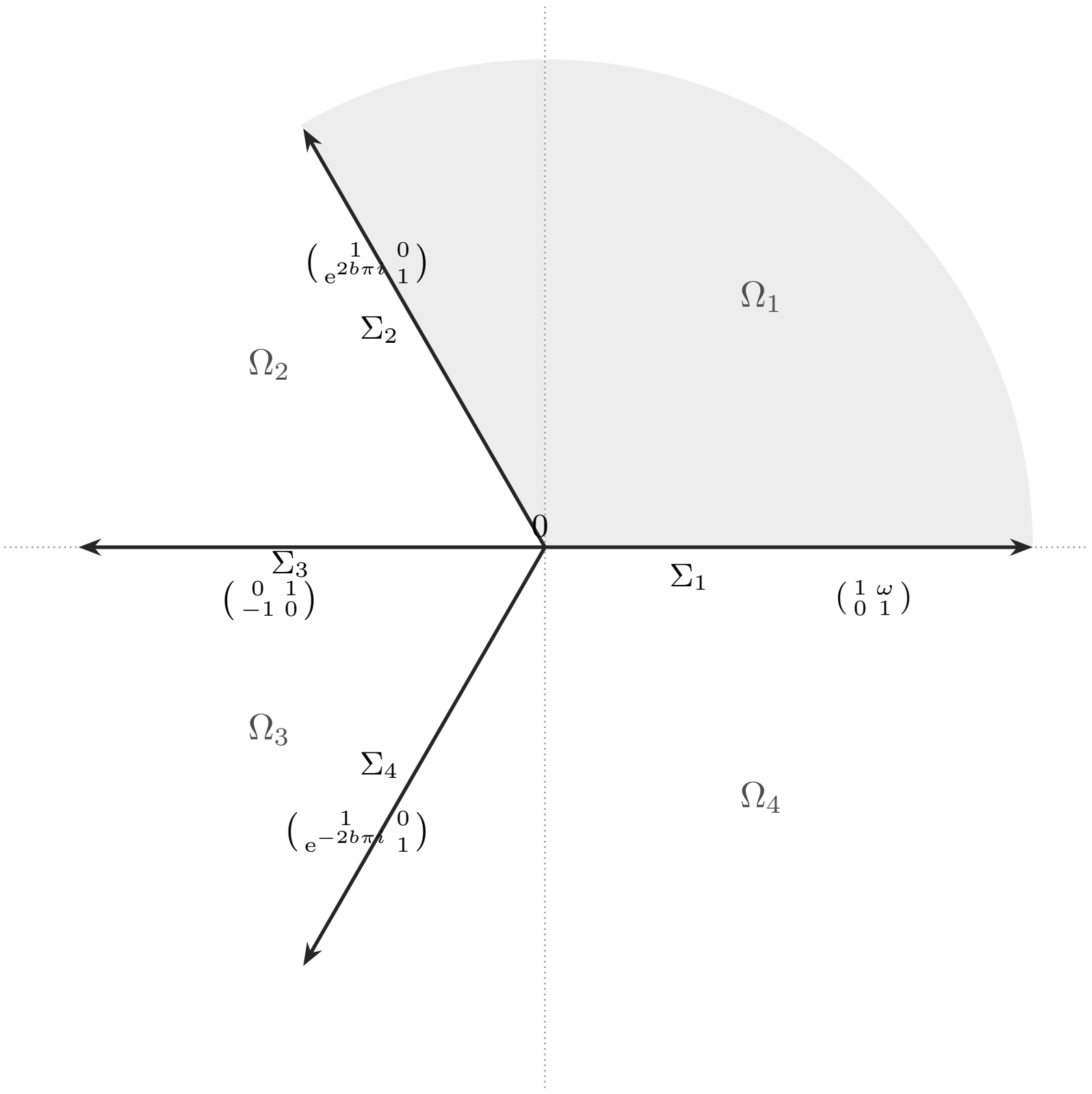

The contour of RH problem~\ref{rp:P34} is shown in Figure~\ref{fig:P34}.

\begin{proposition}\label{prop:P34-solv}
Let $b>-\frac12$, $\omega\in\C\setminus(-\infty,0)$ and $s\in\R$. Then RH problem~\ref{rp:P34}
is uniquely solvable. Moreover,
\begin{gather}
\det M^{P34}\equiv-1,\label{eq:P34-det}\\
M^{P34}(\zeta;b,\omega,s)=\sigma_1\overline{M^{P34}(\bar\zeta;b,\omega,s)}\sigma_1,\qquad
s\in\R,\ \omega\in\R,\label{eq:P34-sym}
\end{gather}
and the function
\begin{equation}\label{eq:P34-u}
u(s)=u(s;b,\omega):=a'(s;b,\omega)-\frac s2
\end{equation}
is real for real $s$ and solves the Painlev\'e XXXIV equation
\begin{equation}\label{eq:P34-P34}
u''(s)=4u(s)^{2}+2su(s)+\frac{u'(s)^{2}-(2b)^{2}}{2u(s)} .
\end{equation}
\end{proposition}

In particular, for $\omega=0$ one has
\begin{equation}\label{eq:P34-u-asym}
u(s;b,0)=\begin{cases}\dfrac{b}{\sqrt s}+\cO(s^{-2}),& s\to+\infty,\\[2mm]
-\dfrac s2+\cO(s^{-2}),& s\to-\infty,\end{cases}
\end{equation}
and, since $a(s;b,0)\to0$ as $s\to-\infty$,
\begin{equation}\label{eq:A3}
a(s;b,0)=\int_{-\infty}^{s}\Big(u(z;b,0)+\frac z2\Big)\di z .
\end{equation}

\begin{lemma}\label{lem:M1}
Let $q=q(s)$ be the solution of the Painlev\'e II equation
\begin{equation}\label{eq:P34-PII}
q''(s)=sq(s)+2q(s)^{3}-\nu,\qquad \nu=2b+\tfrac12,
\end{equation}
characterized by the asymptotic conditions
\begin{equation}\label{eq:P34-q-asym}
q(s)=\begin{cases}\dfrac{\nu}{s}+\cO(s^{-3}),& s\to+\infty,\\[2mm]
\sqrt{-\dfrac s2}+\cO(s^{-1}),& s\to-\infty,\end{cases}
\end{equation}
and let $H(s):=\big(q'(s)\big)^{2}-sq(s)^{2}-q(s)^{4}+2\nu q(s)$ be the corresponding
Hamiltonian, so that $H'=-q^{2}$. Then, with $x:=-2^{1/3}s$,
\begin{align}
\big(M_1^{P34}(s)\big)_{12}&=i\,2^{-\frac23}\big(H-q\big)(x)=:ia(s),\label{eq:M1-12}\\
\big(M_1^{P34}(s)\big)_{11}&=2^{-\frac73}\big(H^{2}-q^{2}\big)(x)-2^{-\frac43}\big(q'+qH\big)(x),\label{eq:M1-11}\\
\big(M_1^{P34}(s)\big)_{22}&=-2^{-\frac73}\big(H^{2}-q^{2}\big)(x)+2^{-\frac43}\big(q'+qH\big)(x).\label{eq:M1-22}
\end{align}
In particular,
\begin{equation}\label{eq:M1-diff}
\operatorname{tr}M_1^{P34}=0,\qquad
\big(M_1^{P34}\big)_{11}-\big(M_1^{P34}\big)_{22}=a(s)^{2}-a'(s),\qquad a'(s)=u(s)+\frac s2 .
\end{equation}
\end{lemma}

\begin{lemma}\label{lem:uniform}
Fix $b\in(-\frac12,\frac12)$ and $\delta\in(0,\frac{\pi}{12}]$. There exist $s_0>0$ and $C>0$ such
that, for every $s$ in the sector
\begin{equation*}
\Omega_{\delta,s_0}:=\{s\in\C:\ |s|\ge s_0,\ |\arg(-s)|\le\delta\},
\end{equation*}
RH problem~\ref{rp:P34} with $\omega=0$ is uniquely solvable, and
\begin{equation}\label{eq:uniform}
\begin{pmatrix}1&0\\ ia(s)&1\end{pmatrix}M^{P34}(\zeta;b,0,s)
=\big(I+\mathcal E(\zeta;s)\big)N(\zeta)\e^{-\varphi(\zeta;s)\sigma_3},\qquad
|\mathcal E(\zeta;s)|\le\frac{C}{|s|\,(1+|\zeta|)},
\end{equation}
where
\begin{equation}\label{eq:P34-N}
N(\zeta):=\zeta^{-\frac14\sigma_3}U,\qquad U:=\frac{1}{\sqrt2}\begin{pmatrix}1&i\\ i&1\end{pmatrix},
\end{equation}
uniformly for $s\in\Omega_{\delta,s_0}$ and $|\zeta|\ge1$, $\zeta\notin\Sigma_2\cup\Sigma_3\cup\Sigma_4$
(with continuous boundary values on the rays). In particular,
\begin{equation}\label{eq:uniform-M1}
M_1^{P34}(s)=\cO(|s|^{-1}),\qquad s\to-\infty,\ s\in\Omega_{\delta,s_0} .
\end{equation}
\end{lemma}

\begin{proof}
Since $\omega=0$, the jump on $\Sigma_1=\R^{+}$ is the identity and RH problem~\ref{rp:P34}
reduces to a three-ray problem with vertex at $0$. We perform the standard normalization of the
exponentials. Write $s=-|s|\e^{i\psi}$ with $|\psi|\le\delta$. For $\zeta\in\Sigma_2\cup\Sigma_4$
we have $\arg\zeta=\pm\frac{2\pi}{3}$ and hence
\begin{equation}\label{eq:Re-phi}
\operatorname{Re}\varphi(\zeta;s)=\operatorname{Re}\Big(\tfrac23\zeta^{\frac32}+s\zeta^{\frac12}\Big)
\le-\tfrac23|\zeta|^{\frac32}+\tfrac12|s||\zeta|^{\frac12}\le-c_0|s||\zeta|^{\frac12},
\end{equation}
with $c_0>0$, for $|\zeta|\le\frac34|s|$; here we used $|\zeta|^{\frac32}\ge\frac34|s||\zeta|^{\frac12}$
and $\operatorname{Re}\big(\zeta^{\frac32}\big)\le-\frac23|\zeta|^{\frac32}$ on $\Sigma_2\cup\Sigma_4$.
Let $\Phi:=M^{P34}\e^{\varphi\sigma_3}N^{-1}$. Then $\Phi\to I$ at $\zeta=\infty$. In the disk $|\zeta|\le\frac12|s|^{-1}$ we use the
Bessel parametrix built from the Bessel model RH problem of order $2b>-1$ \cite{KMVV04} in the
local variable
\begin{equation}\label{eq:P34-w}
w(\zeta;s):=\tfrac14\zeta\Big(s+\tfrac23\zeta\Big)^{2},\qquad 2w^{1/2}=-\varphi(\zeta;s),
\end{equation}
which is conformal in $\{|\zeta|<\frac12\}$ for $|s|\ge s_0\gg1$, with $w'(0)=s^{2}/4\neq0$, and
satisfies $|w|\ge c|s|^{2}$ on $|\zeta|=\frac12$. The matching on $|\zeta|=\frac12$ is then $I+\cO(|s|^{-1})$,
and the standard small norm argument \cite{DZ02} yields unique solvability and \eqref{eq:uniform}
for $|\zeta|\ge1$. Since all estimates depend only on \eqref{eq:Re-phi} and on the conformality of
$w(\cdot\,;s)$ near $\zeta=0$, they are uniform for $s$ in the sector $\Omega_{\delta,s_0}$. Finally, expanding
\eqref{eq:uniform} at $\zeta=\infty$ gives \eqref{eq:uniform-M1}.
\end{proof}

We next record the special case $b=0$, $\omega=1$, which corresponds to the soft edge $c_l$ in
Section~\ref{sec:TIp}. By \eqref{eq:P34-PII}, $\nu=\frac12$; the corresponding solution of the
Painlev\'e II equation is given explicitly by a logarithmic derivative of the Airy function,
\begin{equation}\label{eq:P34-q-Airy}
q(s)=-2^{-\frac13}\,\frac{\mathrm{Ai}'\big(-2^{-\frac13}s\big)}{\mathrm{Ai}\big(-2^{-\frac13}s\big)},
\end{equation}
and, by the asymptotics of $\mathrm{Ai}$ and $\mathrm{Ai}'$,
\begin{equation}\label{eq:P34-q-Airy-asym}
q(s)\sim\frac{1}{2}\sqrt{2}\,(-s)^{\frac12},\qquad s\to-\infty .
\end{equation}

\begin{corollary}\label{cor:P34-Airy}
Let $b=0$ and $\omega=1$. Then RH problem~\ref{rp:P34} is solved by
\begin{equation}\label{eq:P34-Airy-sol}
M^{P34}(\zeta;0,1,s)=\begin{pmatrix}\sqrt{2\pi}&0\\0&-i\sqrt{2\pi}\end{pmatrix}
\times\begin{cases}
\Phi_1(\zeta+s)\,\sigma_1, & 0<\arg\zeta<\frac{2\pi}{3},\\[1mm]
\Phi_2(\zeta+s)\,\sigma_1, & \frac{2\pi}{3}<\arg\zeta<\pi,\\[1mm]
\Phi_2(\zeta+s)\,\sigma_3, & -\pi<\arg\zeta<-\frac{2\pi}{3},\\[1mm]
\Phi_3(\zeta+s)\,\sigma_3, & -\frac{2\pi}{3}<\arg\zeta<0,
\end{cases}
\end{equation}
where $\varpi:=\e^{2\pi i/3}$ and
\begin{equation*}
\begin{gathered}
\Phi_1(z)=\begin{pmatrix}-\varpi^{2}\mathrm{Ai}(\varpi^{2}z)&\mathrm{Ai}(z)\\-\varpi\,\mathrm{Ai}'(\varpi^{2}z)&\mathrm{Ai}'(z)\end{pmatrix},\qquad
\Phi_2(z)=\begin{pmatrix}-\varpi^{2}\mathrm{Ai}(\varpi^{2}z)&-\varpi\,\mathrm{Ai}(\varpi z)\\-\varpi\,\mathrm{Ai}'(\varpi^{2}z)&-\varpi^{2}\mathrm{Ai}'(\varpi z)\end{pmatrix},\\
\Phi_3(z)=\begin{pmatrix}\mathrm{Ai}(z)&-\varpi\,\mathrm{Ai}(\varpi z)\\\mathrm{Ai}'(z)&-\varpi^{2}\mathrm{Ai}'(\varpi z)\end{pmatrix}.
\end{gathered}
\end{equation*}
Moreover, \eqref{eq:A1} holds with the coefficients
\begin{equation}\label{eq:M1-P34-12}
\big(M_1^{P34}(s)\big)_{12}=\frac{i\,s^{2}}{4},
\end{equation}
\begin{equation}\label{eq:M1-P34-11}
\big(M_1^{P34}(s)\big)_{11}=\frac{s^{4}}{32}-\frac s4,
\end{equation}
\begin{equation}\label{eq:M1-P34-22}
\big(M_1^{P34}(s)\big)_{22}=-\frac{s^{4}}{32}+\frac s4 .
\end{equation}
\end{corollary}

\end{document}